\documentclass[sigplan,screen,authorversion=true]{acmart}
\pdfoutput=1
\setcopyright{acmlicensed}
\acmPrice{15.00}
\acmDOI{10.1145/3358504.3361232}
\acmYear{2019}
\copyrightyear{2019}
\acmISBN{978-1-4503-6987-9/19/10}
\acmConference[VMIL '19]{Proceedings of the 11th ACM SIGPLAN International Workshop on Virtual Machines and Intermediate Languages}{October 22, 2019}{Athens, Greece}
\acmBooktitle{Proceedings of the 11th ACM SIGPLAN International Workshop on Virtual Machines and Intermediate Languages (VMIL '19), October 22, 2019, Athens, Greece}

\def\AWFY{Are\,We\,Fast\,Yet\xspace}
\newcommand{\code}[1]{\texttt{#1}}
\newcommand{\ie}{i.e.\xspace}
\newcommand{\eg}{e.g.\xspace}

\usepackage{listings}
\usepackage{xspace}

\definecolor{codegreen}{rgb}{0,0.6,0}
\definecolor{codegray}{rgb}{0.5,0.5,0.5}
\definecolor{codepurple}{rgb}{0.58,0,0.82}
\definecolor{codered}{rgb}{0.82,0.15,0.23}
\definecolor{backcolour}{rgb}{0.95,0.95,0.92}

\lstdefinestyle{mystyle}{
    language=SAS,
    breakatwhitespace=true,
    breaklines=true,
    captionpos=b,
    keepspaces=true,
    numbers=left,
    numbersep=5pt,
    showspaces=false,
    showstringspaces=false,
    showtabs=false,
    tabsize=2,
    aboveskip=20pt,
    belowskip=10pt,
    xleftmargin=0.5cm,
    basicstyle=\footnotesize\ttfamily,
    commentstyle=\itshape\color{codegray},
    numberstyle=\footnotesize\color{codegray},
    stringstyle=\color{codegreen},
    keywordstyle = {\color{codepurple}},
    keywordstyle = [2]{\color{codered}},
    keywordstyle = [3]{\color{codered}},
    keywordstyle = [4]{\color{codegreen}},
    keywords={method,object,interface,type,var,def,return,class,for,in,if},
    otherkeywords = {:,->},
    morekeywords = [2]{:},
    morekeywords = [3]{->},
    morekeywords = [4]{true,false},
    morestring=*[d]{"},
    backgroundcolor={}
}

\usepackage{pgfplots}
\pgfplotsset{compat=1.16}
\colorlet{GraphNormal}{ACMOrange} %

\colorlet{GraphBase}{gray} %
\colorlet{GraphA}{ACMRed}
\colorlet{GraphB}{ACMBlue}
\colorlet{GraphAB}{ACMPurple}

\pgfplotsset{%
	every axis/.append style = {%
		cycle list={}, tick align=center, xtick pos = bottom, ymin=0, axis on top,%
		axis y line*=left, try min ticks=5,%
		every tick label/.append style={font=\tiny},%
		every axis label/.append style={font=\tiny},%
	},%
	Full/.style = { %
		enlarge x limits={abs=1.5pt}, ymax={#1}, height=3.2cm, width=5.9cm,%
		every axis plot/.style={very thin, mark size=0.75pt, draw=GraphNormal, fill=GraphNormal!40}},%
	Right Axis/.style = { %
		axis y line*=right, axis x line=none, xmin=0, xmax=100},%
	Scatter/.style = { %
		xticklabel={$\pgfmathprintnumber{\tick}$\%},%
		every axis plot/.append style={only marks}, mark=o},%
	Scatter Left Label/.style   = {ylabel={Average Time (ms)}},%
	Scatter Bottom Label/.style = {xlabel={Proportion of Type Annotations}},%
	Scatter Right Label/.style  = {ylabel={Relative to 0\% Typed}},%
	Column/.style = {%
		enlarge x limits=false, xtick distance={#1}, const plot,%
		every axis plot/.append style={no markers}},%
	Column Left Label/.style   = {ylabel={Average Time (ms)}},%
	Column Bottom Label/.style = {xlabel={Type Annotation Index}},%
	Column Right Label/.style  = {ylabel={Relative to Untyped}},%
	Pattern/.style = { %
		enlarge x limits={abs=1pt}, ymax={#1}, height=3.2cm, width=4.4cm, %
		every axis plot/.style={very thin, mark size=0.5pt, draw=GraphBase, fill=GraphBase!40}},%
	Pattern A/.style = {draw=GraphA, fill=GraphA!40, mark=*}, %
	Pattern B/.style = {draw=GraphB, fill=GraphB!40, mark=*}, %
	Pattern AB/.style = {draw=GraphAB, fill=GraphAB!40, mark=*}, %
}

\newcommand\DOUBLEAXIS[3]{%
	\begin{axis}[#1]#3\end{axis}%
	\begin{axis}[#2]\end{axis}}

\newcommand\GRAPHS[1]{\null\hfill\begin{tikzpicture}[ampersand replacement=\&]\matrix{#1\\};\end{tikzpicture}\null\hfill}
\begin{document}
\title{Which of My Transient Type Checks Are Not (Almost) Free?}

\author{Isaac Oscar Gariano}
\affiliation{
  \department{Engineering and Computer Science} %
  \institution{Victoria University of Wellington}
  \country{New Zealand}
}
\email{Isaac@ecs.vuw.ac.nz} %

\author{Richard Roberts}
\affiliation{
  \department{Computational Media Innovation Centre} %
  \institution{Victoria University of Wellington}
  \country{New Zealand}
}
\email{rykardo.r@gmail.com} %

\author{Stefan Marr}
\orcid{0000-0001-9059-5180}
\affiliation{
  \department{School of Computing} %
  \institution{University of Kent}
  \country{United Kingdom}
}
\email{s.marr@kent.ac.uk} %

\author{Michael Homer}
\affiliation{
  \department{Engineering and Computer Science} %
  \institution{Victoria University of Wellington}
  \country{New Zealand}
}
\email{mwh@ecs.vuw.ac.nz} %

\author{James Noble}
\orcid{0000-0001-9036-5692}             %
\affiliation{
  \department{Engineering and Computer Science} %
  \institution{Victoria University of Wellington}
  \country{New Zealand}
}
\email{kjx@ecs.vuw.ac.nz} %

\begin{abstract}
One form of type checking used in gradually typed language is
\emph{transient type checking}: whenever an object `flows' through
code with a type annotation, the object is dynamically checked to
ensure it has the methods required by the annotation. 
Just-in-time compilation and optimisation in virtual machines can eliminate much of the overhead of run-time transient type checks. 
Unfortunately this optimisation is not uniform: some type checks will significantly decrease, or even increase, a program's performance.

In this paper, we refine the so called ``Takikawa'' protocol, and use it to identify which type annotations have the greatest effects on performance. In particular, we show how graphing the performance of such benchmarks when varying which type annotations are present in the source code can be used to discern potential patterns in performance. We demonstrate our approach by testing the Moth virtual machine: for many of the benchmarks where Moth's transient type checking impacts performance, we have been able to identify one or two specific type annotations that are the likely cause. Without these type annotations, the performance impact of transient type checking becomes negligible.

Using our technique programmers can optimise programs by removing expensive type checks, and VM engineers can identify new opportunities for compiler optimisation.\pagebreak
\end{abstract}


\maketitle

\section{Introduction}
Gradual typing aims to add static type annotations to dynamic languages, increasing their safety while maintaining flexibility \citep{GiladPluggable2004,Siek2006,XXXSiek2015}, and/or, permitting dynamic type annotations within static languages, increasing flexibility whilst maintaining some safety \citep{AbadiTOPLAS1991}.

There is a vast spectrum of different approaches to gradual typing
\cite{kafka18,bensurvey18icfp}. Here we measure the performance of
``transient'' or ``type-tag'' checks (as in Reticulated Python), which
offer first-order semantics: they check that an object's type
constructor or names of supported methods match any static types that
the object flows through, but not the return types or argument types of those methods \cite{Siek2007,Bloom2009,concrete15,reticPython2014,Greenman2018}.

Unfortunately most gradual systems with run-time semantics, as opposed to type erasure as in TypeScript, impose significant run-time performance overheads. This has lead to a significant body of research to develop techniques to optimise gradual typing \citep{Vitousek2017,Muehlboeck2017,Bauman2017,Richards2017,Greenman2018}.

This has also lead to a technique to evaluate the performance of gradual typing, the ``Takikawa'' protocol \cite{Takikawa2016,Greenman2019jfp,vitousek-transient-arXive-2019}. The Takikawa protocol was created to measure the run-time cost of gradual typing by testing various configurations of typed and untyped code. This approach was designed to characterise how the amount of typed and untyped code influences performance.
In particular, Takikawa protocol evaluations often show that simply adding more type annotations does not always produce a uniform effect on performance. Here we adapt their approach in order to identify individual type annotations that may be responsible for performance effects.

This paper builds upon our recent work on optimising transient type checks \cite{Roberts2017,roberts-and-co-ecoop-2019} to make the following contributions:
\begin{itemize}
  \item an approach to identifying individual gradual type annotations
        that cause significant performance effects
  \item an observation that the overhead of Moth's transient type checking on small benchmarks (with 10--250 type annotations) is usually caused by only one or two type annotations
\end{itemize}

The next section discusses dynamic type checks and gradual typing in
Moth (an implementation of the Grace language), then section~\ref{s-meth} describes our benchmarking protocol. Section~\ref{s-overall} then presents the overall results of
our benchmarks, while section~\ref{s-individual}
looks at the results of benchmarking individual type checks.
Section~\ref{s-rel} presents some additional related work, and finally \ref{s-concl} summarises our results, and briefly considers threats to validity.

\section{Background}
\label{s-bg}

Our work is based on the Moth virtual machine 
\cite{Roberts2017,roberts-and-co-ecoop-2019},
an implementation
of the Grace programming language 
\citep{graceOnward12,graceSigcse13}.
Moth is based on the Graal and Truffle toolchain
\cite{Wurthinger:2017:PPE,Wurthinger2013},
and developed from a Newspeak implementation based on the  Simple
Object Machine \cite{Daloze2016,SOMns}.

\subsection{Grace and Transient Type Checking}

Grace is an object-oriented, imperative, educational programming
language, with a focus on introductory programming
courses, but also intended for more advanced study and research \citep{graceOnward12,graceSigcse13}.
While Grace's syntax draws
from the so-called ``curly bracket'' traditions of C, Java, and
JavaScript, the structure of the language
is in many ways closer to Smalltalk:
all computation is done via dynamically dispatched  ``method requests''
where the object receiving the request decides what code to run,
control structures are built out of lambda expressions support ``non-local'' returns, i.e. they can return to the point where execution first encountered the lambda \citep{bluebook}.  In
other ways, Grace is closer to JavaScript than Smalltalk: Grace
objects are created from object literals, rather than by
instantiating classes \citep{Black2007-emeraldHOPL,JonesECOOP2016} and
objects and classes can be deeply nested within each 
other \citep{betabook}.

\paragraph{Grace's Typing}
In Grace, all declarations can be annotated with types.
As Grace is designed to support a variety of teaching methods, implementation of Grace are free to check such type annotations  statically, dynamically, or not at all.
The type system of Grace is intrinsically gradual:%
~type annotations should not affect the semantics of a correct
program \citep{XXXSiek2015}. The type system
includes a distinguished ``{Unknown}'' type which matches any other type; this unknown type is the default when type annotations are omitted.

Static typing for the core of Grace's type system has been described
elsewhere \citep{TimJonesThesis};
here we explain how
these types can be understood 
dynamically, from the Grace programmer's point of view.
Grace's types are structural \citep{graceOnward12},
that is, an object conforms to a type whenever it conforms to the ''structural`` requirements of a type,
rather than requiring classes or objects to explicitly declare their intended type.

In Grace, types specify a set of method signatures that an object must provide. A type expresses the requests an object can respond to, for example whether a particular accessor is available, rather than a location in a class hierarchy.

\paragraph{Moth's Transient Type Checking}
Moth's implementation of transient type checks are only only first-order.
Moth only checks dynamically that an object has methods of the same name and arity as are required by a type:  any argument and return types of such methods are not checked.

In particular, Moth performs the following type checks at run time:
\begin{itemize}
\item when a method is requested, arguments that are passed are checked against the corresponding parameter type annotations of the called method, this is done before the body of the method is executed;
\item when the body of a method has finished executing, but before it returns to its caller, the method's return value is checked against the return type annotation of the called method;
\item whenever a variable is read or written to, its value is checked against the type specified by the variables declaration.
\end{itemize}

To see how this works in practice, consider this piece of Grace code:

\begin{minipage}{\linewidth}
\begin{lstlisting}
def o = object {
   method three -> Number {3}
}
type ThreeString = interface {
   three -> String
}
def t : ThreeString = o
printNumber (t.three)
\end{lstlisting}
\end{minipage}

Moth will perform dynamic type checks:

\begin{itemize}

\item on line 7,
when the \code{o} object initialises the variable \code{t},
Moth checks that \code{o} has a 0-argument method called \code{three};

\item on line 8,
when the value of \code{t} is read,
Moth checks that its value (\code{o}) still has a \code{three} method;

\item on line 2,
when the method requested by ``\code{t.three}'' returns,
Moth checks that returned value conforms to the \code{Number} type;
and (presumably) within the definition of
\code{printNumber(n :   Number)}
(not shown), Moth will again check that the value is a \code{Number}.
\end{itemize}

Note that we never check
whether the result of requesting ``\code{t.three}'' is actually
a \code{String} (as one may expect from line 5) because Moth only performs first-order type checks
(it checks whether objects have conforming methods) not higher-order
checks (whether the argument and result types of methods' conform). In addition, Moth
only checks when values flow through explicit type annotations.
This is why the type declared in lines 4-6 is checked only on line 7
(where it is mentioned explicitly); and the check only requires the
presence of a method called \code{three}, regardless of the method's
declared return type.

\paragraph{Moth's Optimisation}
We are developing Moth as a
research platform \cite{roberts-and-co-ecoop-2019}. Like other VMs
based on the Truffle and Graal toolchain, Moth is a self-optimising
AST interpreter \cite{Wurthinger:2012:SelfOptAST}. 
The key idea is that an AST rewrites itself based on a program's run time values
to reflect the minimal set of operations needed to execute the program
correctly. The rewritten AST is then compiled into efficient machine
code. This rewriting often depends on the dynamic types of the
objects involved. In the simplest case, a ``self'' call (when one method
on an object requests a second method on the exact same object) will
always result in executing the exact same method. Thus the called method can be inlined into
the callee, avoiding overhead of a machine-level subroutine
invocation and an object-oriented dynamic dispatch.

Moth relies on a number of standard techniques for optimising
object-oriented programs.
``Shapes'' \citep{woss2014object} capture information about objects'
structures and (run time) 
field types, allowing a just-in-time compiler to
represent objects in memory similarly to C structs and, consequently,
can generate highly efficient code.
``Polymorphic inline caches''
\citep{Hoelzle:91:PIC} use object shapes to cache the results of
method lookups, avoiding expensive class hierarchy searches or
indirect jumps through virtual method tables. 
Since Moth is built on the Truffle framework,
Graal comes with  additional support for partial evaluation,
which enables efficient native code generation for
Truffle interpreters \citep{Wurthinger:2017:PPE}.

\section{Experimental Methodology}
\label{s-meth}

Our goal is to identify which type annotations in Grace programs
cause performance effects.
To this end, we built upon the so-called ``Takikawa'' or ``Takikawa-Greenman'' evaluation protocol \cite{Takikawa2016,Greenman2019jfp}.
It uses $2^N$ \emph{configurations} of each benchmark.
A configuration is a particular mix of static and dynamically typed code, forming a lattice of configurations.
We only test a relatively small sample of this lattice,
which is in our experience sufficient to pinpoint performance anomalies caused
by type annotations.

\subsection{The Takikawa Protocol}
The Takikawa evaluation protocol was originally proposed for Typed
Racket, where static vs dynamic typing is set per-module, so $N$ is
the number of modules. The original Takikawa protocol also suggested a sampling strategy, where $10N$ configurations are randomly chosen from the lattice, however the lattice is a binomial distribution, meaning the majority of chosen benchmarks will have around $N/2$ type annotations.

Grace allows programmers to choose whether each individual declaration should be type-checked, and thus follows languages such as
Reticulated Python \cite{reticPython2014,monotonic2015,Vitousek2017}.
This means in Grace $N$ is the number of type annotations in the program, so it is
infeasible to check an entire lattice, even for a moderately sized benchmark.
Vitousek et al.\ therefore modified the Takikawa protocol for
these kinds of languages by using a different form of sampling
\cite{vitousek-transient-arXive-2019}.  The Takikawa-Vitousek protocol
divides the number of type annotations in a fully-typed program into
a maximum of 100 intervals, and then randomly generates ten programs within
each interval by erasing type annotations.
However, this approach was designed for benchmarks with large numbers of type annotations, as well as for a larger sample size than our work.

\paragraph{Refined Takikawa Protocol.}
Unlike prior work, we wish to identify which type annotations cause anomalies,
and thus we adapted the Takikawa protocol
and took inspiration from the Takikawa-Vitousek variant.
For each benchmark, we generated 100 partially typed versions,
or fewer if the benchmark has less than 11 types.
We did an even split so that for each $i \ge 1$ and $i$ < $N$, we generated roughly the same number of configurations with $i$ type annotations. We used Robert Floyd's sampling algorithm \cite{Bentley:1987:PPS:30401.315746} to randomly choose the type annotations each configuration contained, and we ensured that no duplicate configurations were generated. In addition to these, we tested fully untyped and typed versions, for a total of 102 configurations per benchmark (or 97 in the case of our Storage benchmark, since it only has 10 type annotations).

\subsection{The Benchmarks}

For this work, we rely on the benchmark suite compiled for previous work
\cite{roberts-and-co-ecoop-2019}.
It is a collection of 21 benchmarks in total,
derived from the \AWFY
benchmark suite \cite{Marr2016} and other benchmarks
from the gradual-typing literature.

We added type annotations to every benchmark, aiming to use the most appropriate (\ie most specific) annotation for each declaration.

In our previous work, we \cite{roberts-and-co-ecoop-2019}
we determined that the overhead of type checking on Moth is on average
of 5\% (min.\ -13\%, max. 79\%).
This compares the peak performance of Moth with all
checks disabled against an execution that has all checks enabled. 

\subsection{Experimental Set Up}
To account for the complex warmup behaviour of modern systems \citep{Barrett:2017:VMW} as well as
the non-determinism caused by \eg garbage collection and cache effects, we ran each benchmark for 1000 iterations in the same invocation of Moth, and discard the first 350 iterations to ignore warmup JIT compilation. Our previous work  identified this as a suitable cut off \cite{roberts-and-co-ecoop-2019}.

Though outliers remain visible in the plots for each individual benchmark, the largest 95\% confidence interval we obtained (over the mean time after warmup) for any of experiments was $\pm8.3\%$ (for the PyStone benchmark).

All our experiments used the same machine, Graal, and Moth as previously; the machine has two Intel Xeon E5-2620 v3 2.40GHz, with 6 cores each, for a total of 24 hyperthreads.
The machine was running Ubuntu Linux 16.04.6, with Kernel 4.4, and we used ReBench 1.0 \citep{ReBench:2018} and Java 1.8.0\_191 Graal 0.43. Benchmarks were executed one by one to avoid interference between them. The analysis of the results and plots where generated using Python 3.7.3 and PGFPLOTS 1.16. To enable reproductions, the scripts we used to generate and run our experiments, including the source code for all the configurations tested, are available online.\footnote{\url{https://gitlab.ecs.vuw.ac.nz/isaac/Moth-Takikawa}}

In our previous work \cite{roberts-and-co-ecoop-2019}, we also compared the performance of untyped code on Moth against state-of-the-art VMs: Java, Node.js using the V8 JavaScript VM, and the Higgs JavaScript VM. Java was the fastest of these, and on average V8 was about 1.8x slower than Java, Moth was 2.3x slower, and Higgs was 10.4x slower.
We believe this makes Moth suitable for assessing the impact of type checking,
because Moth's performance is close enough to state-of-the-art VMs,
which should make it harder to hide type checking overheads in a slow baseline.

\section{Performance of Benchmark Configurations}
\label{s-overall}

\begin{figure*}
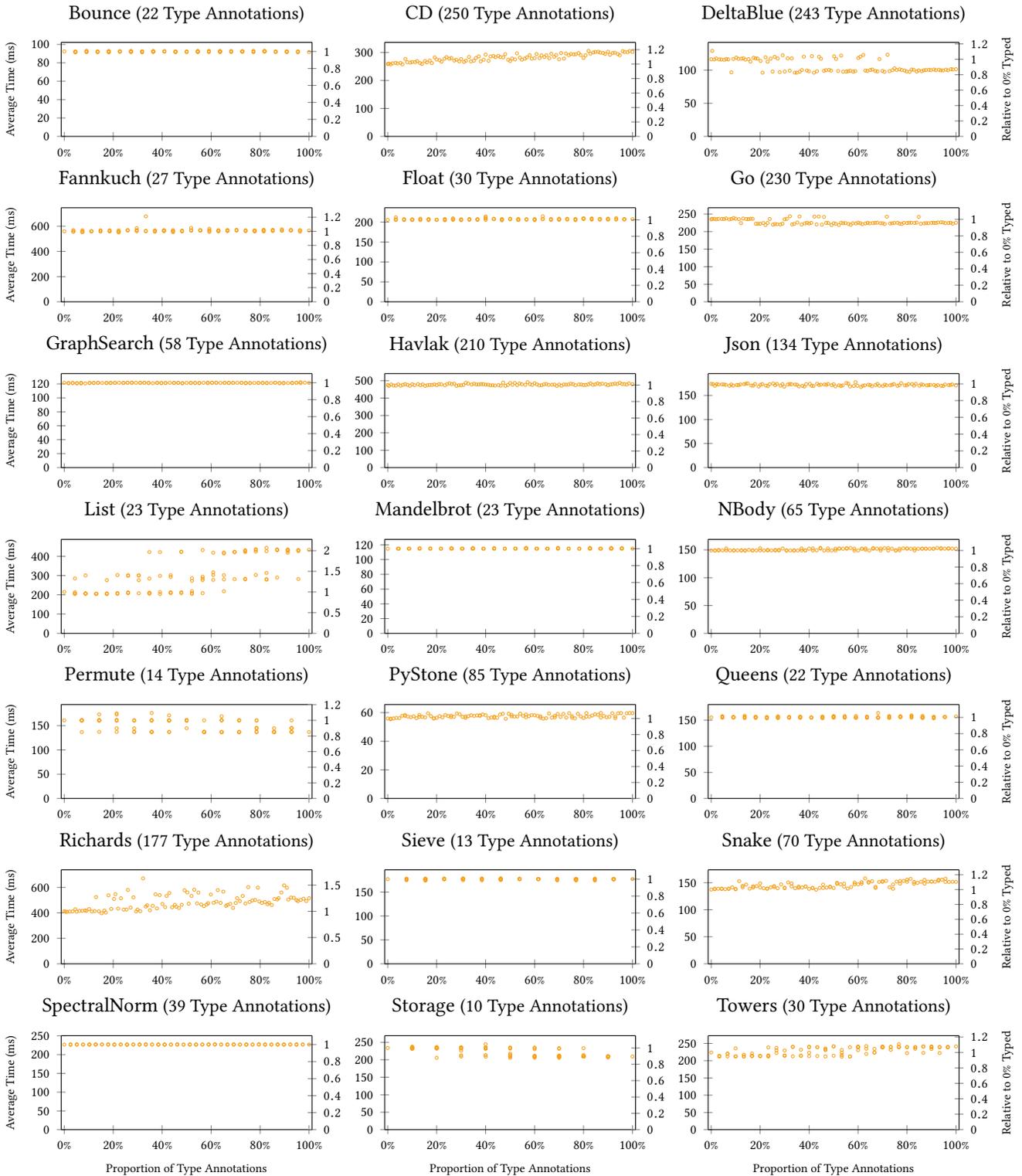

\GRAPHS{
\DOUBLEAXIS{Full = {102.16147446153846}, Scatter, Scatter Left Label, title={Bounce {\small(22 Type Annotations)}}}{Full = {1.1060173075452897}, Scatter, Right Axis}{
	\addplot+[] coordinates{
		(22.727272727272727, 92.50315846153846)
		(9.090909090909092, 92.5867923076923)
		(50.0, 92.22278)
		(18.181818181818183, 91.72483384615384)
		(81.81818181818183, 92.83337692307693)
		(68.18181818181817, 92.29010153846154)
		(4.545454545454546, 91.79665384615384)
		(27.27272727272727, 92.53775538461538)
		(9.090909090909092, 92.13658153846154)
		(77.27272727272727, 92.24595076923077)
		(72.72727272727273, 92.44826)
		(40.909090909090914, 92.65180615384615)
		(81.81818181818183, 92.87406769230769)
		(40.909090909090914, 92.37356307692308)
		(4.545454545454546, 91.30154153846154)
		(90.9090909090909, 91.69153076923077)
		(90.9090909090909, 92.46175538461539)
		(50.0, 91.94288769230769)
		(18.181818181818183, 92.64567538461539)
		(4.545454545454546, 92.1652676923077)
		(13.636363636363635, 92.0306476923077)
		(45.45454545454545, 91.70456307692308)
		(86.36363636363636, 91.63611846153846)
		(68.18181818181817, 92.74865538461539)
		(81.81818181818183, 92.35940923076923)
		(59.09090909090909, 92.61192923076923)
		(27.27272727272727, 91.86559538461539)
		(95.45454545454545, 91.92962923076924)
		(63.63636363636363, 92.3422)
		(63.63636363636363, 91.51009692307693)
		(95.45454545454545, 92.42988153846154)
		(18.181818181818183, 92.34711538461538)
		(50.0, 91.65564153846154)
		(31.818181818181817, 92.42853846153847)
		(36.36363636363637, 92.48369846153847)
		(86.36363636363636, 92.17964461538462)
		(77.27272727272727, 92.86742153846154)
		(36.36363636363637, 91.97473384615385)
		(72.72727272727273, 92.35785076923077)
		(54.54545454545454, 92.73940461538461)
		(77.27272727272727, 91.80059076923077)
		(72.72727272727273, 92.37158923076923)
		(90.9090909090909, 91.73021230769231)
		(86.36363636363636, 92.1904876923077)
		(54.54545454545454, 92.53796615384616)
		(27.27272727272727, 91.78174307692308)
		(86.36363636363636, 92.09407538461538)
		(68.18181818181817, 91.91345846153847)
		(31.818181818181817, 92.36450923076923)
		(0.0, 92.36878461538461)
		(13.636363636363635, 91.55165692307692)
		(22.727272727272727, 92.46737384615385)
		(72.72727272727273, 92.45159230769231)
		(40.909090909090914, 92.11349846153846)
		(45.45454545454545, 92.24439846153847)
		(18.181818181818183, 92.58076461538461)
		(95.45454545454545, 92.15198)
		(63.63636363636363, 92.0794923076923)
		(9.090909090909092, 91.55657384615385)
		(95.45454545454545, 91.45464923076923)
		(81.81818181818183, 92.05032153846153)
		(54.54545454545454, 91.7360476923077)
		(50.0, 92.14829846153846)
		(4.545454545454546, 92.33144769230769)
		(9.090909090909092, 92.75685846153846)
		(45.45454545454545, 91.91870307692308)
		(63.63636363636363, 92.36040923076924)
		(59.09090909090909, 91.64267846153847)
		(68.18181818181817, 91.91452)
		(100.0, 91.67891538461538)
		(36.36363636363637, 91.76644)
		(31.818181818181817, 91.43001384615384)
		(54.54545454545454, 92.12309692307693)
		(77.27272727272727, 91.84244615384615)
		(9.090909090909092, 92.48079538461539)
		(63.63636363636363, 92.84179692307693)
		(72.72727272727273, 91.87307076923076)
		(13.636363636363635, 91.98735384615385)
		(31.818181818181817, 92.20850923076924)
		(90.9090909090909, 91.99286153846154)
		(59.09090909090909, 92.62479692307693)
		(13.636363636363635, 92.45101384615384)
		(40.909090909090914, 92.20646)
		(18.181818181818183, 91.40344923076923)
		(50.0, 92.20810461538461)
		(13.636363636363635, 91.85647538461538)
		(45.45454545454545, 92.37612461538461)
		(68.18181818181817, 92.1461123076923)
		(59.09090909090909, 92.52936923076923)
		(4.545454545454546, 91.65451538461538)
		(27.27272727272727, 92.7852923076923)
		(31.818181818181817, 91.95004923076922)
		(45.45454545454545, 91.58694769230769)
		(36.36363636363637, 92.55119692307693)
		(86.36363636363636, 92.3482876923077)
		(22.727272727272727, 92.31151076923076)
		(90.9090909090909, 92.17395692307693)
		(54.54545454545454, 91.74789692307692)
		(22.727272727272727, 91.9176476923077)
		(81.81818181818183, 92.32270615384616)
		(36.36363636363637, 92.6147276923077)
		(27.27272727272727, 91.90881384615385)
	};

}
\&
\DOUBLEAXIS{Full = {336.1811403076924}, Scatter, title={CD {\small(250 Type Annotations)}}}{Full = {1.3002110091965295}, Scatter, Right Axis}{
	\addplot+[] coordinates{
		(43.2, 287.6336830769231)
		(31.2, 266.2493323076923)
		(41.199999999999996, 270.38674)
		(44.0, 283.2179061538462)
		(13.200000000000001, 262.11080615384617)
		(64.0, 283.7796276923077)
		(8.0, 265.36509538461536)
		(96.0, 304.83217692307693)
		(50.0, 269.28614153846155)
		(86.0, 302.5626092307692)
		(68.0, 284.52114615384613)
		(40.0, 292.44052153846155)
		(32.0, 270.6832830769231)
		(9.2, 264.0376276923077)
		(23.200000000000003, 277.6082492307692)
		(19.2, 282.4954723076923)
		(0.4, 260.32499384615386)
		(96.8, 301.8342)
		(27.200000000000003, 271.03388)
		(94.8, 290.4944123076923)
		(24.0, 283.25242000000003)
		(26.0, 280.4902446153846)
		(68.8, 290.98762923076924)
		(74.0, 290.59594)
		(92.80000000000001, 299.2719476923077)
		(36.0, 280.2516753846154)
		(6.0, 257.55894461538463)
		(74.8, 294.5701876923077)
		(39.2, 274.09852)
		(17.2, 271.1908661538462)
		(98.8, 305.2704876923077)
		(47.199999999999996, 282.68546)
		(48.0, 293.9522630769231)
		(38.0, 268.4655492307692)
		(7.199999999999999, 276.86243538461537)
		(82.8, 295.8310707692308)
		(98.0, 301.25852153846154)
		(54.0, 281.56948153846156)
		(28.000000000000004, 274.2415384615385)
		(62.0, 279.9813323076923)
		(22.0, 267.38671384615384)
		(66.0, 279.49088461538463)
		(72.0, 278.2168938461538)
		(46.0, 283.8635615384615)
		(94.0, 294.4267246153846)
		(58.8, 278.15867692307694)
		(18.0, 262.2581553846154)
		(72.8, 296.2695076923077)
		(78.8, 290.94370153846154)
		(12.0, 269.1274846153846)
		(35.199999999999996, 270.49714153846156)
		(66.8, 295.6016907692308)
		(57.99999999999999, 290.69096615384615)
		(88.8, 294.95501538461536)
		(30.0, 276.62688615384616)
		(45.2, 288.01250923076924)
		(84.8, 302.1272923076923)
		(90.0, 302.78504153846154)
		(78.0, 297.94097384615384)
		(90.8, 292.0766184615385)
		(37.2, 286.171)
		(52.800000000000004, 296.95732)
		(34.0, 268.33237846153844)
		(14.000000000000002, 259.02849692307694)
		(0.0, 258.55890923076925)
		(70.8, 297.7896723076923)
		(25.2, 278.0338446153846)
		(62.8, 289.86449846153846)
		(56.8, 282.2968076923077)
		(11.200000000000001, 265.8217584615385)
		(88.0, 298.08496)
		(70.0, 275.90223692307694)
		(5.2, 262.79028923076925)
		(3.2, 262.8688230769231)
		(60.0, 279.6458)
		(33.2, 283.93842615384614)
		(64.8, 290.33412)
		(80.0, 280.71709384615383)
		(60.8, 294.4023569230769)
		(21.2, 272.35324153846153)
		(86.8, 299.77858615384616)
		(76.8, 293.8412369230769)
		(42.0, 286.10308615384616)
		(15.2, 269.46322)
		(92.0, 298.38041076923076)
		(54.800000000000004, 274.41198)
		(29.2, 268.41496461538463)
		(56.00000000000001, 278.07138615384616)
		(2.0, 261.20972)
		(76.0, 293.4293784615385)
		(80.80000000000001, 290.08761384615383)
		(52.0, 275.33246153846153)
		(82.0, 305.6192184615385)
		(1.2, 257.4973830769231)
		(4.0, 260.95384615384614)
		(49.2, 274.2494723076923)
		(84.0, 300.23830615384617)
		(100.0, 302.12799384615386)
		(20.0, 274.7902107692308)
		(10.0, 265.90955692307693)
		(16.0, 264.49719230769233)
		(50.8, 278.43575076923076)
	};

}
\&
\DOUBLEAXIS{Full = {141.963998}, Scatter, title={DeltaBlue {\small(243 Type Annotations)}}}{Full = {1.2190709449214199}, Scatter, Right Axis, Scatter Right Label}{
	\addplot+[] coordinates{
		(74.8971193415638, 102.79937538461539)
		(51.85185185185185, 98.77684)
		(36.21399176954733, 97.7412476923077)
		(60.08230452674898, 118.35273384615385)
		(34.97942386831276, 96.24038461538461)
		(15.22633744855967, 116.68366)
		(81.06995884773663, 97.43668307692307)
		(9.053497942386832, 116.8360676923077)
		(3.292181069958848, 116.24710615384615)
		(87.65432098765432, 100.17044)
		(24.279835390946502, 117.42726307692308)
		(82.71604938271605, 100.61789846153846)
		(55.96707818930041, 98.38092307692308)
		(98.76543209876543, 101.01187692307693)
		(37.86008230452675, 120.4092676923077)
		(4.11522633744856, 116.04187538461538)
		(65.02057613168725, 100.26148153846154)
		(1.2345679012345678, 116.21091846153845)
		(55.144032921810705, 97.86580615384615)
		(20.16460905349794, 113.94469384615384)
		(63.78600823045267, 98.84524)
		(96.70781893004116, 99.71338153846153)
		(43.20987654320987, 98.36232615384615)
		(6.995884773662551, 116.20899692307692)
		(44.03292181069959, 120.24432)
		(13.168724279835391, 117.88778615384615)
		(5.349794238683128, 115.64011846153846)
		(29.218106995884774, 122.61527230769231)
		(18.106995884773664, 117.83492)
		(16.049382716049383, 112.95672307692308)
		(13.991769547325102, 115.96123230769231)
		(89.7119341563786, 100.00995692307693)
		(30.04115226337449, 97.88481076923077)
		(8.23045267489712, 96.70032615384615)
		(70.78189300411523, 97.49270769230769)
		(46.913580246913575, 99.95576153846154)
		(2.05761316872428, 117.56572153846153)
		(77.77777777777779, 100.27120923076923)
		(31.275720164609055, 117.8457323076923)
		(84.77366255144034, 99.67058)
		(88.88888888888889, 100.48995692307692)
		(81.89300411522635, 100.9443323076923)
		(23.045267489711936, 112.8896)
		(91.76954732510289, 100.00746615384615)
		(79.83539094650206, 98.08506461538461)
		(97.94238683127571, 101.46202615384615)
		(69.95884773662551, 98.82914769230769)
		(72.8395061728395, 98.60330153846154)
		(86.0082304526749, 98.42036769230769)
		(86.83127572016461, 99.74354307692307)
		(41.9753086419753, 96.1938876923077)
		(67.90123456790124, 98.11441692307692)
		(58.0246913580247, 97.80864769230769)
		(12.345679012345679, 117.35646923076924)
		(68.72427983539094, 116.75647538461538)
		(93.82716049382715, 99.72120153846154)
		(44.8559670781893, 117.20996461538462)
		(50.20576131687243, 121.00435076923077)
		(46.09053497942387, 98.90593076923076)
		(75.7201646090535, 100.33826923076923)
		(83.9506172839506, 97.88067846153847)
		(39.09465020576132, 99.50878307692308)
		(62.96296296296296, 99.41400307692308)
		(18.930041152263374, 117.54514153846154)
		(53.90946502057613, 97.73607384615384)
		(0.411522633744856, 129.05818)
		(72.0164609053498, 123.35539692307692)
		(93.00411522633745, 100.62641538461538)
		(39.91769547325103, 98.46678461538461)
		(41.1522633744856, 121.17123538461539)
		(27.983539094650205, 97.0733923076923)
		(0.0, 116.45261384615385)
		(60.90534979423868, 120.45114615384615)
		(25.102880658436217, 97.87683538461539)
		(67.07818930041152, 98.66171230769231)
		(51.028806584362144, 116.96270923076924)
		(74.07407407407408, 99.99728461538461)
		(37.03703703703704, 100.16418615384616)
		(22.22222222222222, 119.40240615384616)
		(58.8477366255144, 98.54415076923077)
		(94.65020576131687, 99.56739384615385)
		(79.01234567901234, 99.42199538461539)
		(10.2880658436214, 118.9864)
		(90.94650205761316, 100.99156307692307)
		(6.172839506172839, 116.8913276923077)
		(62.139917695473244, 122.98940461538461)
		(53.086419753086425, 122.03403230769231)
		(48.971193415637856, 99.28898615384615)
		(32.92181069958848, 117.86491846153847)
		(34.15637860082305, 96.27592307692308)
		(17.28395061728395, 118.69502461538461)
		(65.84362139917695, 100.93916307692308)
		(20.98765432098765, 96.25279846153846)
		(11.11111111111111, 117.14130769230769)
		(95.88477366255144, 100.80978615384615)
		(25.925925925925924, 120.04929692307692)
		(48.148148148148145, 100.4532)
		(56.79012345679012, 99.00476461538462)
		(27.160493827160494, 117.62664615384615)
		(76.95473251028807, 101.22075538461539)
		(32.098765432098766, 99.20771692307693)
		(100.0, 101.4503723076923)
	};

}
\\
\DOUBLEAXIS{Full = {746.5459963076923}, Scatter, Scatter Left Label, title={Fannkuch {\small(27 Type Annotations)}}}{Full = {1.333979249325812}, Scatter, Right Axis}{
	\addplot+[] coordinates{
		(33.33333333333333, 562.5478723076923)
		(11.11111111111111, 560.7467476923077)
		(59.25925925925925, 567.9427876923077)
		(62.96296296296296, 559.8181723076923)
		(22.22222222222222, 550.7505984615385)
		(62.96296296296296, 561.1061169230769)
		(51.85185185185185, 586.6807446153846)
		(96.29629629629629, 568.483103076923)
		(48.148148148148145, 559.24876)
		(70.37037037037037, 568.5381153846154)
		(74.07407407407408, 560.4505292307692)
		(33.33333333333333, 561.7925046153846)
		(51.85185185185185, 566.5944461538462)
		(0.0, 559.6383876923077)
		(44.44444444444444, 552.6590430769231)
		(11.11111111111111, 560.969756923077)
		(74.07407407407408, 566.0547153846154)
		(62.96296296296296, 563.97616)
		(18.51851851851852, 560.2420123076923)
		(85.18518518518519, 568.8415107692308)
		(85.18518518518519, 567.0249338461539)
		(33.33333333333333, 678.6781784615384)
		(40.74074074074074, 559.9933338461539)
		(40.74074074074074, 568.8272184615385)
		(85.18518518518519, 570.5369230769231)
		(66.66666666666666, 565.0244246153846)
		(7.4074074074074066, 553.8153338461539)
		(92.5925925925926, 565.8834338461538)
		(70.37037037037037, 569.753976923077)
		(81.48148148148148, 558.4385061538461)
		(92.5925925925926, 567.70704)
		(44.44444444444444, 568.7137661538461)
		(74.07407407407408, 567.603383076923)
		(66.66666666666666, 567.1327907692307)
		(59.25925925925925, 580.04086)
		(48.148148148148145, 565.8750646153846)
		(37.03703703703704, 556.3353584615385)
		(96.29629629629629, 566.8859461538461)
		(3.7037037037037033, 569.4189953846154)
		(81.48148148148148, 565.9369969230769)
		(25.925925925925924, 568.3671953846153)
		(3.7037037037037033, 556.4456015384616)
		(37.03703703703704, 560.9114076923076)
		(85.18518518518519, 561.8913523076923)
		(25.925925925925924, 566.8584538461538)
		(3.7037037037037033, 558.5150692307692)
		(48.148148148148145, 560.4192953846153)
		(59.25925925925925, 561.2265046153847)
		(18.51851851851852, 562.2615846153847)
		(62.96296296296296, 568.4773107692308)
		(55.55555555555556, 567.3785630769231)
		(33.33333333333333, 560.9299369230769)
		(88.88888888888889, 575.5292246153846)
		(22.22222222222222, 555.2616646153846)
		(77.77777777777779, 567.02466)
		(81.48148148148148, 567.3332123076923)
		(66.66666666666666, 569.4687984615384)
		(55.55555555555556, 569.1453015384616)
		(29.629629629629626, 567.7458138461539)
		(88.88888888888889, 566.954616923077)
		(44.44444444444444, 560.8693923076924)
		(88.88888888888889, 565.0222630769231)
		(66.66666666666666, 560.9895092307693)
		(59.25925925925925, 555.3154461538461)
		(29.629629629629626, 584.8821646153846)
		(96.29629629629629, 556.7471553846153)
		(18.51851851851852, 568.1499846153846)
		(81.48148148148148, 571.93924)
		(88.88888888888889, 566.0443061538462)
		(7.4074074074074066, 560.8887215384615)
		(77.77777777777779, 558.7697661538461)
		(7.4074074074074066, 552.3626753846154)
		(29.629629629629626, 568.2363323076924)
		(77.77777777777779, 565.7075184615385)
		(40.74074074074074, 564.9062046153846)
		(11.11111111111111, 558.2049415384615)
		(51.85185185185185, 568.641383076923)
		(37.03703703703704, 563.8243061538461)
		(92.5925925925926, 569.4632415384615)
		(51.85185185185185, 565.0400984615385)
		(3.7037037037037033, 559.1660338461538)
		(14.814814814814813, 559.7719923076924)
		(22.22222222222222, 567.0576153846154)
		(55.55555555555556, 566.0495661538462)
		(14.814814814814813, 559.7368938461539)
		(7.4074074074074066, 567.7332661538461)
		(55.55555555555556, 565.9189615384615)
		(92.5925925925926, 569.4176430769231)
		(11.11111111111111, 560.4443707692308)
		(100.0, 566.2207676923077)
		(40.74074074074074, 565.9712307692307)
		(18.51851851851852, 559.6528461538461)
		(44.44444444444444, 561.7516476923076)
		(70.37037037037037, 569.5718461538462)
		(14.814814814814813, 568.4336446153847)
		(37.03703703703704, 566.7041138461539)
		(77.77777777777779, 560.2538784615384)
		(25.925925925925924, 568.23122)
		(70.37037037037037, 566.559203076923)
		(29.629629629629626, 560.0896738461538)
		(14.814814814814813, 559.1624246153846)
		(22.22222222222222, 559.0915676923076)
	};

}
\&
\DOUBLEAXIS{Full = {235.77577353846158}, Scatter, title={Float {\small(30 Type Annotations)}}}{Full = {1.1478900052736314}, Scatter, Right Axis}{
	\addplot+[] coordinates{
		(13.333333333333334, 208.38691076923078)
		(66.66666666666666, 207.56433692307692)
		(23.333333333333332, 206.61682769230768)
		(46.666666666666664, 204.2462153846154)
		(86.66666666666667, 207.82839846153846)
		(33.33333333333333, 205.59759076923078)
		(53.333333333333336, 205.28947846153847)
		(3.3333333333333335, 211.80416461538462)
		(6.666666666666667, 205.19015384615383)
		(23.333333333333332, 207.46678)
		(53.333333333333336, 205.4621476923077)
		(66.66666666666666, 205.50731076923077)
		(63.33333333333333, 208.18058615384615)
		(50.0, 208.08298153846152)
		(6.666666666666667, 207.31165076923077)
		(96.66666666666667, 207.54590615384615)
		(80.0, 207.28970923076923)
		(90.0, 206.48116769230768)
		(90.0, 206.35482307692308)
		(100.0, 207.02298307692308)
		(70.0, 206.31562153846153)
		(96.66666666666667, 206.19989384615386)
		(63.33333333333333, 214.34161230769232)
		(40.0, 205.44146923076923)
		(3.3333333333333335, 205.45548615384615)
		(36.666666666666664, 205.19838923076924)
		(73.33333333333333, 206.49426153846153)
		(50.0, 206.7805953846154)
		(60.0, 205.83950153846155)
		(76.66666666666667, 204.7997523076923)
		(63.33333333333333, 206.8901676923077)
		(83.33333333333334, 206.9219646153846)
		(90.0, 208.60411384615384)
		(50.0, 207.94476)
		(83.33333333333334, 207.6514523076923)
		(10.0, 206.37606153846153)
		(30.0, 206.91107692307693)
		(0.0, 205.39927384615385)
		(93.33333333333333, 206.64258307692307)
		(10.0, 205.89902307692307)
		(56.666666666666664, 207.06618153846154)
		(33.33333333333333, 206.94646)
		(70.0, 206.59335384615386)
		(46.666666666666664, 207.03076153846155)
		(26.666666666666668, 205.41030615384616)
		(6.666666666666667, 205.66680769230769)
		(16.666666666666664, 206.83882461538462)
		(23.333333333333332, 205.46131692307694)
		(73.33333333333333, 208.28817692307692)
		(40.0, 210.95396461538462)
		(3.3333333333333335, 205.68119384615386)
		(70.0, 208.40413230769232)
		(30.0, 205.66301692307692)
		(20.0, 205.26703230769232)
		(76.66666666666667, 207.39510153846155)
		(56.666666666666664, 206.17463846153845)
		(93.33333333333333, 206.49019692307692)
		(16.666666666666664, 207.10979230769232)
		(56.666666666666664, 206.41390615384614)
		(36.666666666666664, 206.87996307692308)
		(20.0, 205.6168076923077)
		(83.33333333333334, 206.2345523076923)
		(16.666666666666664, 204.7775323076923)
		(36.666666666666664, 207.19907076923076)
		(23.333333333333332, 204.46936)
		(60.0, 204.96222461538463)
		(80.0, 208.94253692307691)
		(16.666666666666664, 206.73112153846154)
		(3.3333333333333335, 206.8956969230769)
		(10.0, 205.12643076923078)
		(53.333333333333336, 207.86056923076924)
		(93.33333333333333, 209.85441384615385)
		(53.333333333333336, 205.41668923076924)
		(86.66666666666667, 207.46494307692308)
		(80.0, 209.67568307692306)
		(13.333333333333334, 206.57753692307693)
		(60.0, 206.97980923076923)
		(83.33333333333334, 209.01205692307693)
		(66.66666666666666, 206.51365384615386)
		(40.0, 213.91392)
		(43.333333333333336, 208.26980461538463)
		(43.333333333333336, 207.6039353846154)
		(13.333333333333334, 204.9162876923077)
		(46.666666666666664, 206.0468753846154)
		(33.33333333333333, 207.70658615384616)
		(46.666666666666664, 207.4696846153846)
		(73.33333333333333, 206.58227846153846)
		(30.0, 205.18153846153845)
		(30.0, 204.73488307692307)
		(70.0, 205.83247846153847)
		(40.0, 208.52000307692308)
		(86.66666666666667, 206.03441384615385)
		(10.0, 206.38428153846155)
		(26.666666666666668, 206.96754153846155)
		(76.66666666666667, 206.44443076923076)
		(96.66666666666667, 207.24155076923077)
		(90.0, 207.53179076923078)
		(43.333333333333336, 207.08910461538463)
		(60.0, 207.16342923076922)
		(20.0, 205.80636615384614)
		(76.66666666666667, 206.48951538461537)
		(26.666666666666668, 209.1494923076923)
	};

}
\&
\DOUBLEAXIS{Full = {267.975460923077}, Scatter, title={Go {\small(230 Type Annotations)}}}{Full = {1.1403910642505535}, Scatter, Right Axis, Scatter Right Label}{
	\addplot+[] coordinates{
		(98.69565217391305, 222.84408461538462)
		(5.217391304347826, 235.86017692307692)
		(46.95652173913044, 224.79925846153847)
		(100.0, 225.30320615384616)
		(23.91304347826087, 236.05612153846153)
		(40.869565217391305, 224.95568923076922)
		(12.173913043478262, 224.07898923076922)
		(93.91304347826087, 226.74027846153845)
		(42.173913043478265, 242.67882)
		(32.17391304347826, 243.6140553846154)
		(23.043478260869566, 224.7242169230769)
		(52.17391304347826, 218.8292153846154)
		(96.95652173913044, 224.5267723076923)
		(77.82608695652173, 221.80032153846153)
		(36.08695652173913, 243.04395384615384)
		(3.0434782608695654, 236.27922923076923)
		(76.95652173913044, 222.79824461538462)
		(43.913043478260875, 242.9471523076923)
		(78.69565217391305, 225.05610153846155)
		(38.26086956521739, 223.37118)
		(9.130434782608695, 238.63282153846154)
		(6.086956521739131, 237.62395846153845)
		(50.0, 223.77532461538462)
		(49.130434782608695, 218.7477323076923)
		(74.78260869565217, 224.6892969230769)
		(90.8695652173913, 225.25820153846155)
		(20.0, 222.25060307692308)
		(14.347826086956522, 235.75727076923076)
		(1.3043478260869565, 236.12947384615384)
		(80.8695652173913, 225.48148615384616)
		(70.0, 223.85701076923078)
		(73.91304347826086, 223.7416323076923)
		(33.91304347826087, 221.95980153846153)
		(61.73913043478261, 221.25410923076922)
		(39.130434782608695, 225.27353692307693)
		(53.04347826086957, 225.3541153846154)
		(87.82608695652175, 224.17702)
		(81.73913043478261, 225.42404307692308)
		(28.26086956521739, 220.4592953846154)
		(63.91304347826087, 224.82604615384616)
		(29.130434782608695, 222.4939123076923)
		(83.04347826086956, 224.75323076923078)
		(40.0, 223.17001384615384)
		(88.69565217391305, 224.03955384615384)
		(50.8695652173913, 222.3208753846154)
		(53.91304347826087, 223.88956307692308)
		(56.086956521739125, 221.89136923076924)
		(0.0, 234.98558461538462)
		(91.73913043478261, 224.75514)
		(90.0, 225.0864723076923)
		(21.304347826086957, 224.52041076923078)
		(11.304347826086957, 235.50840461538462)
		(10.0, 236.74045384615386)
		(66.08695652173913, 224.30836153846153)
		(84.78260869565217, 242.00618461538463)
		(76.08695652173914, 225.25789076923076)
		(2.1739130434782608, 235.1877123076923)
		(54.78260869565217, 224.2157923076923)
		(71.73913043478261, 242.5008553846154)
		(60.0, 225.49804153846154)
		(0.43478260869565216, 236.22696923076924)
		(30.0, 235.20751076923077)
		(63.04347826086957, 223.35454615384614)
		(36.95652173913043, 220.1883523076923)
		(33.04347826086956, 220.54861076923078)
		(86.95652173913044, 223.56521230769232)
		(4.3478260869565215, 236.2188323076923)
		(67.82608695652173, 225.48838615384616)
		(46.08695652173913, 241.89788153846155)
		(59.130434782608695, 223.21814307692307)
		(85.65217391304348, 222.55481230769232)
		(13.043478260869565, 237.3012646153846)
		(57.826086956521735, 223.8374476923077)
		(22.17391304347826, 218.91295692307693)
		(56.95652173913044, 219.44589692307693)
		(45.21739130434783, 219.87266615384615)
		(60.86956521739131, 223.84949230769232)
		(19.130434782608695, 222.27543384615385)
		(66.95652173913044, 225.12337076923077)
		(94.78260869565217, 226.03764)
		(47.82608695652174, 224.68816615384614)
		(25.217391304347824, 220.0284276923077)
		(18.26086956521739, 222.47198153846153)
		(83.91304347826087, 225.64304)
		(64.78260869565217, 223.6128276923077)
		(16.956521739130434, 236.42621384615384)
		(70.86956521739131, 221.65042)
		(95.65217391304348, 224.3283953846154)
		(16.08695652173913, 236.76494461538462)
		(97.82608695652173, 225.22459384615385)
		(26.08695652173913, 221.65232923076923)
		(69.1304347826087, 223.90974)
		(30.869565217391305, 236.26694923076923)
		(35.21739130434783, 226.05219076923078)
		(15.217391304347828, 235.57470153846154)
		(80.0, 224.5386630769231)
		(7.391304347826087, 235.00985384615385)
		(8.26086956521739, 236.12082615384617)
		(92.6086956521739, 226.14022615384616)
		(43.04347826086957, 220.54209384615385)
		(26.956521739130434, 221.11829538461538)
		(73.04347826086956, 223.19428)
	};

}
\\
\DOUBLEAXIS{Full = {134.70518430769232}, Scatter, Scatter Left Label, title={GraphSearch {\small(58 Type Annotations)}}}{Full = {1.1078076844467712}, Scatter, Right Axis}{
	\addplot+[] coordinates{
		(5.172413793103448, 121.12530307692307)
		(86.20689655172413, 121.71748)
		(82.75862068965517, 121.23325538461539)
		(81.03448275862068, 121.49617846153846)
		(29.310344827586203, 121.6468276923077)
		(84.48275862068965, 121.31679230769231)
		(6.896551724137931, 121.59647230769231)
		(87.93103448275862, 121.15588461538462)
		(15.517241379310345, 121.28068461538462)
		(12.068965517241379, 121.34382307692307)
		(68.96551724137932, 121.52742)
		(79.3103448275862, 121.35873076923077)
		(43.103448275862064, 121.63098307692307)
		(53.44827586206896, 121.22320615384615)
		(6.896551724137931, 120.70822307692308)
		(96.55172413793103, 122.45925846153847)
		(25.862068965517242, 121.51848923076923)
		(87.93103448275862, 121.18307692307692)
		(81.03448275862068, 121.17972307692308)
		(72.41379310344827, 121.57305230769231)
		(96.55172413793103, 121.15590461538461)
		(98.27586206896551, 121.87624461538462)
		(20.689655172413794, 121.60125384615385)
		(58.620689655172406, 121.86501384615384)
		(94.82758620689656, 121.52657538461538)
		(53.44827586206896, 121.41052461538462)
		(93.10344827586206, 121.59723076923076)
		(24.137931034482758, 121.2263876923077)
		(39.6551724137931, 121.27630307692307)
		(75.86206896551724, 121.19793692307692)
		(5.172413793103448, 120.76778769230769)
		(0.0, 121.59618153846154)
		(34.48275862068966, 121.20703692307693)
		(89.65517241379311, 121.04582307692307)
		(12.068965517241379, 121.38799538461538)
		(20.689655172413794, 121.51600923076923)
		(36.206896551724135, 120.98754615384615)
		(31.03448275862069, 121.32126769230769)
		(62.06896551724138, 121.85304615384615)
		(25.862068965517242, 121.79435692307692)
		(77.58620689655173, 121.41647846153846)
		(18.96551724137931, 121.59325384615384)
		(48.275862068965516, 120.84548769230769)
		(44.827586206896555, 121.67183692307692)
		(31.03448275862069, 121.72963384615385)
		(89.65517241379311, 121.35669076923077)
		(27.586206896551722, 121.98511076923077)
		(34.48275862068966, 121.59258923076924)
		(46.55172413793103, 121.65005846153846)
		(3.4482758620689653, 121.25262923076923)
		(17.24137931034483, 121.24723846153846)
		(93.10344827586206, 121.03182615384615)
		(24.137931034482758, 121.50080923076924)
		(79.3103448275862, 121.44910923076922)
		(48.275862068965516, 121.19351538461538)
		(63.793103448275865, 121.45301846153846)
		(60.3448275862069, 121.73447538461538)
		(58.620689655172406, 121.05658615384615)
		(67.24137931034483, 121.44672307692308)
		(100.0, 121.47516615384616)
		(32.758620689655174, 121.57794615384616)
		(10.344827586206897, 121.4122)
		(10.344827586206897, 121.23628461538462)
		(55.172413793103445, 121.41724615384615)
		(51.724137931034484, 121.19939846153846)
		(44.827586206896555, 121.22566)
		(70.6896551724138, 121.63275384615385)
		(8.620689655172415, 120.70519692307693)
		(13.793103448275861, 121.72043384615385)
		(18.96551724137931, 121.19159076923077)
		(41.37931034482759, 121.25772923076923)
		(17.24137931034483, 121.43738615384615)
		(62.06896551724138, 121.52574153846153)
		(75.86206896551724, 121.26665230769231)
		(65.51724137931035, 121.68096461538461)
		(74.13793103448276, 120.98846615384615)
		(67.24137931034483, 121.29350769230768)
		(41.37931034482759, 121.45135384615385)
		(46.55172413793103, 121.4397676923077)
		(37.93103448275862, 121.23744923076923)
		(86.20689655172413, 121.7655523076923)
		(91.37931034482759, 121.17742615384616)
		(51.724137931034484, 121.60493846153847)
		(50.0, 121.50193846153846)
		(82.75862068965517, 121.26928307692307)
		(13.793103448275861, 121.3209876923077)
		(55.172413793103445, 121.19675846153847)
		(60.3448275862069, 121.47329692307693)
		(3.4482758620689653, 121.24196923076923)
		(72.41379310344827, 121.52926461538462)
		(94.82758620689656, 121.73941692307692)
		(27.586206896551722, 121.34798461538462)
		(56.896551724137936, 121.61828307692308)
		(65.51724137931035, 121.37912307692308)
		(39.6551724137931, 121.21116923076923)
		(74.13793103448276, 121.43341230769231)
		(1.7241379310344827, 121.3152076923077)
		(32.758620689655174, 121.59353538461538)
		(37.93103448275862, 121.69894153846154)
		(1.7241379310344827, 120.85520461538462)
		(22.413793103448278, 121.66412769230769)
		(68.96551724137932, 121.45788461538461)
	};

}
\&
\DOUBLEAXIS{Full = {540.6993595384615}, Scatter, title={Havlak {\small(210 Type Annotations)}}}{Full = {1.137997147256934}, Scatter, Right Axis}{
	\addplot+[] coordinates{
		(17.142857142857142, 475.81677076923074)
		(56.19047619047619, 476.4693384615385)
		(35.23809523809524, 482.52513076923077)
		(20.952380952380953, 480.1778553846154)
		(84.76190476190476, 478.6680369230769)
		(65.71428571428571, 476.05300923076925)
		(79.04761904761905, 478.3449)
		(76.66666666666667, 473.98662)
		(48.095238095238095, 470.71437076923075)
		(19.047619047619047, 478.74909846153844)
		(74.76190476190476, 482.3941753846154)
		(60.952380952380956, 478.87623538461537)
		(10.952380952380953, 474.14507846153845)
		(12.380952380952381, 480.1689846153846)
		(50.95238095238095, 477.1539646153846)
		(92.85714285714286, 483.0715215384615)
		(75.71428571428571, 480.53142153846153)
		(36.19047619047619, 478.6237876923077)
		(81.9047619047619, 479.1836923076923)
		(43.80952380952381, 472.54087692307695)
		(15.238095238095239, 470.5337846153846)
		(37.142857142857146, 478.52563846153845)
		(6.190476190476191, 476.37336)
		(70.95238095238095, 471.67498461538463)
		(69.04761904761905, 470.04300923076926)
		(8.095238095238095, 476.3642046153846)
		(30.952380952380953, 476.1283861538462)
		(61.904761904761905, 473.13450923076925)
		(32.857142857142854, 483.82162)
		(0.4761904761904762, 470.49335846153843)
		(31.9047619047619, 487.95160923076924)
		(100.0, 479.7917)
		(72.85714285714285, 477.57642153846155)
		(83.80952380952381, 476.7844892307692)
		(10.0, 479.8344492307692)
		(94.76190476190476, 487.6253461538462)
		(97.61904761904762, 478.14547384615383)
		(85.71428571428571, 484.8252523076923)
		(71.9047619047619, 478.34222923076925)
		(14.285714285714285, 478.37926923076924)
		(53.80952380952381, 484.2431676923077)
		(0.0, 475.13243846153847)
		(80.95238095238095, 475.8861138461538)
		(91.9047619047619, 480.3763569230769)
		(73.80952380952381, 478.0933123076923)
		(45.23809523809524, 475.06821384615387)
		(2.380952380952381, 472.6865923076923)
		(34.285714285714285, 478.13438615384615)
		(82.85714285714286, 484.55173692307693)
		(98.57142857142858, 484.7225507692308)
		(18.095238095238095, 475.9124276923077)
		(59.04761904761905, 476.21401692307694)
		(86.66666666666667, 479.69071076923075)
		(90.0, 474.17252615384615)
		(27.142857142857142, 482.8051046153846)
		(93.80952380952381, 481.3174246153846)
		(24.285714285714285, 481.10267384615383)
		(63.8095238095238, 479.03792923076924)
		(20.0, 474.8084353846154)
		(40.95238095238095, 477.28601692307694)
		(52.85714285714286, 475.20169384615383)
		(54.761904761904766, 479.88986307692306)
		(40.0, 477.48688)
		(4.285714285714286, 478.8741153846154)
		(70.0, 477.8750953846154)
		(39.04761904761905, 480.1275353846154)
		(51.90476190476191, 486.8542123076923)
		(78.0952380952381, 478.56088)
		(13.333333333333334, 473.95328461538463)
		(42.857142857142854, 474.15598153846156)
		(25.238095238095237, 485.09013846153846)
		(80.0, 475.4012)
		(23.333333333333332, 473.7148876923077)
		(41.904761904761905, 473.71412153846154)
		(7.142857142857142, 480.40487384615386)
		(62.857142857142854, 473.96555538461536)
		(89.04761904761904, 479.87380461538464)
		(1.4285714285714286, 478.36176461538463)
		(26.190476190476193, 480.3659215384615)
		(90.95238095238095, 478.5262784615385)
		(5.238095238095238, 471.51821384615386)
		(16.19047619047619, 474.03736923076923)
		(58.0952380952381, 480.21205692307694)
		(95.71428571428572, 480.94632615384614)
		(9.047619047619047, 471.07358923076924)
		(46.19047619047619, 471.73838461538463)
		(64.76190476190476, 481.1933123076923)
		(57.14285714285714, 491.5448723076923)
		(96.66666666666667, 480.66367846153844)
		(67.14285714285714, 477.1530430769231)
		(29.04761904761905, 473.95092307692306)
		(3.3333333333333335, 472.17782923076925)
		(28.095238095238095, 470.5633861538462)
		(87.61904761904762, 482.79979692307694)
		(30.0, 473.8225753846154)
		(21.904761904761905, 479.3486846153846)
		(60.0, 485.5712169230769)
		(68.0952380952381, 477.91927846153845)
		(49.047619047619044, 475.7533430769231)
		(47.14285714285714, 488.91988923076923)
		(50.0, 486.81440615384616)
		(38.095238095238095, 480.5950184615385)
	};

}
\&
\DOUBLEAXIS{Full = {195.2798996923077}, Scatter, title={Json {\small(134 Type Annotations)}}}{Full = {1.1225294198732045}, Scatter, Right Axis, Scatter Right Label}{
	\addplot+[] coordinates{
		(55.223880597014926, 175.71049384615384)
		(92.53731343283582, 171.53264153846155)
		(35.07462686567165, 172.28333692307692)
		(89.55223880597015, 171.94824307692306)
		(5.223880597014925, 172.68620153846155)
		(10.44776119402985, 171.83396307692308)
		(46.26865671641791, 172.77547384615386)
		(97.76119402985076, 171.90346153846153)
		(85.07462686567165, 170.21673846153845)
		(50.74626865671642, 168.83480307692307)
		(50.0, 173.22181076923076)
		(62.68656716417911, 170.99116153846154)
		(71.64179104477611, 171.15657692307693)
		(20.149253731343283, 172.25893076923077)
		(37.3134328358209, 170.0299553846154)
		(58.95522388059702, 177.52718153846155)
		(44.02985074626866, 171.7505676923077)
		(73.13432835820896, 171.08624615384616)
		(23.134328358208954, 169.47055384615385)
		(20.8955223880597, 173.48855230769232)
		(70.8955223880597, 172.49303076923076)
		(88.80597014925374, 171.37408615384615)
		(5.970149253731343, 172.10772307692307)
		(59.70149253731343, 170.16109230769231)
		(17.16417910447761, 170.55124615384617)
		(67.16417910447761, 171.68151230769232)
		(76.86567164179104, 171.12082615384617)
		(61.940298507462686, 170.19189846153847)
		(64.92537313432835, 171.6184246153846)
		(70.1492537313433, 169.4971476923077)
		(86.56716417910447, 172.13789538461538)
		(2.2388059701492535, 173.38218307692307)
		(73.88059701492537, 172.22969846153845)
		(85.82089552238806, 172.88388923076923)
		(17.91044776119403, 173.72397230769232)
		(77.61194029850746, 175.41718307692307)
		(8.955223880597014, 170.38774615384617)
		(80.59701492537313, 173.47786923076924)
		(47.76119402985074, 173.77804615384616)
		(43.28358208955223, 174.24379846153846)
		(83.5820895522388, 174.09923076923076)
		(61.19402985074627, 167.54895692307693)
		(82.08955223880598, 173.48941846153846)
		(94.77611940298507, 169.68675384615383)
		(28.35820895522388, 170.3916046153846)
		(23.88059701492537, 173.2905153846154)
		(53.73134328358209, 170.83348923076923)
		(65.67164179104478, 169.16839846153846)
		(47.01492537313433, 172.11491076923076)
		(7.462686567164178, 170.00080923076922)
		(95.52238805970148, 172.71231538461538)
		(29.850746268656714, 168.18070153846153)
		(0.7462686567164178, 173.58076461538462)
		(14.925373134328357, 174.84640769230768)
		(4.477611940298507, 172.9343353846154)
		(94.02985074626866, 170.01346)
		(44.776119402985074, 172.36557846153846)
		(49.25373134328358, 171.3912476923077)
		(25.37313432835821, 174.68568153846155)
		(35.82089552238806, 168.75591076923078)
		(1.4925373134328357, 170.19082)
		(64.17910447761194, 172.26682615384615)
		(40.298507462686565, 174.0615723076923)
		(91.04477611940298, 171.71981384615384)
		(82.83582089552239, 170.4570369230769)
		(11.194029850746269, 171.60086615384614)
		(79.8507462686567, 171.07154461538462)
		(13.432835820895523, 173.4628646153846)
		(38.059701492537314, 168.46174923076924)
		(38.80597014925373, 173.3625676923077)
		(97.01492537313433, 171.6187276923077)
		(56.71641791044776, 168.76458)
		(91.7910447761194, 169.78995076923076)
		(11.940298507462686, 172.90889846153846)
		(19.402985074626866, 168.58201846153847)
		(76.11940298507463, 169.97938461538462)
		(68.65671641791045, 170.31847692307693)
		(26.119402985074625, 171.41591384615384)
		(2.9850746268656714, 172.6440276923077)
		(98.50746268656717, 172.76446)
		(29.1044776119403, 173.28739076923077)
		(52.98507462686567, 169.87476307692307)
		(32.83582089552239, 173.28686461538462)
		(67.91044776119402, 173.86280923076924)
		(100.0, 170.72024615384615)
		(58.2089552238806, 169.83102461538462)
		(0.0, 173.96417076923078)
		(41.04477611940299, 170.14858615384614)
		(16.417910447761194, 170.57913384615384)
		(26.865671641791046, 173.92400923076923)
		(52.23880597014925, 171.1361446153846)
		(74.6268656716418, 173.55657076923077)
		(41.7910447761194, 169.5297676923077)
		(31.343283582089555, 172.63255846153845)
		(88.05970149253731, 170.97635076923078)
		(22.388059701492537, 170.43658307692309)
		(14.17910447761194, 174.31824307692307)
		(8.208955223880597, 172.6991046153846)
		(79.1044776119403, 170.96677692307694)
		(55.970149253731336, 174.57297692307694)
		(34.32835820895522, 172.73087384615386)
		(32.08955223880597, 173.75718307692307)
	};

}
\\
\DOUBLEAXIS{Full = {487.8821284615385}, Scatter, Scatter Left Label, title={List {\small(23 Type Annotations)}}}{Full = {2.2729455968334378}, Scatter, Right Axis}{
	\addplot+[] coordinates{
		(39.130434782608695, 203.76336153846154)
		(56.52173913043478, 294.11142615384614)
		(65.21739130434783, 414.2744723076923)
		(95.65217391304348, 432.19901846153846)
		(30.434782608695656, 301.67285692307695)
		(13.043478260869565, 204.96308923076924)
		(82.6086956521739, 443.5292076923077)
		(73.91304347826086, 422.14082)
		(4.3478260869565215, 205.62648615384614)
		(13.043478260869565, 205.5540553846154)
		(30.434782608695656, 299.45758615384614)
		(4.3478260869565215, 204.70628923076924)
		(34.78260869565217, 205.61570461538463)
		(43.47826086956522, 213.50244307692307)
		(69.56521739130434, 422.4920584615385)
		(34.78260869565217, 209.87120307692308)
		(4.3478260869565215, 285.2347246153846)
		(65.21739130434783, 279.92569692307694)
		(26.08695652173913, 298.4554676923077)
		(82.6086956521739, 426.7074030769231)
		(52.17391304347826, 218.45641076923076)
		(4.3478260869565215, 212.5166123076923)
		(39.130434782608695, 421.6899507692308)
		(34.78260869565217, 211.06125076923078)
		(82.6086956521739, 313.68094)
		(8.695652173913043, 205.6064076923077)
		(4.3478260869565215, 205.85359692307694)
		(13.043478260869565, 204.50179076923078)
		(91.30434782608695, 434.93765076923074)
		(56.52173913043478, 276.1561923076923)
		(26.08695652173913, 209.40336153846152)
		(21.73913043478261, 302.4377123076923)
		(69.56521739130434, 422.15309692307693)
		(47.82608695652174, 213.70174615384616)
		(95.65217391304348, 281.86659692307694)
		(95.65217391304348, 432.1258323076923)
		(73.91304347826086, 283.2683430769231)
		(21.73913043478261, 209.31984615384616)
		(86.95652173913044, 432.1335446153846)
		(39.130434782608695, 213.87380153846155)
		(8.695652173913043, 208.15599076923078)
		(73.91304347826086, 427.2881246153846)
		(91.30434782608695, 432.2348984615385)
		(43.47826086956522, 209.81009076923078)
		(13.043478260869565, 205.26062461538461)
		(65.21739130434783, 218.41380307692307)
		(95.65217391304348, 426.98387692307693)
		(30.434782608695656, 208.08602769230768)
		(78.26086956521739, 432.3689123076923)
		(43.47826086956522, 292.5950323076923)
		(86.95652173913044, 432.3203169230769)
		(52.17391304347826, 272.7143384615385)
		(91.30434782608695, 418.42740153846154)
		(17.391304347826086, 205.4463323076923)
		(100.0, 433.9553723076923)
		(43.47826086956522, 301.66147384615385)
		(34.78260869565217, 422.1372430769231)
		(91.30434782608695, 426.43911692307694)
		(52.17391304347826, 204.90848615384616)
		(17.391304347826086, 275.9406153846154)
		(82.6086956521739, 281.6327923076923)
		(0.0, 214.64751692307692)
		(52.17391304347826, 209.32162923076922)
		(78.26086956521739, 424.6643523076923)
		(78.26086956521739, 303.14302615384617)
		(26.08695652173913, 212.80659846153847)
		(60.86956521739131, 277.5932507692308)
		(91.30434782608695, 433.4522969230769)
		(56.52173913043478, 284.55373692307694)
		(56.52173913043478, 432.34235384615386)
		(21.73913043478261, 205.4845723076923)
		(52.17391304347826, 286.3656384615385)
		(17.391304347826086, 205.39220153846153)
		(47.82608695652174, 424.3952830769231)
		(30.434782608695656, 275.90955384615387)
		(8.695652173913043, 205.6112523076923)
		(39.130434782608695, 298.39793692307694)
		(78.26086956521739, 436.7692215384615)
		(65.21739130434783, 302.0162723076923)
		(21.73913043478261, 206.71018)
		(73.91304347826086, 431.49852923076924)
		(60.86956521739131, 302.72734153846153)
		(60.86956521739131, 418.4613846153846)
		(82.6086956521739, 278.40741846153844)
		(26.08695652173913, 302.15180923076923)
		(43.47826086956522, 209.51488153846154)
		(8.695652173913043, 205.51459384615384)
		(17.391304347826086, 207.17219076923078)
		(47.82608695652174, 421.9823030769231)
		(73.91304347826086, 280.3063446153846)
		(69.56521739130434, 281.52176)
		(69.56521739130434, 422.20660615384617)
		(34.78260869565217, 284.3918261538461)
		(17.391304347826086, 203.89332923076924)
		(8.695652173913043, 300.79152153846155)
		(26.08695652173913, 209.65139846153846)
		(86.95652173913044, 289.7698753846154)
		(60.86956521739131, 316.606)
		(65.21739130434783, 419.67258923076923)
		(56.52173913043478, 208.76297692307693)
		(86.95652173913044, 435.5484276923077)
		(47.82608695652174, 208.63745384615385)
	};

}
\&
\DOUBLEAXIS{Full = {127.40397830769233}, Scatter, title={Mandelbrot {\small(23 Type Annotations)}}}{Full = {1.1101808475183856}, Scatter, Right Axis}{
	\addplot+[] coordinates{
		(0.0, 114.75966153846154)
		(47.82608695652174, 114.80745384615385)
		(56.52173913043478, 114.76316769230769)
		(13.043478260869565, 114.84070461538461)
		(17.391304347826086, 114.79131846153847)
		(34.78260869565217, 114.79011538461539)
		(69.56521739130434, 114.88319076923077)
		(91.30434782608695, 114.72386923076922)
		(65.21739130434783, 114.76876461538461)
		(21.73913043478261, 114.76086615384615)
		(60.86956521739131, 114.83480923076922)
		(34.78260869565217, 114.80201076923078)
		(78.26086956521739, 114.7886123076923)
		(73.91304347826086, 114.78426923076923)
		(43.47826086956522, 114.73466461538462)
		(26.08695652173913, 114.85406307692308)
		(82.6086956521739, 115.60138153846154)
		(69.56521739130434, 115.61753692307693)
		(52.17391304347826, 114.80274461538461)
		(13.043478260869565, 114.83623538461538)
		(91.30434782608695, 115.56441230769231)
		(26.08695652173913, 114.77404923076924)
		(60.86956521739131, 114.7482)
		(73.91304347826086, 114.80876153846154)
		(43.47826086956522, 114.75819846153846)
		(65.21739130434783, 114.81584615384615)
		(73.91304347826086, 114.8743123076923)
		(17.391304347826086, 114.73391076923077)
		(52.17391304347826, 114.74636923076923)
		(34.78260869565217, 115.73033230769231)
		(8.695652173913043, 114.80016461538462)
		(52.17391304347826, 114.94336461538461)
		(91.30434782608695, 114.73569384615385)
		(4.3478260869565215, 114.75517692307692)
		(95.65217391304348, 114.79689692307693)
		(34.78260869565217, 114.80873384615384)
		(73.91304347826086, 114.72432461538462)
		(21.73913043478261, 114.76657230769231)
		(39.130434782608695, 114.77938153846154)
		(47.82608695652174, 114.80545538461539)
		(65.21739130434783, 114.75986)
		(95.65217391304348, 115.66740461538461)
		(21.73913043478261, 114.78067076923077)
		(69.56521739130434, 115.82179846153846)
		(43.47826086956522, 115.62028461538462)
		(8.695652173913043, 114.76792461538462)
		(26.08695652173913, 115.54483538461538)
		(8.695652173913043, 114.8675476923077)
		(86.95652173913044, 114.76670615384616)
		(91.30434782608695, 114.76526)
		(78.26086956521739, 114.74334307692308)
		(78.26086956521739, 114.74876307692308)
		(4.3478260869565215, 114.9046523076923)
		(17.391304347826086, 115.20829076923077)
		(43.47826086956522, 114.78182153846154)
		(4.3478260869565215, 114.77255384615384)
		(52.17391304347826, 114.82604923076923)
		(56.52173913043478, 115.62255846153846)
		(26.08695652173913, 114.76161846153846)
		(4.3478260869565215, 114.78106769230769)
		(73.91304347826086, 114.78515538461538)
		(56.52173913043478, 115.65486153846155)
		(56.52173913043478, 115.54803230769231)
		(78.26086956521739, 114.80889538461538)
		(100.0, 114.85802)
		(30.434782608695656, 114.87329230769231)
		(17.391304347826086, 114.76866923076923)
		(65.21739130434783, 114.76824307692307)
		(91.30434782608695, 115.67563384615384)
		(82.6086956521739, 114.83298923076923)
		(86.95652173913044, 114.75546307692308)
		(17.391304347826086, 114.73942461538462)
		(39.130434782608695, 114.83399230769231)
		(95.65217391304348, 114.79814307692308)
		(8.695652173913043, 114.8314676923077)
		(30.434782608695656, 114.94535538461538)
		(30.434782608695656, 114.86509384615384)
		(47.82608695652174, 114.76278)
		(69.56521739130434, 114.82913692307692)
		(30.434782608695656, 114.73137230769231)
		(60.86956521739131, 114.76616769230769)
		(65.21739130434783, 114.76756)
		(82.6086956521739, 114.80056923076923)
		(8.695652173913043, 114.85468615384616)
		(52.17391304347826, 114.81798)
		(21.73913043478261, 114.77947384615385)
		(13.043478260869565, 114.85407076923077)
		(47.82608695652174, 114.76607846153846)
		(39.130434782608695, 114.86322615384616)
		(26.08695652173913, 114.82585076923077)
		(13.043478260869565, 115.45183384615385)
		(34.78260869565217, 114.78708307692308)
		(60.86956521739131, 114.77685384615384)
		(82.6086956521739, 115.61261846153846)
		(86.95652173913044, 114.73528)
		(95.65217391304348, 114.77652)
		(43.47826086956522, 114.73172923076923)
		(86.95652173913044, 115.02547230769231)
		(4.3478260869565215, 114.88323384615384)
		(56.52173913043478, 114.72702307692307)
		(82.6086956521739, 114.83935846153847)
		(39.130434782608695, 114.77747692307692)
	};

}
\&
\DOUBLEAXIS{Full = {168.8449856923077}, Scatter, title={NBody {\small(65 Type Annotations)}}}{Full = {1.1344301796858876}, Scatter, Right Axis, Scatter Right Label}{
	\addplot+[] coordinates{
		(10.76923076923077, 148.9934676923077)
		(83.07692307692308, 152.40088153846153)
		(38.46153846153847, 148.98205076923077)
		(50.76923076923077, 148.9055876923077)
		(80.0, 152.2100676923077)
		(58.46153846153847, 152.03244923076923)
		(35.38461538461539, 148.81251384615385)
		(9.230769230769232, 148.9061523076923)
		(93.84615384615384, 152.04551538461538)
		(69.23076923076923, 149.0216353846154)
		(72.3076923076923, 153.16925538461538)
		(6.153846153846154, 152.43478769230768)
		(40.0, 152.07649230769232)
		(76.92307692307693, 151.98408)
		(24.615384615384617, 148.93517384615384)
		(95.38461538461539, 152.21770307692307)
		(33.84615384615385, 148.92519076923077)
		(52.307692307692314, 152.17553076923076)
		(47.69230769230769, 152.50781384615385)
		(69.23076923076923, 152.09637692307692)
		(83.07692307692308, 148.87494153846154)
		(36.92307692307693, 148.96049846153846)
		(23.076923076923077, 150.04322769230768)
		(92.3076923076923, 153.30205846153845)
		(16.923076923076923, 148.94449384615385)
		(20.0, 152.91784615384614)
		(87.6923076923077, 152.07820615384617)
		(26.153846153846157, 148.8413923076923)
		(100.0, 152.07173230769232)
		(36.92307692307693, 152.44871230769232)
		(27.692307692307693, 150.00729846153845)
		(63.07692307692307, 152.57259076923077)
		(13.846153846153847, 148.81941692307691)
		(96.92307692307692, 152.8623)
		(16.923076923076923, 149.04642307692308)
		(21.53846153846154, 148.8652953846154)
		(90.76923076923077, 152.01118923076922)
		(41.53846153846154, 152.57619076923078)
		(38.46153846153847, 148.98866153846154)
		(66.15384615384615, 152.13796769230768)
		(23.076923076923077, 148.90548153846154)
		(81.53846153846153, 152.03838615384615)
		(98.46153846153847, 152.13379384615385)
		(50.76923076923077, 151.99954461538462)
		(84.61538461538461, 152.04791076923078)
		(33.84615384615385, 148.88693692307692)
		(53.84615384615385, 152.11450307692309)
		(76.92307692307693, 148.96806)
		(29.230769230769234, 152.47762153846153)
		(1.5384615384615385, 148.79225076923078)
		(30.76923076923077, 148.89050461538463)
		(1.5384615384615385, 148.93906307692308)
		(80.0, 152.56712615384615)
		(61.53846153846154, 152.02468153846155)
		(46.15384615384615, 148.8707276923077)
		(61.53846153846154, 152.03587846153846)
		(12.307692307692308, 149.08258923076923)
		(66.15384615384615, 153.18455538461538)
		(55.38461538461539, 152.17752615384615)
		(75.38461538461539, 148.92492307692308)
		(70.76923076923077, 153.12410923076922)
		(96.92307692307692, 152.18487846153846)
		(93.84615384615384, 153.10723846153846)
		(52.307692307692314, 152.50945384615383)
		(58.46153846153847, 148.81516615384615)
		(20.0, 148.9122076923077)
		(64.61538461538461, 148.93609230769232)
		(3.076923076923077, 148.88537846153847)
		(60.0, 148.8611646153846)
		(89.23076923076924, 151.99509076923076)
		(0.0, 148.83682461538461)
		(32.30769230769231, 148.86797384615386)
		(75.38461538461539, 148.9761876923077)
		(6.153846153846154, 148.94526307692308)
		(41.53846153846154, 148.9055753846154)
		(18.461538461538463, 148.90876461538463)
		(7.6923076923076925, 148.93461384615384)
		(26.153846153846157, 149.95932153846155)
		(55.38461538461539, 153.05352)
		(90.76923076923077, 152.05859692307692)
		(87.6923076923077, 152.15724615384616)
		(27.692307692307693, 148.88219692307692)
		(63.07692307692307, 150.05554307692307)
		(78.46153846153847, 152.82133076923077)
		(86.15384615384616, 152.27032615384616)
		(67.6923076923077, 152.13224461538462)
		(3.076923076923077, 149.86198153846155)
		(15.384615384615385, 149.9609846153846)
		(44.61538461538462, 153.49544153846153)
		(72.3076923076923, 148.90402307692307)
		(47.69230769230769, 148.8730876923077)
		(73.84615384615385, 152.0368076923077)
		(4.615384615384616, 149.00476923076923)
		(43.07692307692308, 149.19817692307691)
		(86.15384615384616, 152.06196)
		(56.92307692307692, 153.22585384615385)
		(13.846153846153847, 148.81347692307693)
		(44.61538461538462, 152.38190461538463)
		(30.76923076923077, 148.81860461538463)
		(49.23076923076923, 148.91515384615386)
		(9.230769230769232, 148.8158169230769)
		(12.307692307692308, 148.9312846153846)
	};

}
\\
\DOUBLEAXIS{Full = {193.7184513846154}, Scatter, Scatter Left Label, title={Permute {\small(14 Type Annotations)}}}{Full = {1.203410245209007}, Scatter, Right Axis}{
	\addplot+[] coordinates{
		(35.714285714285715, 176.10768307692308)
		(71.42857142857143, 136.8766553846154)
		(64.28571428571429, 161.1371353846154)
		(100.0, 136.88462615384614)
		(57.14285714285714, 136.88589692307693)
		(64.28571428571429, 160.9886)
		(35.714285714285715, 137.28191692307692)
		(14.285714285714285, 161.2795)
		(71.42857142857143, 136.72838307692308)
		(85.71428571428571, 137.21357538461538)
		(78.57142857142857, 137.14604)
		(78.57142857142857, 137.21827076923077)
		(85.71428571428571, 136.98489692307692)
		(28.57142857142857, 137.25917692307692)
		(7.142857142857142, 161.12510615384616)
		(21.428571428571427, 137.21247076923078)
		(21.428571428571427, 175.24994923076923)
		(42.857142857142854, 160.75281076923076)
		(50.0, 160.87430307692307)
		(42.857142857142854, 137.32260153846153)
		(7.142857142857142, 160.67028769230768)
		(21.428571428571427, 144.55693538461537)
		(14.285714285714285, 160.7906046153846)
		(92.85714285714286, 144.98229692307692)
		(50.0, 160.45328923076923)
		(64.28571428571429, 169.27450615384615)
		(35.714285714285715, 136.91826923076923)
		(7.142857142857142, 160.97658307692308)
		(35.714285714285715, 160.67992153846154)
		(14.285714285714285, 161.15290307692308)
		(57.14285714285714, 136.97136)
		(21.428571428571427, 173.2810076923077)
		(78.57142857142857, 160.88316307692307)
		(50.0, 160.79721846153845)
		(85.71428571428571, 136.7328076923077)
		(42.857142857142854, 160.69878923076922)
		(35.714285714285715, 161.48660615384617)
		(57.14285714285714, 160.96273846153846)
		(21.428571428571427, 160.85313076923077)
		(92.85714285714286, 160.85056615384616)
		(78.57142857142857, 144.43697230769232)
		(35.714285714285715, 137.38470923076923)
		(50.0, 160.8291323076923)
		(28.57142857142857, 137.26133384615383)
		(21.428571428571427, 160.93747076923077)
		(42.857142857142854, 161.11040615384616)
		(85.71428571428571, 144.7620246153846)
		(35.714285714285715, 136.72496)
		(71.42857142857143, 161.11022461538462)
		(14.285714285714285, 173.44653076923078)
		(78.57142857142857, 145.50936153846155)
		(92.85714285714286, 136.72719230769232)
		(42.857142857142854, 160.72576923076923)
		(50.0, 160.89916923076922)
		(64.28571428571429, 161.1835153846154)
		(14.285714285714285, 161.02518307692307)
		(92.85714285714286, 137.0094723076923)
		(42.857142857142854, 171.16817846153847)
		(14.285714285714285, 137.1746753846154)
		(64.28571428571429, 136.68120769230768)
		(57.14285714285714, 137.11639846153847)
		(28.57142857142857, 144.59462307692309)
		(7.142857142857142, 160.97060307692308)
		(14.285714285714285, 160.67462153846154)
		(7.142857142857142, 161.44650307692308)
		(64.28571428571429, 137.14062)
		(71.42857142857143, 136.88532153846154)
		(78.57142857142857, 160.8726153846154)
		(71.42857142857143, 136.936)
		(64.28571428571429, 136.79855538461538)
		(50.0, 144.82084615384616)
		(78.57142857142857, 144.6558476923077)
		(28.57142857142857, 160.86786923076923)
		(35.714285714285715, 160.5618523076923)
		(7.142857142857142, 136.80124615384617)
		(28.57142857142857, 160.81333538461539)
		(28.57142857142857, 160.68494153846154)
		(57.14285714285714, 136.61441384615384)
		(28.57142857142857, 137.1363630769231)
		(78.57142857142857, 137.09567692307692)
		(92.85714285714286, 145.10071846153846)
		(50.0, 161.30881384615384)
		(7.142857142857142, 160.66321076923077)
		(64.28571428571429, 136.63045076923078)
		(85.71428571428571, 136.79644769230768)
		(0.0, 160.97457384615385)
		(57.14285714285714, 136.91577846153845)
		(42.857142857142854, 161.0521923076923)
		(21.428571428571427, 160.89698307692308)
		(7.142857142857142, 160.8899953846154)
		(85.71428571428571, 136.91848153846155)
		(71.42857142857143, 160.83399384615385)
		(57.14285714285714, 137.0270123076923)
		(71.42857142857143, 160.81938615384615)
		(92.85714285714286, 137.47786923076924)
		(85.71428571428571, 136.95242)
		(57.14285714285714, 136.95814153846155)
		(21.428571428571427, 161.5591276923077)
		(92.85714285714286, 145.07678307692308)
		(85.71428571428571, 136.85939384615384)
		(14.285714285714285, 160.88954615384614)
		(42.857142857142854, 136.66572)
	};

}
\&
\DOUBLEAXIS{Full = {65.71047492307693}, Scatter, title={PyStone {\small(85 Type Annotations)}}}{Full = {1.1760959870666312}, Scatter, Right Axis}{
	\addplot+[] coordinates{
		(15.294117647058824, 56.660033846153844)
		(84.70588235294117, 58.87713230769231)
		(68.23529411764706, 56.657747692307694)
		(63.52941176470588, 55.79986615384615)
		(96.47058823529412, 56.51936769230769)
		(43.529411764705884, 56.40105692307692)
		(36.470588235294116, 58.00651538461538)
		(29.411764705882355, 58.848078461538464)
		(34.11764705882353, 58.13297076923077)
		(56.470588235294116, 58.69872615384615)
		(24.705882352941178, 57.57697846153846)
		(28.235294117647058, 59.37404)
		(30.58823529411765, 57.064676923076924)
		(49.411764705882355, 59.18167846153846)
		(50.588235294117645, 58.45326615384615)
		(75.29411764705883, 56.50503846153846)
		(9.411764705882353, 56.80349846153846)
		(88.23529411764706, 57.69730615384616)
		(78.82352941176471, 57.33108615384615)
		(74.11764705882354, 59.55146)
		(12.941176470588237, 57.07210307692308)
		(40.0, 58.21980153846154)
		(75.29411764705883, 57.503483076923075)
		(18.823529411764707, 56.09611384615385)
		(62.35294117647059, 57.423864615384616)
		(25.882352941176475, 56.89674307692308)
		(7.0588235294117645, 57.661316923076924)
		(37.64705882352941, 57.744843076923075)
		(43.529411764705884, 57.19272153846154)
		(76.47058823529412, 56.03307538461539)
		(32.94117647058823, 57.49186923076923)
		(41.17647058823529, 56.14072923076923)
		(18.823529411764707, 55.878126153846154)
		(17.647058823529413, 57.535556923076925)
		(56.470588235294116, 56.744883076923074)
		(67.05882352941175, 56.51126769230769)
		(12.941176470588237, 58.41391076923077)
		(85.88235294117646, 58.50896461538461)
		(27.058823529411764, 57.073107692307694)
		(54.11764705882353, 59.248584615384615)
		(87.05882352941177, 59.736795384615384)
		(44.70588235294118, 56.42778153846154)
		(10.588235294117647, 57.410586153846154)
		(31.76470588235294, 56.39568461538462)
		(72.94117647058823, 56.57508769230769)
		(55.294117647058826, 57.73047230769231)
		(23.52941176470588, 58.36691230769231)
		(45.88235294117647, 58.68052615384615)
		(98.82352941176471, 59.69126923076923)
		(4.705882352941177, 56.16973230769231)
		(91.76470588235294, 59.54349846153846)
		(95.29411764705881, 59.66524153846154)
		(87.05882352941177, 56.599369230769234)
		(25.882352941176475, 57.873004615384616)
		(61.1764705882353, 57.718583076923075)
		(1.1764705882352942, 55.84576461538462)
		(0.0, 55.871693846153846)
		(68.23529411764706, 57.794570769230766)
		(52.94117647058824, 57.28815384615385)
		(3.5294117647058822, 56.30575384615385)
		(2.3529411764705883, 55.91780307692308)
		(89.41176470588236, 56.764738461538464)
		(60.0, 58.14136307692308)
		(11.76470588235294, 56.12747230769231)
		(42.35294117647059, 56.60390769230769)
		(5.88235294117647, 57.76413230769231)
		(97.6470588235294, 59.53561538461538)
		(92.94117647058823, 57.245404615384615)
		(77.64705882352942, 59.20310615384616)
		(92.94117647058823, 56.10924461538462)
		(51.76470588235295, 57.21867692307692)
		(58.82352941176471, 58.29754153846154)
		(94.11764705882352, 59.46309230769231)
		(80.0, 56.24801076923077)
		(70.58823529411765, 57.02556153846154)
		(35.294117647058826, 57.688324615384616)
		(37.64705882352941, 58.10487538461538)
		(14.117647058823529, 57.05501692307692)
		(65.88235294117646, 59.39486)
		(47.05882352941176, 59.19590615384615)
		(48.23529411764706, 55.70439538461538)
		(100.0, 59.63024769230769)
		(83.52941176470588, 57.94184461538462)
		(7.0588235294117645, 58.35161230769231)
		(81.17647058823529, 57.17773692307692)
		(50.588235294117645, 57.31403076923077)
		(38.82352941176471, 58.694367692307694)
		(20.0, 56.35365846153846)
		(1.1764705882352942, 55.33909538461538)
		(71.76470588235294, 59.10808)
		(57.647058823529406, 57.51346461538461)
		(81.17647058823529, 59.62351230769231)
		(64.70588235294117, 56.14512)
		(31.76470588235294, 57.342815384615385)
		(22.35294117647059, 56.72327846153846)
		(21.176470588235293, 57.50318769230769)
		(8.235294117647058, 57.202804615384615)
		(69.41176470588235, 59.0571)
		(16.470588235294116, 59.458395384615386)
		(90.58823529411765, 57.11436)
		(82.35294117647058, 57.1653)
		(62.35294117647059, 56.44223846153846)
	};

}
\&
\DOUBLEAXIS{Full = {180.24034600000002}, Scatter, title={Queens {\small(22 Type Annotations)}}}{Full = {1.159170175019064}, Scatter, Right Axis, Scatter Right Label}{
	\addplot+[] coordinates{
		(22.727272727272727, 155.85902615384614)
		(4.545454545454546, 156.71204153846153)
		(13.636363636363635, 155.10634)
		(50.0, 157.6307676923077)
		(68.18181818181817, 163.85486)
		(31.818181818181817, 155.77659538461538)
		(86.36363636363636, 155.02911076923078)
		(86.36363636363636, 156.13924923076922)
		(81.81818181818183, 157.23244615384615)
		(18.181818181818183, 154.31588923076924)
		(40.909090909090914, 155.06026307692306)
		(18.181818181818183, 156.49715846153845)
		(68.18181818181817, 155.4959676923077)
		(59.09090909090909, 154.35058307692307)
		(63.63636363636363, 154.93988153846155)
		(63.63636363636363, 155.4574)
		(81.81818181818183, 156.75371846153845)
		(22.727272727272727, 154.33512615384615)
		(95.45454545454545, 156.4158876923077)
		(36.36363636363637, 156.24374307692307)
		(86.36363636363636, 158.29866153846154)
		(50.0, 155.57588615384614)
		(68.18181818181817, 155.18016)
		(18.181818181818183, 156.40935384615383)
		(31.818181818181817, 156.08736923076924)
		(90.9090909090909, 156.65146923076924)
		(36.36363636363637, 156.22121076923077)
		(45.45454545454545, 156.78964)
		(68.18181818181817, 155.6391953846154)
		(27.27272727272727, 157.15407076923077)
		(27.27272727272727, 155.70982)
		(4.545454545454546, 155.32848307692308)
		(22.727272727272727, 153.80339230769232)
		(9.090909090909092, 155.5964846153846)
		(63.63636363636363, 156.70224153846155)
		(36.36363636363637, 157.02371538461537)
		(4.545454545454546, 155.11598923076923)
		(13.636363636363635, 155.12386307692307)
		(31.818181818181817, 155.74355384615384)
		(54.54545454545454, 155.4074723076923)
		(90.9090909090909, 154.71228615384615)
		(81.81818181818183, 159.1688)
		(4.545454545454546, 157.74750461538463)
		(27.27272727272727, 154.13345076923076)
		(4.545454545454546, 156.02130153846153)
		(95.45454545454545, 156.08573384615386)
		(40.909090909090914, 155.34785538461537)
		(45.45454545454545, 155.7795569230769)
		(9.090909090909092, 156.35406615384616)
		(36.36363636363637, 155.6675276923077)
		(68.18181818181817, 154.4035353846154)
		(77.27272727272727, 157.72632)
		(50.0, 154.91770153846153)
		(27.27272727272727, 154.62784769230768)
		(31.818181818181817, 154.70917384615385)
		(36.36363636363637, 154.48784153846154)
		(40.909090909090914, 154.8538046153846)
		(54.54545454545454, 158.0056476923077)
		(77.27272727272727, 155.74706)
		(72.72727272727273, 157.0037923076923)
		(50.0, 157.77252153846155)
		(54.54545454545454, 155.66060307692308)
		(81.81818181818183, 156.73675384615385)
		(59.09090909090909, 157.8963553846154)
		(0.0, 155.49084153846155)
		(90.9090909090909, 153.75602153846154)
		(45.45454545454545, 154.20675538461538)
		(45.45454545454545, 156.86692923076924)
		(13.636363636363635, 154.1972953846154)
		(86.36363636363636, 154.19752615384616)
		(72.72727272727273, 157.8732830769231)
		(63.63636363636363, 154.89938)
		(9.090909090909092, 156.33306615384615)
		(90.9090909090909, 155.78196615384616)
		(9.090909090909092, 155.16765076923076)
		(63.63636363636363, 156.70844615384615)
		(9.090909090909092, 155.45107692307693)
		(27.27272727272727, 157.16421692307694)
		(72.72727272727273, 154.4151923076923)
		(77.27272727272727, 156.01707076923077)
		(95.45454545454545, 156.64557692307693)
		(90.9090909090909, 155.88913384615384)
		(40.909090909090914, 155.65627846153845)
		(50.0, 156.47540153846154)
		(22.727272727272727, 155.97404923076923)
		(81.81818181818183, 155.17679384615386)
		(95.45454545454545, 155.80678153846154)
		(59.09090909090909, 155.48408)
		(59.09090909090909, 155.05360153846155)
		(100.0, 157.1969476923077)
		(54.54545454545454, 155.5574353846154)
		(54.54545454545454, 155.19898615384616)
		(13.636363636363635, 156.47118615384616)
		(45.45454545454545, 154.20575384615384)
		(13.636363636363635, 158.5739953846154)
		(72.72727272727273, 154.8326046153846)
		(77.27272727272727, 157.01979384615385)
		(18.181818181818183, 156.02094153846153)
		(31.818181818181817, 155.71007076923078)
		(86.36363636363636, 155.30976)
		(18.181818181818183, 155.70562615384614)
		(72.72727272727273, 155.45935384615385)
	};

}
\\
\DOUBLEAXIS{Full = {737.2800803076924}, Scatter, Scatter Left Label, title={Richards {\small(177 Type Annotations)}}}{Full = {1.7896774262069879}, Scatter, Right Axis}{
	\addplot+[] coordinates{
		(91.52542372881356, 504.05951230769233)
		(44.06779661016949, 443.7914476923077)
		(19.2090395480226, 430.73608461538464)
		(62.14689265536724, 488.02977384615383)
		(20.903954802259886, 546.2992246153847)
		(76.8361581920904, 497.25998153846155)
		(54.80225988700565, 558.5911630769231)
		(59.887005649717516, 543.8835415384615)
		(12.429378531073446, 414.0604769230769)
		(100.0, 514.5723876923076)
		(33.33333333333333, 449.57924615384616)
		(8.47457627118644, 418.9058923076923)
		(31.07344632768362, 413.02338153846154)
		(51.9774011299435, 532.7223984615384)
		(22.033898305084744, 433.82622769230767)
		(92.65536723163842, 519.7150861538462)
		(61.016949152542374, 478.88476)
		(66.66666666666666, 462.14758153846157)
		(15.254237288135593, 397.02209230769233)
		(29.943502824858758, 425.8875123076923)
		(68.92655367231639, 438.82138)
		(2.2598870056497176, 409.80076153846153)
		(58.75706214689266, 475.82073384615387)
		(23.163841807909606, 514.4926307692308)
		(88.70056497175142, 559.4826923076923)
		(16.94915254237288, 402.7624246153846)
		(71.75141242937853, 492.34334615384614)
		(29.37853107344633, 411.3420446153846)
		(38.983050847457626, 538.9156415384615)
		(41.24293785310734, 444.98053076923077)
		(16.38418079096045, 417.8806030769231)
		(36.15819209039548, 455.2175646153846)
		(85.87570621468926, 508.7282784615385)
		(32.20338983050847, 670.2546184615385)
		(80.7909604519774, 477.85305999999997)
		(35.02824858757062, 420.3649461538462)
		(41.80790960451977, 532.1499538461538)
		(33.89830508474576, 459.4782)
		(37.85310734463277, 434.8993261538462)
		(97.74011299435028, 510.97612153846154)
		(87.00564971751412, 504.8343953846154)
		(20.33898305084746, 508.6414323076923)
		(0.5649717514124294, 409.27085384615384)
		(4.519774011299435, 428.37446923076925)
		(25.423728813559322, 425.7896507692308)
		(64.97175141242938, 566.348076923077)
		(48.0225988700565, 463.5241446153846)
		(24.293785310734464, 425.57146153846156)
		(67.79661016949152, 470.47513692307695)
		(70.05649717514125, 477.4141107692308)
		(3.389830508474576, 411.21206615384614)
		(62.71186440677966, 578.1479938461539)
		(6.214689265536723, 414.99601384615386)
		(53.10734463276836, 581.0554507692308)
		(0.0, 411.9625523076923)
		(75.70621468926554, 485.0088123076923)
		(83.05084745762711, 468.17091384615384)
		(89.83050847457628, 615.2387015384616)
		(70.62146892655367, 515.2256676923076)
		(75.14124293785311, 601.6462923076923)
		(46.89265536723164, 437.62101538461536)
		(77.96610169491525, 500.93047538461536)
		(74.01129943502825, 473.71236153846155)
		(95.48022598870057, 491.86513384615387)
		(7.344632768361582, 416.0288676923077)
		(50.282485875706215, 560.1288384615384)
		(14.124293785310735, 412.0237507692308)
		(66.10169491525424, 455.1298061538462)
		(79.66101694915254, 489.4486076923077)
		(87.57062146892656, 515.3765123076923)
		(58.19209039548022, 465.6170569230769)
		(50.847457627118644, 472.2157446153846)
		(40.11299435028249, 469.9168246153846)
		(27.11864406779661, 440.8407476923077)
		(63.84180790960452, 478.30451846153846)
		(55.932203389830505, 465.5353276923077)
		(9.03954802259887, 415.4103215384615)
		(49.152542372881356, 576.8831292307692)
		(25.98870056497175, 579.5762184615385)
		(45.19774011299435, 463.3425846153846)
		(42.93785310734463, 555.3567876923076)
		(11.299435028248588, 410.50579076923077)
		(72.88135593220339, 532.7232446153846)
		(54.23728813559322, 474.4570907692308)
		(81.92090395480226, 485.3327815384615)
		(98.87005649717514, 492.9808323076923)
		(57.06214689265536, 452.8882323076923)
		(96.61016949152543, 501.57664923076925)
		(28.24858757062147, 521.8492446153846)
		(90.96045197740112, 596.9399169230769)
		(94.91525423728814, 499.26756)
		(5.084745762711865, 409.3059153846154)
		(37.28813559322034, 545.5940215384616)
		(10.16949152542373, 429.6085953846154)
		(83.61581920903954, 479.42617538461536)
		(79.09604519774011, 597.7494830769231)
		(1.1299435028248588, 407.94353538461536)
		(45.76271186440678, 447.66892461538464)
		(84.7457627118644, 461.5028123076923)
		(93.78531073446328, 517.9437307692308)
		(12.994350282485875, 523.7277953846154)
		(18.07909604519774, 537.4592184615385)
	};

}
\&
\DOUBLEAXIS{Full = {196.51061353846154}, Scatter, title={Sieve {\small(13 Type Annotations)}}}{Full = {1.1103929110329531}, Scatter, Right Axis}{
	\addplot+[] coordinates{
		(76.92307692307693, 177.71983846153847)
		(53.84615384615385, 177.43605384615384)
		(7.6923076923076925, 176.70619692307693)
		(76.92307692307693, 176.75793076923077)
		(15.384615384615385, 177.37116307692307)
		(38.46153846153847, 177.7785846153846)
		(15.384615384615385, 174.26084461538463)
		(92.3076923076923, 176.58615846153847)
		(84.61538461538461, 174.73045384615384)
		(15.384615384615385, 175.40210769230768)
		(38.46153846153847, 177.1869276923077)
		(7.6923076923076925, 175.52610923076924)
		(92.3076923076923, 177.6417076923077)
		(53.84615384615385, 177.2099353846154)
		(92.3076923076923, 177.6831676923077)
		(84.61538461538461, 176.30277384615385)
		(23.076923076923077, 176.91848153846155)
		(76.92307692307693, 174.68833384615385)
		(61.53846153846154, 177.54876923076924)
		(69.23076923076923, 176.1644723076923)
		(92.3076923076923, 176.50733076923078)
		(76.92307692307693, 174.84331230769232)
		(30.76923076923077, 177.4180276923077)
		(7.6923076923076925, 177.7717353846154)
		(46.15384615384615, 176.43904615384616)
		(84.61538461538461, 174.66923692307694)
		(69.23076923076923, 174.71807692307692)
		(46.15384615384615, 175.07548307692306)
		(92.3076923076923, 177.03922)
		(53.84615384615385, 178.01152923076924)
		(7.6923076923076925, 174.82770923076924)
		(23.076923076923077, 176.83801384615384)
		(69.23076923076923, 177.13440461538463)
		(15.384615384615385, 178.10349076923077)
		(38.46153846153847, 174.58110153846152)
		(53.84615384615385, 177.15658307692308)
		(38.46153846153847, 177.51114615384614)
		(23.076923076923077, 177.62555076923076)
		(61.53846153846154, 177.13825538461538)
		(92.3076923076923, 176.82738307692307)
		(46.15384615384615, 177.8657276923077)
		(61.53846153846154, 177.01818307692307)
		(61.53846153846154, 177.0578753846154)
		(23.076923076923077, 176.68521384615386)
		(84.61538461538461, 178.00904615384616)
		(84.61538461538461, 176.77285230769232)
		(38.46153846153847, 177.29462153846154)
		(53.84615384615385, 176.85863384615385)
		(23.076923076923077, 177.24140461538462)
		(7.6923076923076925, 177.11699846153846)
		(7.6923076923076925, 177.52588461538463)
		(15.384615384615385, 177.0629430769231)
		(30.76923076923077, 177.65095384615384)
		(76.92307692307693, 177.29825692307693)
		(92.3076923076923, 176.88639538461538)
		(30.76923076923077, 177.6488)
		(61.53846153846154, 177.04074307692306)
		(76.92307692307693, 177.49860615384617)
		(53.84615384615385, 177.04481846153845)
		(84.61538461538461, 177.72016769230768)
		(69.23076923076923, 177.3570246153846)
		(15.384615384615385, 177.49091692307692)
		(84.61538461538461, 176.83666461538462)
		(84.61538461538461, 177.0063046153846)
		(38.46153846153847, 176.17093846153847)
		(46.15384615384615, 177.7225923076923)
		(7.6923076923076925, 177.1969553846154)
		(69.23076923076923, 176.48508615384614)
		(61.53846153846154, 177.11935384615384)
		(76.92307692307693, 177.2118046153846)
		(30.76923076923077, 178.32304153846155)
		(46.15384615384615, 177.60156)
		(92.3076923076923, 177.07165230769232)
		(38.46153846153847, 177.1539646153846)
		(61.53846153846154, 177.02236615384615)
		(30.76923076923077, 178.31851230769232)
		(69.23076923076923, 176.31854153846155)
		(69.23076923076923, 177.81437076923078)
		(7.6923076923076925, 177.33685384615384)
		(0.0, 176.97394461538462)
		(30.76923076923077, 174.6858830769231)
		(23.076923076923077, 177.11656615384615)
		(100.0, 177.26395538461537)
		(76.92307692307693, 175.08827538461537)
		(61.53846153846154, 176.89889384615384)
		(15.384615384615385, 174.26774615384616)
		(7.6923076923076925, 177.80912923076923)
		(69.23076923076923, 175.12682615384617)
		(30.76923076923077, 177.02141384615385)
		(30.76923076923077, 177.30498)
		(23.076923076923077, 177.65096153846153)
		(53.84615384615385, 177.03480615384615)
		(46.15384615384615, 176.01060153846154)
		(46.15384615384615, 176.76984307692308)
		(15.384615384615385, 177.74065384615383)
		(53.84615384615385, 177.19623076923077)
		(38.46153846153847, 178.52592615384614)
		(46.15384615384615, 178.6460123076923)
		(53.84615384615385, 177.74910615384616)
		(30.76923076923077, 177.0140876923077)
		(76.92307692307693, 176.72023384615383)
		(23.076923076923077, 178.37599538461538)
	};

}
\&
\DOUBLEAXIS{Full = {173.78632615384615}, Scatter, title={Snake {\small(70 Type Annotations)}}}{Full = {1.2660815714786908}, Scatter, Right Axis, Scatter Right Label}{
	\addplot+[] coordinates{
		(77.14285714285715, 148.24031076923077)
		(65.71428571428571, 152.56996615384617)
		(97.14285714285714, 151.35855846153845)
		(4.285714285714286, 139.15833846153845)
		(48.57142857142857, 141.85866153846155)
		(70.0, 152.92828615384616)
		(15.714285714285714, 138.97562769230768)
		(24.285714285714285, 141.44269846153847)
		(22.857142857142858, 138.7841523076923)
		(17.142857142857142, 141.24251692307692)
		(52.85714285714286, 146.9992523076923)
		(38.57142857142858, 142.9218276923077)
		(98.57142857142858, 151.7106723076923)
		(37.142857142857146, 141.23192307692307)
		(61.42857142857143, 150.98125384615383)
		(75.71428571428571, 152.06031076923077)
		(40.0, 138.64231076923076)
		(32.857142857142854, 148.49848615384616)
		(87.14285714285714, 148.21643846153847)
		(30.0, 140.74615538461538)
		(58.57142857142858, 146.93851538461539)
		(22.857142857142858, 139.23705846153845)
		(82.85714285714286, 148.0190769230769)
		(14.285714285714285, 146.5344553846154)
		(42.857142857142854, 137.04412461538462)
		(50.0, 140.26587230769232)
		(90.0, 151.96715846153847)
		(87.14285714285714, 155.16162461538462)
		(71.42857142857143, 153.41660153846155)
		(95.71428571428572, 151.42418307692307)
		(1.4285714285714286, 137.8740676923077)
		(80.0, 151.7230523076923)
		(30.0, 139.8903553846154)
		(20.0, 141.29144615384615)
		(84.28571428571429, 149.74332307692308)
		(51.42857142857142, 149.8898153846154)
		(54.285714285714285, 142.7319753846154)
		(82.85714285714286, 151.32469384615385)
		(81.42857142857143, 157.14972153846153)
		(61.42857142857143, 148.19296153846153)
		(68.57142857142857, 151.40998769230768)
		(48.57142857142857, 142.72091076923076)
		(8.571428571428571, 141.50776615384615)
		(45.714285714285715, 140.23538)
		(88.57142857142857, 151.88347692307693)
		(55.714285714285715, 139.14624923076923)
		(21.428571428571427, 139.75436)
		(4.285714285714286, 138.37035076923078)
		(78.57142857142857, 153.13199846153847)
		(64.28571428571429, 150.07781846153847)
		(35.714285714285715, 140.64610769230768)
		(17.142857142857142, 143.38611384615385)
		(72.85714285714285, 141.62324923076923)
		(95.71428571428572, 157.45564769230768)
		(51.42857142857142, 147.82239692307692)
		(11.428571428571429, 152.9018876923077)
		(67.14285714285714, 142.24117846153845)
		(41.42857142857143, 140.31450923076923)
		(60.0, 146.94792307692308)
		(1.4285714285714286, 138.99359076923076)
		(47.14285714285714, 140.97602)
		(31.428571428571427, 150.5234846153846)
		(74.28571428571429, 146.70612461538462)
		(2.857142857142857, 138.8188430769231)
		(12.857142857142856, 141.58609384615386)
		(67.14285714285714, 140.3740446153846)
		(7.142857142857142, 139.09550615384615)
		(85.71428571428571, 151.6125569230769)
		(20.0, 148.88160153846155)
		(7.142857142857142, 138.46062307692307)
		(35.714285714285715, 142.76799538461538)
		(100.0, 151.56647076923076)
		(34.285714285714285, 139.76908615384616)
		(58.57142857142858, 150.2810769230769)
		(0.0, 137.2631353846154)
		(41.42857142857143, 140.23929076923076)
		(44.285714285714285, 140.33402307692307)
		(90.0, 152.1695246153846)
		(27.142857142857142, 139.69166153846155)
		(25.71428571428571, 143.80101692307693)
		(5.714285714285714, 137.9061153846154)
		(45.714285714285715, 146.5614753846154)
		(28.57142857142857, 146.85259076923077)
		(57.14285714285714, 141.95314153846155)
		(10.0, 138.5602046153846)
		(64.28571428571429, 151.6493876923077)
		(77.14285714285715, 153.28524307692308)
		(25.71428571428571, 141.75535538461537)
		(91.42857142857143, 151.70387846153847)
		(74.28571428571429, 138.72943076923076)
		(10.0, 137.80947846153848)
		(38.57142857142858, 142.92664307692309)
		(92.85714285714286, 155.46339384615385)
		(54.285714285714285, 148.13312153846155)
		(12.857142857142856, 144.13727076923078)
		(70.0, 151.57587846153845)
		(94.28571428571428, 155.46103692307693)
		(18.571428571428573, 143.60359846153847)
		(80.0, 155.44901692307693)
		(62.857142857142854, 157.98756923076922)
		(92.85714285714286, 153.45888769230768)
		(32.857142857142854, 143.19363076923077)
	};

}
\\
\DOUBLEAXIS{Full = {250.39650646153848}, Scatter, Scatter Left Label, Scatter Bottom Label, title={SpectralNorm {\small(39 Type Annotations)}}}{Full = {1.104028130236953}, Scatter, Right Axis}{
	\addplot+[] coordinates{
		(89.74358974358975, 226.82492153846155)
		(74.35897435897436, 226.58053076923076)
		(23.076923076923077, 227.15347384615384)
		(38.46153846153847, 227.04938923076924)
		(51.28205128205128, 226.57457076923077)
		(28.205128205128204, 226.63353076923076)
		(66.66666666666666, 226.83652615384617)
		(28.205128205128204, 227.23826615384615)
		(51.28205128205128, 227.02276307692307)
		(17.94871794871795, 227.09706923076922)
		(74.35897435897436, 226.73946)
		(7.6923076923076925, 226.7561876923077)
		(33.33333333333333, 226.90816307692307)
		(12.82051282051282, 227.18390615384615)
		(35.8974358974359, 226.9544446153846)
		(20.51282051282051, 226.95772923076922)
		(82.05128205128204, 226.62398615384615)
		(92.3076923076923, 226.73751076923077)
		(20.51282051282051, 226.69440923076922)
		(56.41025641025641, 226.7377646153846)
		(5.128205128205128, 227.00809384615386)
		(74.35897435897436, 227.10038461538463)
		(89.74358974358975, 227.02226615384615)
		(79.48717948717949, 227.07069384615386)
		(82.05128205128204, 227.01605538461538)
		(7.6923076923076925, 226.92597692307692)
		(2.564102564102564, 226.80759076923076)
		(10.256410256410255, 227.02845692307693)
		(64.1025641025641, 226.78377230769232)
		(38.46153846153847, 226.59366307692306)
		(33.33333333333333, 227.36286615384614)
		(43.58974358974359, 226.76580307692308)
		(23.076923076923077, 226.9082553846154)
		(71.7948717948718, 226.7334123076923)
		(25.64102564102564, 226.8610953846154)
		(79.48717948717949, 226.8877553846154)
		(94.87179487179486, 227.03368307692307)
		(94.87179487179486, 226.5139323076923)
		(58.97435897435898, 226.95309230769232)
		(46.15384615384615, 226.85103384615385)
		(94.87179487179486, 227.02801692307693)
		(38.46153846153847, 226.9835276923077)
		(0.0, 226.80265076923078)
		(87.17948717948718, 226.97202615384614)
		(69.23076923076923, 226.84096153846153)
		(84.61538461538461, 226.97115384615384)
		(56.41025641025641, 227.02202461538462)
		(79.48717948717949, 227.6331876923077)
		(48.717948717948715, 226.8613246153846)
		(30.76923076923077, 227.25885692307693)
		(76.92307692307693, 226.71115230769232)
		(61.53846153846154, 227.52407076923078)
		(61.53846153846154, 227.12770307692307)
		(66.66666666666666, 226.91137384615385)
		(25.64102564102564, 226.71591538461539)
		(20.51282051282051, 226.6186676923077)
		(69.23076923076923, 227.1197276923077)
		(89.74358974358975, 226.77347076923076)
		(58.97435897435898, 226.99096923076922)
		(48.717948717948715, 226.56870307692307)
		(12.82051282051282, 226.75885076923078)
		(5.128205128205128, 226.8497923076923)
		(25.64102564102564, 226.46503076923076)
		(46.15384615384615, 227.1843076923077)
		(2.564102564102564, 226.67950153846155)
		(30.76923076923077, 226.70180307692308)
		(5.128205128205128, 227.01085846153845)
		(92.3076923076923, 227.36036615384614)
		(2.564102564102564, 226.77119076923077)
		(10.256410256410255, 226.60431692307694)
		(87.17948717948718, 226.89827076923078)
		(43.58974358974359, 226.84273076923077)
		(12.82051282051282, 227.05401692307692)
		(76.92307692307693, 226.8348723076923)
		(33.33333333333333, 226.9889353846154)
		(46.15384615384615, 227.1965846153846)
		(71.7948717948718, 226.79121384615385)
		(41.02564102564102, 226.69188615384616)
		(17.94871794871795, 226.99216153846154)
		(58.97435897435898, 227.11883538461538)
		(64.1025641025641, 226.97715846153847)
		(97.43589743589743, 226.8894753846154)
		(41.02564102564102, 226.7412676923077)
		(35.8974358974359, 226.89766923076922)
		(87.17948717948718, 226.90666153846155)
		(69.23076923076923, 227.13126615384616)
		(82.05128205128204, 226.9521476923077)
		(10.256410256410255, 227.08288615384615)
		(15.384615384615385, 227.17519384615386)
		(30.76923076923077, 226.99680307692307)
		(17.94871794871795, 226.84930153846153)
		(15.384615384615385, 226.9622523076923)
		(53.84615384615385, 226.61151846153845)
		(51.28205128205128, 226.9589323076923)
		(97.43589743589743, 227.17191384615384)
		(84.61538461538461, 226.6657)
		(41.02564102564102, 227.0159846153846)
		(100.0, 226.84829846153846)
		(66.66666666666666, 226.7480676923077)
		(61.53846153846154, 226.99118615384614)
		(53.84615384615385, 226.74774923076924)
		(53.84615384615385, 226.8682723076923)
	};

}
\&
\DOUBLEAXIS{Full = {269.00966230769234}, Scatter, Scatter Bottom Label, title={Storage {\small(10 Type Annotations)}}}{Full = {1.1497947639102746}, Scatter, Right Axis}{
	\addplot+[] coordinates{
		(80.0, 209.42293538461539)
		(70.0, 231.67321384615386)
		(80.0, 209.51715846153846)
		(60.0, 210.56156615384614)
		(90.0, 209.40012615384614)
		(40.0, 231.73102461538463)
		(60.0, 211.32083846153847)
		(20.0, 232.88442923076923)
		(80.0, 209.28538307692307)
		(30.0, 233.42917846153847)
		(80.0, 209.31805076923078)
		(40.0, 234.19968923076922)
		(10.0, 234.07856)
		(60.0, 206.24904615384617)
		(60.0, 210.34022153846155)
		(30.0, 226.3926323076923)
		(90.0, 206.6323476923077)
		(40.0, 244.55423846153846)
		(80.0, 209.9479353846154)
		(80.0, 209.74799692307693)
		(0.0, 233.9632)
		(50.0, 217.73828461538463)
		(50.0, 232.83632153846153)
		(20.0, 232.74974307692307)
		(60.0, 208.13414307692307)
		(50.0, 213.83179076923076)
		(20.0, 233.24686307692306)
		(70.0, 233.6284523076923)
		(80.0, 209.49720615384615)
		(80.0, 213.60555384615384)
		(20.0, 205.87155076923077)
		(10.0, 233.63935384615385)
		(30.0, 235.93422)
		(90.0, 209.89741692307692)
		(30.0, 233.73963076923076)
		(30.0, 234.80595846153847)
		(60.0, 209.58012461538462)
		(80.0, 210.97568923076923)
		(70.0, 209.84974461538462)
		(60.0, 234.1275753846154)
		(50.0, 210.48473076923077)
		(30.0, 232.73002)
		(60.0, 209.51862307692306)
		(30.0, 212.88276)
		(30.0, 235.2026323076923)
		(70.0, 207.3595276923077)
		(60.0, 209.8039076923077)
		(70.0, 209.67627384615383)
		(90.0, 209.65978)
		(50.0, 205.8122276923077)
		(100.0, 209.5058123076923)
		(50.0, 232.5969076923077)
		(40.0, 213.6570430769231)
		(10.0, 232.63717076923078)
		(10.0, 233.5437276923077)
		(40.0, 233.96808153846155)
		(50.0, 209.56650307692308)
		(70.0, 209.5801676923077)
		(20.0, 233.39773538461537)
		(50.0, 209.9609769230769)
		(20.0, 235.68544)
		(70.0, 212.43143846153845)
		(90.0, 209.9727353846154)
		(10.0, 234.1343646153846)
		(20.0, 232.93066923076924)
		(60.0, 230.45046769230768)
		(90.0, 210.54276307692308)
		(90.0, 209.53164)
		(40.0, 233.06428923076922)
		(40.0, 232.7180753846154)
		(10.0, 231.01360153846153)
		(20.0, 232.70123076923076)
		(80.0, 233.38258615384615)
		(40.0, 235.44752923076922)
		(20.0, 236.54592)
		(30.0, 232.55863846153846)
		(50.0, 233.56723538461537)
		(40.0, 233.66649384615386)
		(40.0, 234.4443723076923)
		(70.0, 209.79225538461537)
		(10.0, 232.23065076923078)
		(30.0, 235.05510153846154)
		(30.0, 209.47526769230768)
		(80.0, 209.51779076923077)
		(50.0, 235.16251846153847)
		(40.0, 209.33778615384617)
		(50.0, 209.68088153846153)
		(70.0, 205.90892615384615)
		(60.0, 234.77026307692307)
		(90.0, 209.29608461538461)
		(90.0, 209.6274123076923)
		(20.0, 234.35426615384614)
		(70.0, 210.06052461538462)
		(70.0, 211.44169384615384)
		(10.0, 237.7799276923077)
		(10.0, 235.93800307692308)
		(20.0, 233.7396523076923)
	};

}
\&
\DOUBLEAXIS{Full = {272.46147923076927}, Scatter, Scatter Bottom Label, title={Towers {\small(30 Type Annotations)}}}{Full = {1.2168193290864888}, Scatter, Right Axis, Scatter Right Label}{
	\addplot+[] coordinates{
		(0.0, 223.91284615384615)
		(13.333333333333334, 217.49595846153846)
		(40.0, 239.02276307692307)
		(86.66666666666667, 241.50708923076922)
		(96.66666666666667, 240.40972153846153)
		(23.333333333333332, 212.18949076923076)
		(83.33333333333334, 239.17597692307692)
		(16.666666666666664, 213.0431846153846)
		(50.0, 241.30826923076924)
		(6.666666666666667, 219.6328969230769)
		(30.0, 213.94085384615386)
		(23.333333333333332, 238.51851692307693)
		(93.33333333333333, 222.21682153846155)
		(86.66666666666667, 222.66867692307693)
		(40.0, 240.46041846153847)
		(96.66666666666667, 239.1412753846154)
		(16.666666666666664, 214.29190923076922)
		(60.0, 238.75083230769232)
		(80.0, 239.97922461538462)
		(36.666666666666664, 214.92647538461537)
		(76.66666666666667, 240.44008461538462)
		(90.0, 239.15996153846154)
		(50.0, 214.5379523076923)
		(3.3333333333333335, 213.5648723076923)
		(66.66666666666666, 223.5774076923077)
		(66.66666666666666, 223.30894615384616)
		(36.666666666666664, 213.65870615384614)
		(73.33333333333333, 240.53868307692306)
		(33.33333333333333, 240.2524369230769)
		(36.666666666666664, 231.6886846153846)
		(90.0, 238.90554923076922)
		(50.0, 241.73956923076923)
		(20.0, 213.0876153846154)
		(3.3333333333333335, 212.8868723076923)
		(40.0, 213.23785846153845)
		(70.0, 238.02116)
		(3.3333333333333335, 213.44038923076923)
		(30.0, 215.1937276923077)
		(60.0, 239.42718)
		(6.666666666666667, 220.7634369230769)
		(20.0, 213.87923384615385)
		(46.666666666666664, 230.58668923076922)
		(70.0, 239.74950307692308)
		(73.33333333333333, 239.1098723076923)
		(53.333333333333336, 231.15082153846154)
		(30.0, 229.5979123076923)
		(26.666666666666668, 222.11592)
		(60.0, 241.0847246153846)
		(30.0, 214.71219846153846)
		(83.33333333333334, 239.0887953846154)
		(33.33333333333333, 239.96599846153845)
		(73.33333333333333, 231.44867076923077)
		(10.0, 212.41576153846154)
		(66.66666666666666, 234.5037846153846)
		(3.3333333333333335, 212.0765046153846)
		(13.333333333333334, 212.67780153846155)
		(23.333333333333332, 215.50087692307693)
		(70.0, 241.08880923076924)
		(56.666666666666664, 239.25831538461537)
		(43.333333333333336, 239.21967538461539)
		(63.33333333333333, 242.29971076923076)
		(76.66666666666667, 247.69225384615385)
		(20.0, 213.47112923076924)
		(10.0, 212.44271538461538)
		(53.333333333333336, 216.41477846153848)
		(76.66666666666667, 239.0761)
		(83.33333333333334, 238.77756615384615)
		(6.666666666666667, 212.36898769230768)
		(46.666666666666664, 212.26865384615385)
		(80.0, 221.88496307692307)
		(33.33333333333333, 212.53464615384615)
		(90.0, 241.25777076923077)
		(80.0, 238.7341153846154)
		(40.0, 223.1754646153846)
		(13.333333333333334, 211.6085923076923)
		(10.0, 212.49878923076923)
		(70.0, 238.96836615384615)
		(90.0, 234.27347538461538)
		(46.666666666666664, 232.8668846153846)
		(16.666666666666664, 213.6557276923077)
		(93.33333333333333, 238.5782246153846)
		(53.333333333333336, 223.39519384615386)
		(26.666666666666668, 230.04194923076923)
		(16.666666666666664, 220.77631538461537)
		(53.333333333333336, 211.90716307692307)
		(86.66666666666667, 240.40315076923076)
		(43.333333333333336, 213.51231692307692)
		(60.0, 230.69135538461538)
		(96.66666666666667, 238.44070153846152)
		(43.333333333333336, 238.91382307692308)
		(63.33333333333333, 219.7595353846154)
		(93.33333333333333, 240.04356923076924)
		(10.0, 235.8168446153846)
		(63.33333333333333, 234.73901538461539)
		(100.0, 241.3871323076923)
		(26.666666666666668, 237.2334723076923)
		(83.33333333333334, 234.82378307692306)
		(46.666666666666664, 235.36269384615386)
		(23.333333333333332, 213.5879323076923)
		(56.666666666666664, 211.94837076923076)
		(56.666666666666664, 212.40291846153846)
		(76.66666666666667, 239.01577230769232)
	};

}
}
 	\caption{Graphs of (at most) 102 configurations in the typing lattices for each benchmark. Time is measured as the mean of the 351\textsuperscript{st} to the 1,000\textsuperscript{th} benchmark iteration under a single invocation of Moth (lower is better).}
	\label{f:full}
\end{figure*}

Before we start to investigate specific type annotations,
we present the performance measurements of our sample of the typing lattice configurations in figure~\ref{f:full}.
These results are the foundation for a more detailed analysis.

Following \cite{vitousek-transient-arXive-2019}, the points on each graph in figure~\ref{f:full} show the average execution of each individual configuration. The x-axis represents the proportion of type annotations for each configuration, with the left- and right-most points showing the times for the fully untyped and typed configurations respectively. The execution time in milliseconds is shown on the left y-axes, and time relative to the fully untyped configuration is shown on the right y-axes.

Most of the graphs are essentially horizontal lines, indicating that the overhead of including type annotations is negligible. The plots for CD and Richards show a roughly linear increase, i.e.\ for these two benchmarks, adding type annotations reduces performance linearly. On the other hand, the plots for Go, Permute, DeltaBlue, and Storage show \emph{decreases}: i.e. adding more type annotations \emph{improves} performance of these benchmarks.

By inspecting the scatterplots, we observe that the performance of
Moth in almost any configuration of a benchmark is bounded by
the performance in the untyped and fully-typed configuration.
That is, for these benchmarks on transient type checks,
measuring just the untyped and fully-typed configurations
would provide excellent estimates of a benchmark's
performance bounds.  This is different to the experience
of other kinds of gradual typing, where the best, and most
importantly worst, configurations are not always those
fully typed or fully untyped \cite{Greenman2019jfp}.
However, the Richards benchmark \textit{does} have
sections outside these bounds, and isolated executions of
a couple of others (Fannkuch and DeltaBlue) are also outliers.
We believe that the 0\% and 100\% bounds are nonetheless a
reasonable heuristic estimator in most cases, and we will
examine the outlying benchmarks further.

Of particular note is that some of the graphs (Permute, Storage, Towers, and Richards) show bimodal performance profiles, that is, two separate roughly-horizontal lines. Presumably Moth can remove all the overhead from some configurations, but in others there must be one or more type checks that could not be optimised away. List shows three performance modes: the graph consists of three mostly flat lines, at about 1 times slower, 1.5 times slower, and 1.8 times slower than untyped code.

\section{Identifying Type Annotations With Signification Performance Impact}
\label{s-individual}
\begin{figure*}
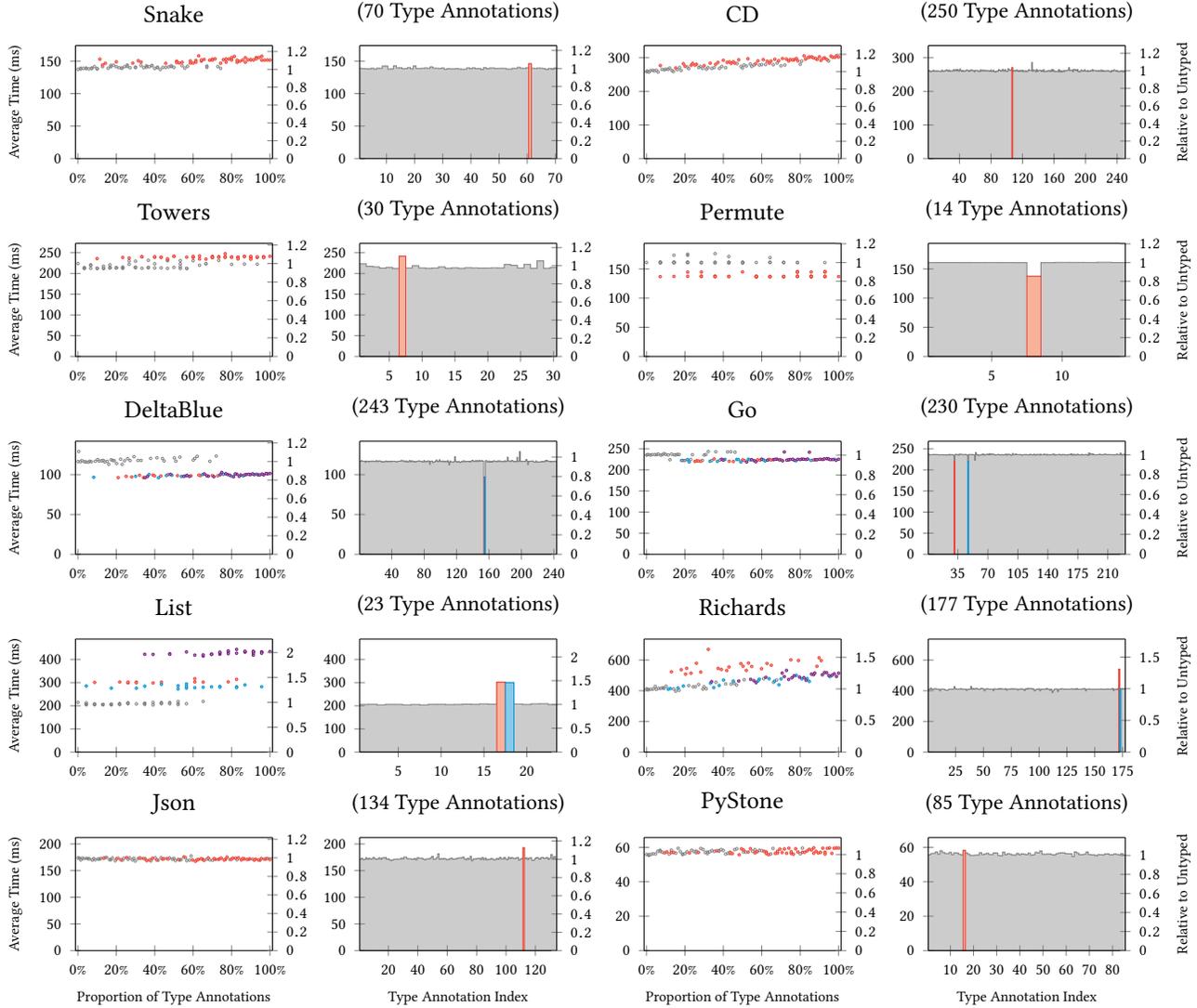

\GRAPHS{
\DOUBLEAXIS{Pattern = {173.78632615384615}, Scatter, Scatter Left Label = {173.78632615384615}, title={\null\hfill Snake}}{Pattern = {1.2660815714786908}, Scatter, Right Axis = {1.2660815714786908}}{
	\addplot+[] coordinates{
		(4.285714285714286, 139.15833846153845)
		(48.57142857142857, 141.85866153846155)
		(15.714285714285714, 138.97562769230768)
		(24.285714285714285, 141.44269846153847)
		(22.857142857142858, 138.7841523076923)
		(17.142857142857142, 141.24251692307692)
		(38.57142857142858, 142.9218276923077)
		(37.142857142857146, 141.23192307692307)
		(40.0, 138.64231076923076)
		(30.0, 140.74615538461538)
		(22.857142857142858, 139.23705846153845)
		(42.857142857142854, 137.04412461538462)
		(50.0, 140.26587230769232)
		(1.4285714285714286, 137.8740676923077)
		(30.0, 139.8903553846154)
		(20.0, 141.29144615384615)
		(54.285714285714285, 142.7319753846154)
		(48.57142857142857, 142.72091076923076)
		(8.571428571428571, 141.50776615384615)
		(45.714285714285715, 140.23538)
		(55.714285714285715, 139.14624923076923)
		(21.428571428571427, 139.75436)
		(4.285714285714286, 138.37035076923078)
		(35.714285714285715, 140.64610769230768)
		(17.142857142857142, 143.38611384615385)
		(72.85714285714285, 141.62324923076923)
		(67.14285714285714, 142.24117846153845)
		(41.42857142857143, 140.31450923076923)
		(1.4285714285714286, 138.99359076923076)
		(47.14285714285714, 140.97602)
		(2.857142857142857, 138.8188430769231)
		(12.857142857142856, 141.58609384615386)
		(67.14285714285714, 140.3740446153846)
		(7.142857142857142, 139.09550615384615)
		(7.142857142857142, 138.46062307692307)
		(35.714285714285715, 142.76799538461538)
		(34.285714285714285, 139.76908615384616)
		(0.0, 137.2631353846154)
		(41.42857142857143, 140.23929076923076)
		(44.285714285714285, 140.33402307692307)
		(27.142857142857142, 139.69166153846155)
		(25.71428571428571, 143.80101692307693)
		(5.714285714285714, 137.9061153846154)
		(57.14285714285714, 141.95314153846155)
		(10.0, 138.5602046153846)
		(25.71428571428571, 141.75535538461537)
		(74.28571428571429, 138.72943076923076)
		(10.0, 137.80947846153848)
		(38.57142857142858, 142.92664307692309)
		(18.571428571428573, 143.60359846153847)
		(32.857142857142854, 143.19363076923077)
	};

	\addplot+[Pattern A] coordinates{
		(77.14285714285715, 148.24031076923077)
		(65.71428571428571, 152.56996615384617)
		(97.14285714285714, 151.35855846153845)
		(70.0, 152.92828615384616)
		(52.85714285714286, 146.9992523076923)
		(98.57142857142858, 151.7106723076923)
		(61.42857142857143, 150.98125384615383)
		(75.71428571428571, 152.06031076923077)
		(32.857142857142854, 148.49848615384616)
		(87.14285714285714, 148.21643846153847)
		(58.57142857142858, 146.93851538461539)
		(82.85714285714286, 148.0190769230769)
		(14.285714285714285, 146.5344553846154)
		(90.0, 151.96715846153847)
		(87.14285714285714, 155.16162461538462)
		(71.42857142857143, 153.41660153846155)
		(95.71428571428572, 151.42418307692307)
		(80.0, 151.7230523076923)
		(84.28571428571429, 149.74332307692308)
		(51.42857142857142, 149.8898153846154)
		(82.85714285714286, 151.32469384615385)
		(81.42857142857143, 157.14972153846153)
		(61.42857142857143, 148.19296153846153)
		(68.57142857142857, 151.40998769230768)
		(88.57142857142857, 151.88347692307693)
		(78.57142857142857, 153.13199846153847)
		(64.28571428571429, 150.07781846153847)
		(95.71428571428572, 157.45564769230768)
		(51.42857142857142, 147.82239692307692)
		(11.428571428571429, 152.9018876923077)
		(60.0, 146.94792307692308)
		(31.428571428571427, 150.5234846153846)
		(74.28571428571429, 146.70612461538462)
		(85.71428571428571, 151.6125569230769)
		(20.0, 148.88160153846155)
		(100.0, 151.56647076923076)
		(58.57142857142858, 150.2810769230769)
		(90.0, 152.1695246153846)
		(45.714285714285715, 146.5614753846154)
		(28.57142857142857, 146.85259076923077)
		(64.28571428571429, 151.6493876923077)
		(77.14285714285715, 153.28524307692308)
		(91.42857142857143, 151.70387846153847)
		(92.85714285714286, 155.46339384615385)
		(54.285714285714285, 148.13312153846155)
		(12.857142857142856, 144.13727076923078)
		(70.0, 151.57587846153845)
		(94.28571428571428, 155.46103692307693)
		(80.0, 155.44901692307693)
		(62.857142857142854, 157.98756923076922)
		(92.85714285714286, 153.45888769230768)
	};

	\addplot+[Pattern B] coordinates{
	};

	\addplot+[Pattern AB] coordinates{
	};

}
\&
\DOUBLEAXIS{Pattern = {173.78632615384615}, Column = {10} = {173.78632615384615}{10}, title={\null\hfill{\small(70 Type Annotations)}}}{Pattern = {1.2537865262318229}, Column = {10}, Right Axis = {1.2537865262318229}}{
	\addplot+[] coordinates{
		(0.5, 138.80607384615385)
		(1.5, 138.27838615384616)
		(2.5, 138.10170769230768)
		(3.5, 137.81825692307692)
		(4.5, 139.08932)
		(5.5, 137.5257246153846)
		(6.5, 139.11592769230768)
		(7.5, 138.81304923076922)
		(8.5, 142.27389846153847)
		(9.5, 142.22963076923077)
		(10.5, 137.57934615384616)
		(11.5, 137.98507846153845)
		(12.5, 142.59833846153848)
		(13.5, 139.36599384615386)
		(14.5, 139.58753076923077)
		(15.5, 138.1967446153846)
		(16.5, 138.54532)
		(17.5, 140.15573384615385)
		(18.5, 137.99420615384616)
		(19.5, 142.41830769230768)
		(20.5, 138.5490046153846)
		(21.5, 138.72084923076923)
		(22.5, 138.93402153846154)
		(23.5, 139.32421076923077)
		(24.5, 138.84688615384616)
		(25.5, 140.91193692307692)
		(26.5, 139.0592630769231)
		(27.5, 138.9477723076923)
		(28.5, 138.5279676923077)
		(29.5, 139.3490230769231)
		(30.5, 137.37452923076924)
		(31.5, 138.26954307692307)
		(32.5, 137.98064)
		(33.5, 137.88620461538463)
		(34.5, 138.5932646153846)
		(35.5, 137.81735692307691)
		(36.5, 137.42848615384617)
		(37.5, 137.50793076923077)
		(38.5, 139.40098461538463)
		(39.5, 138.51059076923076)
		(40.5, 139.41079846153846)
		(41.5, 138.42385384615383)
		(42.5, 138.65668461538462)
		(43.5, 136.94454307692308)
		(44.5, 139.83599692307692)
		(45.5, 138.20852461538462)
		(46.5, 139.29564461538462)
		(47.5, 139.19913538461537)
		(48.5, 138.59387384615385)
		(49.5, 138.7757169230769)
		(50.5, 138.87965692307694)
		(51.5, 138.81602153846154)
		(52.5, 137.69192615384614)
		(53.5, 137.56085692307693)
		(54.5, 138.87279692307692)
		(55.5, 138.75921692307693)
		(56.5, 140.43454307692306)
		(57.5, 139.03106461538462)
		(58.5, 138.7214846153846)
		(59.5, 139.44407692307692)
		(60.5, 145.86855538461538)
		(61.5, 139.01669846153845)
		(62.5, 138.57935538461538)
		(63.5, 139.20648461538462)
		(64.5, 139.0097876923077)
		(65.5, 138.47870461538463)
		(66.5, 137.11559846153847)
		(67.5, 138.7116276923077)
		(68.5, 137.47666153846154)
		(69.5, 139.17956)
		(70.5, 0)
		(0.5, 0)
	};

	\addplot+[Pattern A] coordinates{
		(60.5, 0)
		(60.5, 145.86855538461538)
		(61.5, 0)
		(60.5, 0)
	};

}
\&
\DOUBLEAXIS{Pattern = {336.1811403076924}, Scatter = {336.1811403076924}, title={\null\hfill CD}}{Pattern = {1.3002110091965295}, Scatter, Right Axis = {1.3002110091965295}}{
	\addplot+[] coordinates{
		(31.2, 266.2493323076923)
		(41.199999999999996, 270.38674)
		(13.200000000000001, 262.11080615384617)
		(8.0, 265.36509538461536)
		(50.0, 269.28614153846155)
		(68.0, 284.52114615384613)
		(32.0, 270.6832830769231)
		(9.2, 264.0376276923077)
		(0.4, 260.32499384615386)
		(27.200000000000003, 271.03388)
		(94.8, 290.4944123076923)
		(6.0, 257.55894461538463)
		(39.2, 274.09852)
		(17.2, 271.1908661538462)
		(47.199999999999996, 282.68546)
		(38.0, 268.4655492307692)
		(54.0, 281.56948153846156)
		(28.000000000000004, 274.2415384615385)
		(62.0, 279.9813323076923)
		(22.0, 267.38671384615384)
		(66.0, 279.49088461538463)
		(72.0, 278.2168938461538)
		(58.8, 278.15867692307694)
		(18.0, 262.2581553846154)
		(12.0, 269.1274846153846)
		(35.199999999999996, 270.49714153846156)
		(34.0, 268.33237846153844)
		(14.000000000000002, 259.02849692307694)
		(0.0, 258.55890923076925)
		(11.200000000000001, 265.8217584615385)
		(70.0, 275.90223692307694)
		(5.2, 262.79028923076925)
		(3.2, 262.8688230769231)
		(60.0, 279.6458)
		(80.0, 280.71709384615383)
		(21.2, 272.35324153846153)
		(54.800000000000004, 274.41198)
		(29.2, 268.41496461538463)
		(56.00000000000001, 278.07138615384616)
		(2.0, 261.20972)
		(80.80000000000001, 290.08761384615383)
		(52.0, 275.33246153846153)
		(1.2, 257.4973830769231)
		(4.0, 260.95384615384614)
		(49.2, 274.2494723076923)
		(10.0, 265.90955692307693)
		(16.0, 264.49719230769233)
		(50.8, 278.43575076923076)
	};

	\addplot+[Pattern A] coordinates{
		(43.2, 287.6336830769231)
		(44.0, 283.2179061538462)
		(64.0, 283.7796276923077)
		(96.0, 304.83217692307693)
		(86.0, 302.5626092307692)
		(40.0, 292.44052153846155)
		(23.200000000000003, 277.6082492307692)
		(19.2, 282.4954723076923)
		(96.8, 301.8342)
		(24.0, 283.25242000000003)
		(26.0, 280.4902446153846)
		(68.8, 290.98762923076924)
		(74.0, 290.59594)
		(92.80000000000001, 299.2719476923077)
		(36.0, 280.2516753846154)
		(74.8, 294.5701876923077)
		(98.8, 305.2704876923077)
		(48.0, 293.9522630769231)
		(7.199999999999999, 276.86243538461537)
		(82.8, 295.8310707692308)
		(98.0, 301.25852153846154)
		(46.0, 283.8635615384615)
		(94.0, 294.4267246153846)
		(72.8, 296.2695076923077)
		(78.8, 290.94370153846154)
		(66.8, 295.6016907692308)
		(57.99999999999999, 290.69096615384615)
		(88.8, 294.95501538461536)
		(30.0, 276.62688615384616)
		(45.2, 288.01250923076924)
		(84.8, 302.1272923076923)
		(90.0, 302.78504153846154)
		(78.0, 297.94097384615384)
		(90.8, 292.0766184615385)
		(37.2, 286.171)
		(52.800000000000004, 296.95732)
		(70.8, 297.7896723076923)
		(25.2, 278.0338446153846)
		(62.8, 289.86449846153846)
		(56.8, 282.2968076923077)
		(88.0, 298.08496)
		(33.2, 283.93842615384614)
		(64.8, 290.33412)
		(60.8, 294.4023569230769)
		(86.8, 299.77858615384616)
		(76.8, 293.8412369230769)
		(42.0, 286.10308615384616)
		(15.2, 269.46322)
		(92.0, 298.38041076923076)
		(76.0, 293.4293784615385)
		(82.0, 305.6192184615385)
		(84.0, 300.23830615384617)
		(100.0, 302.12799384615386)
		(20.0, 274.7902107692308)
	};

	\addplot+[Pattern B] coordinates{
	};

	\addplot+[Pattern AB] coordinates{
	};

}
\&
\DOUBLEAXIS{Pattern = {336.1811403076924}, Column = {40} = {336.1811403076924}{40}, title={\null\hfill{\small(250 Type Annotations)}}}{Pattern = {1.2877479358811348}, Column = {40}, Right Axis, Column Right Label = {1.2877479358811348}}{
	\addplot+[] coordinates{
		(0.5, 261.7250953846154)
		(1.5, 261.0607692307692)
		(2.5, 258.55636923076923)
		(3.5, 260.6138507692308)
		(4.5, 259.10236615384616)
		(5.5, 259.57930153846155)
		(6.5, 263.1671969230769)
		(7.5, 260.34418307692306)
		(8.5, 260.5794584615385)
		(9.5, 260.74379230769233)
		(10.5, 259.22296153846156)
		(11.5, 260.91009692307694)
		(12.5, 260.31644307692306)
		(13.5, 260.4411553846154)
		(14.5, 259.65093384615386)
		(15.5, 258.73968615384615)
		(16.5, 260.34544)
		(17.5, 263.55194923076925)
		(18.5, 259.5484323076923)
		(19.5, 259.1901092307692)
		(20.5, 261.24210307692306)
		(21.5, 260.8538646153846)
		(22.5, 262.7752123076923)
		(23.5, 264.02183846153844)
		(24.5, 261.2796246153846)
		(25.5, 260.24805076923076)
		(26.5, 263.45480923076923)
		(27.5, 259.58022153846156)
		(28.5, 260.6737476923077)
		(29.5, 260.61053076923076)
		(30.5, 260.6411769230769)
		(31.5, 261.59788615384616)
		(32.5, 265.01936)
		(33.5, 261.05897230769233)
		(34.5, 262.38107846153844)
		(35.5, 264.5881676923077)
		(36.5, 260.96536)
		(37.5, 264.7090076923077)
		(38.5, 261.55404153846155)
		(39.5, 260.3580707692308)
		(40.5, 266.5269784615385)
		(41.5, 261.75022615384614)
		(42.5, 260.32756153846157)
		(43.5, 261.8594276923077)
		(44.5, 259.03314923076925)
		(45.5, 260.9354430769231)
		(46.5, 261.00876153846156)
		(47.5, 260.42949076923077)
		(48.5, 263.52724923076926)
		(49.5, 262.6739292307692)
		(50.5, 262.92650615384616)
		(51.5, 259.13726)
		(52.5, 261.76257846153845)
		(53.5, 260.7106123076923)
		(54.5, 263.90150307692306)
		(55.5, 263.33201846153844)
		(56.5, 261.4874123076923)
		(57.5, 262.35242)
		(58.5, 258.31643076923075)
		(59.5, 260.7665215384615)
		(60.5, 262.77271076923074)
		(61.5, 262.09373384615384)
		(62.5, 261.84189076923076)
		(63.5, 259.5936446153846)
		(64.5, 260.36813538461536)
		(65.5, 258.5436)
		(66.5, 264.06114615384615)
		(67.5, 260.8836984615385)
		(68.5, 263.0377092307692)
		(69.5, 260.43696)
		(70.5, 262.92105692307695)
		(71.5, 263.1372907692308)
		(72.5, 260.2457753846154)
		(73.5, 261.3876861538462)
		(74.5, 262.6000015384615)
		(75.5, 260.80731846153844)
		(76.5, 260.64741692307695)
		(77.5, 261.0330584615385)
		(78.5, 259.33861538461537)
		(79.5, 263.8102723076923)
		(80.5, 259.60852923076925)
		(81.5, 259.5199969230769)
		(82.5, 262.92519384615383)
		(83.5, 258.65128153846155)
		(84.5, 259.97894)
		(85.5, 257.5307430769231)
		(86.5, 259.58474461538464)
		(87.5, 260.7510184615385)
		(88.5, 259.93521846153845)
		(89.5, 264.4961907692308)
		(90.5, 262.9584276923077)
		(91.5, 262.05498)
		(92.5, 259.8925276923077)
		(93.5, 261.77554153846154)
		(94.5, 265.7971769230769)
		(95.5, 261.0979923076923)
		(96.5, 260.7577276923077)
		(97.5, 264.6700246153846)
		(98.5, 259.9643938461538)
		(99.5, 259.90332)
		(100.5, 259.0021030769231)
		(101.5, 267.4224938461538)
		(102.5, 263.79465076923077)
		(103.5, 262.8581461538462)
		(104.5, 262.17973384615385)
		(105.5, 264.1481430769231)
		(106.5, 270.83676153846153)
		(107.5, 258.6313076923077)
		(108.5, 261.46268)
		(109.5, 262.61113384615385)
		(110.5, 260.4249184615385)
		(111.5, 261.4289830769231)
		(112.5, 264.0808169230769)
		(113.5, 260.70320153846154)
		(114.5, 259.11989384615384)
		(115.5, 261.52360153846155)
		(116.5, 259.6403923076923)
		(117.5, 261.8290907692308)
		(118.5, 263.0216646153846)
		(119.5, 258.87958)
		(120.5, 262.5275461538462)
		(121.5, 258.7928830769231)
		(122.5, 259.41085076923076)
		(123.5, 261.32320153846155)
		(124.5, 258.1358230769231)
		(125.5, 260.4830753846154)
		(126.5, 260.2779061538462)
		(127.5, 264.70506615384613)
		(128.5, 262.18895076923076)
		(129.5, 262.06171384615385)
		(130.5, 262.2325430769231)
		(131.5, 285.9759369230769)
		(132.5, 259.18143846153845)
		(133.5, 263.4097061538462)
		(134.5, 263.45266153846154)
		(135.5, 260.80565384615386)
		(136.5, 258.0494323076923)
		(137.5, 265.2024830769231)
		(138.5, 259.42552461538463)
		(139.5, 261.09343384615386)
		(140.5, 262.1138676923077)
		(141.5, 262.6931553846154)
		(142.5, 262.9660846153846)
		(143.5, 259.68836923076924)
		(144.5, 261.9991492307692)
		(145.5, 259.1524646153846)
		(146.5, 264.46004)
		(147.5, 258.7931184615385)
		(148.5, 259.22300461538464)
		(149.5, 260.4305538461538)
		(150.5, 259.00935384615383)
		(151.5, 260.6512430769231)
		(152.5, 261.18598153846153)
		(153.5, 262.1369846153846)
		(154.5, 258.75796)
		(155.5, 260.26739692307694)
		(156.5, 264.4828123076923)
		(157.5, 260.10213076923077)
		(158.5, 264.3383646153846)
		(159.5, 260.6977046153846)
		(160.5, 258.27980615384615)
		(161.5, 260.30607692307694)
		(162.5, 259.39574307692305)
		(163.5, 258.77935692307693)
		(164.5, 260.02580153846156)
		(165.5, 264.69257384615383)
		(166.5, 257.61136153846155)
		(167.5, 261.7832523076923)
		(168.5, 260.28488923076924)
		(169.5, 257.3412846153846)
		(170.5, 260.50106)
		(171.5, 261.14108923076924)
		(172.5, 258.56590923076925)
		(173.5, 261.86016461538463)
		(174.5, 261.8463123076923)
		(175.5, 261.7625615384615)
		(176.5, 261.95256615384613)
		(177.5, 263.3646123076923)
		(178.5, 270.7054046153846)
		(179.5, 261.68932461538463)
		(180.5, 262.91158)
		(181.5, 262.2551169230769)
		(182.5, 263.52810307692306)
		(183.5, 258.52014615384616)
		(184.5, 264.3039815384615)
		(185.5, 265.63548153846153)
		(186.5, 262.56459384615385)
		(187.5, 263.84920307692306)
		(188.5, 261.07484923076925)
		(189.5, 264.58951538461537)
		(190.5, 258.48602)
		(191.5, 259.39127846153843)
		(192.5, 257.99788615384614)
		(193.5, 262.31752153846156)
		(194.5, 263.81489384615384)
		(195.5, 259.9261630769231)
		(196.5, 260.19851384615384)
		(197.5, 261.49238)
		(198.5, 264.1565276923077)
		(199.5, 260.3649923076923)
		(200.5, 260.19609692307694)
		(201.5, 259.1243369230769)
		(202.5, 262.4226738461538)
		(203.5, 259.6087076923077)
		(204.5, 259.5264769230769)
		(205.5, 261.7275107692308)
		(206.5, 260.58961846153846)
		(207.5, 260.96471846153844)
		(208.5, 264.8588876923077)
		(209.5, 261.0808292307692)
		(210.5, 261.72443076923076)
		(211.5, 262.10994307692306)
		(212.5, 262.0715507692308)
		(213.5, 260.24121692307693)
		(214.5, 260.9568830769231)
		(215.5, 260.45068)
		(216.5, 262.0428753846154)
		(217.5, 260.2808430769231)
		(218.5, 260.2677061538462)
		(219.5, 263.81296615384616)
		(220.5, 259.2586630769231)
		(221.5, 260.4863030769231)
		(222.5, 262.5443369230769)
		(223.5, 259.92299076923075)
		(224.5, 259.79096)
		(225.5, 260.8448569230769)
		(226.5, 261.1601276923077)
		(227.5, 261.49986615384614)
		(228.5, 260.2684353846154)
		(229.5, 260.88307846153845)
		(230.5, 259.1456707692308)
		(231.5, 257.52304769230767)
		(232.5, 261.85165538461536)
		(233.5, 260.58038)
		(234.5, 259.4907769230769)
		(235.5, 263.5455276923077)
		(236.5, 263.63054153846156)
		(237.5, 261.9904523076923)
		(238.5, 259.96132615384613)
		(239.5, 261.4462476923077)
		(240.5, 260.64758)
		(241.5, 258.0985830769231)
		(242.5, 259.93255230769233)
		(243.5, 259.1185738461538)
		(244.5, 259.75987692307694)
		(245.5, 259.81572)
		(246.5, 259.86269384615383)
		(247.5, 257.67873076923075)
		(248.5, 260.3241261538462)
		(249.5, 258.9433123076923)
		(250.5, 0)
		(0.5, 0)
	};

	\addplot+[Pattern A] coordinates{
		(106.5, 0)
		(106.5, 270.83676153846153)
		(107.5, 0)
		(106.5, 0)
	};

}
\\
\DOUBLEAXIS{Pattern = {272.46147923076927}, Scatter, Scatter Left Label = {272.46147923076927}, title={\null\hfill Towers}}{Pattern = {1.2168193290864888}, Scatter, Right Axis = {1.2168193290864888}}{
	\addplot+[] coordinates{
		(0.0, 223.91284615384615)
		(13.333333333333334, 217.49595846153846)
		(23.333333333333332, 212.18949076923076)
		(16.666666666666664, 213.0431846153846)
		(6.666666666666667, 219.6328969230769)
		(30.0, 213.94085384615386)
		(93.33333333333333, 222.21682153846155)
		(86.66666666666667, 222.66867692307693)
		(16.666666666666664, 214.29190923076922)
		(36.666666666666664, 214.92647538461537)
		(50.0, 214.5379523076923)
		(3.3333333333333335, 213.5648723076923)
		(66.66666666666666, 223.5774076923077)
		(66.66666666666666, 223.30894615384616)
		(36.666666666666664, 213.65870615384614)
		(36.666666666666664, 231.6886846153846)
		(20.0, 213.0876153846154)
		(3.3333333333333335, 212.8868723076923)
		(40.0, 213.23785846153845)
		(3.3333333333333335, 213.44038923076923)
		(30.0, 215.1937276923077)
		(6.666666666666667, 220.7634369230769)
		(20.0, 213.87923384615385)
		(46.666666666666664, 230.58668923076922)
		(53.333333333333336, 231.15082153846154)
		(30.0, 229.5979123076923)
		(26.666666666666668, 222.11592)
		(30.0, 214.71219846153846)
		(73.33333333333333, 231.44867076923077)
		(10.0, 212.41576153846154)
		(3.3333333333333335, 212.0765046153846)
		(13.333333333333334, 212.67780153846155)
		(23.333333333333332, 215.50087692307693)
		(20.0, 213.47112923076924)
		(10.0, 212.44271538461538)
		(53.333333333333336, 216.41477846153848)
		(6.666666666666667, 212.36898769230768)
		(46.666666666666664, 212.26865384615385)
		(80.0, 221.88496307692307)
		(33.33333333333333, 212.53464615384615)
		(40.0, 223.1754646153846)
		(13.333333333333334, 211.6085923076923)
		(10.0, 212.49878923076923)
		(46.666666666666664, 232.8668846153846)
		(16.666666666666664, 213.6557276923077)
		(53.333333333333336, 223.39519384615386)
		(26.666666666666668, 230.04194923076923)
		(16.666666666666664, 220.77631538461537)
		(53.333333333333336, 211.90716307692307)
		(43.333333333333336, 213.51231692307692)
		(60.0, 230.69135538461538)
		(63.33333333333333, 219.7595353846154)
		(23.333333333333332, 213.5879323076923)
		(56.666666666666664, 211.94837076923076)
		(56.666666666666664, 212.40291846153846)
	};

	\addplot+[Pattern A] coordinates{
		(40.0, 239.02276307692307)
		(86.66666666666667, 241.50708923076922)
		(96.66666666666667, 240.40972153846153)
		(83.33333333333334, 239.17597692307692)
		(50.0, 241.30826923076924)
		(23.333333333333332, 238.51851692307693)
		(40.0, 240.46041846153847)
		(96.66666666666667, 239.1412753846154)
		(60.0, 238.75083230769232)
		(80.0, 239.97922461538462)
		(76.66666666666667, 240.44008461538462)
		(90.0, 239.15996153846154)
		(73.33333333333333, 240.53868307692306)
		(33.33333333333333, 240.2524369230769)
		(90.0, 238.90554923076922)
		(50.0, 241.73956923076923)
		(70.0, 238.02116)
		(60.0, 239.42718)
		(70.0, 239.74950307692308)
		(73.33333333333333, 239.1098723076923)
		(60.0, 241.0847246153846)
		(83.33333333333334, 239.0887953846154)
		(33.33333333333333, 239.96599846153845)
		(66.66666666666666, 234.5037846153846)
		(70.0, 241.08880923076924)
		(56.666666666666664, 239.25831538461537)
		(43.333333333333336, 239.21967538461539)
		(63.33333333333333, 242.29971076923076)
		(76.66666666666667, 247.69225384615385)
		(76.66666666666667, 239.0761)
		(83.33333333333334, 238.77756615384615)
		(90.0, 241.25777076923077)
		(80.0, 238.7341153846154)
		(70.0, 238.96836615384615)
		(90.0, 234.27347538461538)
		(93.33333333333333, 238.5782246153846)
		(86.66666666666667, 240.40315076923076)
		(96.66666666666667, 238.44070153846152)
		(43.333333333333336, 238.91382307692308)
		(93.33333333333333, 240.04356923076924)
		(10.0, 235.8168446153846)
		(63.33333333333333, 234.73901538461539)
		(100.0, 241.3871323076923)
		(26.666666666666668, 237.2334723076923)
		(83.33333333333334, 234.82378307692306)
		(46.666666666666664, 235.36269384615386)
		(76.66666666666667, 239.01577230769232)
	};

	\addplot+[Pattern B] coordinates{
	};

	\addplot+[Pattern AB] coordinates{
	};

}
\&
\DOUBLEAXIS{Pattern = {272.46147923076927}, Column = {5} = {272.46147923076927}{5}, title={\null\hfill{\small(30 Type Annotations)}}}{Pattern = {1.2455920047239388}, Column = {5}, Right Axis = {1.2455920047239388}}{
	\addplot+[] coordinates{
		(0.5, 223.22038307692307)
		(1.5, 217.09938)
		(2.5, 216.03224307692307)
		(3.5, 212.84210923076924)
		(4.5, 214.41119692307691)
		(5.5, 211.92760923076924)
		(6.5, 241.56817076923076)
		(7.5, 213.3020753846154)
		(8.5, 218.6500676923077)
		(9.5, 213.40565692307692)
		(10.5, 212.21063384615385)
		(11.5, 213.23863384615385)
		(12.5, 215.95587692307691)
		(13.5, 212.29518307692308)
		(14.5, 213.71742153846154)
		(15.5, 212.12339538461538)
		(16.5, 213.88684307692307)
		(17.5, 213.41679846153846)
		(18.5, 212.8999246153846)
		(19.5, 213.07731384615386)
		(20.5, 213.57975846153846)
		(21.5, 213.36252153846155)
		(22.5, 221.6126846153846)
		(23.5, 219.53006923076924)
		(24.5, 213.2804676923077)
		(25.5, 222.07854)
		(26.5, 212.37720153846155)
		(27.5, 230.29299230769232)
		(28.5, 212.21944615384615)
		(29.5, 214.45381076923076)
		(30.5, 0)
		(0.5, 0)
	};

	\addplot+[Pattern A] coordinates{
		(6.5, 0)
		(6.5, 241.56817076923076)
		(7.5, 0)
		(6.5, 0)
	};

}
\&
\DOUBLEAXIS{Pattern = {193.7184513846154}, Scatter = {193.7184513846154}, title={\null\hfill Permute}}{Pattern = {1.203410245209007}, Scatter, Right Axis = {1.203410245209007}}{
	\addplot+[] coordinates{
		(35.714285714285715, 176.10768307692308)
		(64.28571428571429, 161.1371353846154)
		(64.28571428571429, 160.9886)
		(14.285714285714285, 161.2795)
		(7.142857142857142, 161.12510615384616)
		(21.428571428571427, 175.24994923076923)
		(42.857142857142854, 160.75281076923076)
		(50.0, 160.87430307692307)
		(7.142857142857142, 160.67028769230768)
		(14.285714285714285, 160.7906046153846)
		(50.0, 160.45328923076923)
		(64.28571428571429, 169.27450615384615)
		(7.142857142857142, 160.97658307692308)
		(35.714285714285715, 160.67992153846154)
		(14.285714285714285, 161.15290307692308)
		(21.428571428571427, 173.2810076923077)
		(78.57142857142857, 160.88316307692307)
		(50.0, 160.79721846153845)
		(42.857142857142854, 160.69878923076922)
		(35.714285714285715, 161.48660615384617)
		(57.14285714285714, 160.96273846153846)
		(21.428571428571427, 160.85313076923077)
		(92.85714285714286, 160.85056615384616)
		(50.0, 160.8291323076923)
		(21.428571428571427, 160.93747076923077)
		(42.857142857142854, 161.11040615384616)
		(71.42857142857143, 161.11022461538462)
		(14.285714285714285, 173.44653076923078)
		(42.857142857142854, 160.72576923076923)
		(50.0, 160.89916923076922)
		(64.28571428571429, 161.1835153846154)
		(14.285714285714285, 161.02518307692307)
		(42.857142857142854, 171.16817846153847)
		(7.142857142857142, 160.97060307692308)
		(14.285714285714285, 160.67462153846154)
		(7.142857142857142, 161.44650307692308)
		(78.57142857142857, 160.8726153846154)
		(28.57142857142857, 160.86786923076923)
		(35.714285714285715, 160.5618523076923)
		(28.57142857142857, 160.81333538461539)
		(28.57142857142857, 160.68494153846154)
		(50.0, 161.30881384615384)
		(7.142857142857142, 160.66321076923077)
		(0.0, 160.97457384615385)
		(42.857142857142854, 161.0521923076923)
		(21.428571428571427, 160.89698307692308)
		(7.142857142857142, 160.8899953846154)
		(71.42857142857143, 160.83399384615385)
		(71.42857142857143, 160.81938615384615)
		(21.428571428571427, 161.5591276923077)
		(14.285714285714285, 160.88954615384614)
	};

	\addplot+[Pattern A] coordinates{
		(71.42857142857143, 136.8766553846154)
		(100.0, 136.88462615384614)
		(57.14285714285714, 136.88589692307693)
		(35.714285714285715, 137.28191692307692)
		(71.42857142857143, 136.72838307692308)
		(85.71428571428571, 137.21357538461538)
		(78.57142857142857, 137.14604)
		(78.57142857142857, 137.21827076923077)
		(85.71428571428571, 136.98489692307692)
		(28.57142857142857, 137.25917692307692)
		(21.428571428571427, 137.21247076923078)
		(42.857142857142854, 137.32260153846153)
		(21.428571428571427, 144.55693538461537)
		(92.85714285714286, 144.98229692307692)
		(35.714285714285715, 136.91826923076923)
		(57.14285714285714, 136.97136)
		(85.71428571428571, 136.7328076923077)
		(78.57142857142857, 144.43697230769232)
		(35.714285714285715, 137.38470923076923)
		(28.57142857142857, 137.26133384615383)
		(85.71428571428571, 144.7620246153846)
		(35.714285714285715, 136.72496)
		(78.57142857142857, 145.50936153846155)
		(92.85714285714286, 136.72719230769232)
		(92.85714285714286, 137.0094723076923)
		(14.285714285714285, 137.1746753846154)
		(64.28571428571429, 136.68120769230768)
		(57.14285714285714, 137.11639846153847)
		(28.57142857142857, 144.59462307692309)
		(64.28571428571429, 137.14062)
		(71.42857142857143, 136.88532153846154)
		(71.42857142857143, 136.936)
		(64.28571428571429, 136.79855538461538)
		(50.0, 144.82084615384616)
		(78.57142857142857, 144.6558476923077)
		(7.142857142857142, 136.80124615384617)
		(57.14285714285714, 136.61441384615384)
		(28.57142857142857, 137.1363630769231)
		(78.57142857142857, 137.09567692307692)
		(92.85714285714286, 145.10071846153846)
		(64.28571428571429, 136.63045076923078)
		(85.71428571428571, 136.79644769230768)
		(57.14285714285714, 136.91577846153845)
		(85.71428571428571, 136.91848153846155)
		(57.14285714285714, 137.0270123076923)
		(92.85714285714286, 137.47786923076924)
		(85.71428571428571, 136.95242)
		(57.14285714285714, 136.95814153846155)
		(92.85714285714286, 145.07678307692308)
		(85.71428571428571, 136.85939384615384)
		(42.857142857142854, 136.66572)
	};

	\addplot+[Pattern B] coordinates{
	};

	\addplot+[Pattern AB] coordinates{
	};

}
\&
\DOUBLEAXIS{Pattern = {193.7184513846154}, Column = {5} = {193.7184513846154}{5}, title={\null\hfill{\small(14 Type Annotations)}}}{Pattern = {1.203918030232163}, Column = {5}, Right Axis, Column Right Label = {1.203918030232163}}{
	\addplot+[] coordinates{
		(0.5, 160.9144276923077)
		(1.5, 160.79358307692308)
		(2.5, 160.74922923076923)
		(3.5, 160.96209692307693)
		(4.5, 160.99839538461538)
		(5.5, 160.93339538461538)
		(6.5, 160.67315076923077)
		(7.5, 137.3947076923077)
		(8.5, 160.69423076923076)
		(9.5, 161.17084)
		(10.5, 161.13486461538463)
		(11.5, 161.06003692307692)
		(12.5, 161.37625538461538)
		(13.5, 160.85249076923077)
		(14.5, 0)
		(0.5, 0)
	};

	\addplot+[Pattern A] coordinates{
		(7.5, 0)
		(7.5, 137.3947076923077)
		(8.5, 0)
		(7.5, 0)
	};

}
\\
\DOUBLEAXIS{Pattern = {141.963998}, Scatter, Scatter Left Label = {141.963998}, title={\null\hfill DeltaBlue}}{Pattern = {1.2190709449214199}, Scatter, Right Axis = {1.2190709449214199}}{
	\addplot+[] coordinates{
		(60.08230452674898, 118.35273384615385)
		(15.22633744855967, 116.68366)
		(9.053497942386832, 116.8360676923077)
		(3.292181069958848, 116.24710615384615)
		(24.279835390946502, 117.42726307692308)
		(37.86008230452675, 120.4092676923077)
		(4.11522633744856, 116.04187538461538)
		(1.2345679012345678, 116.21091846153845)
		(20.16460905349794, 113.94469384615384)
		(6.995884773662551, 116.20899692307692)
		(44.03292181069959, 120.24432)
		(13.168724279835391, 117.88778615384615)
		(5.349794238683128, 115.64011846153846)
		(29.218106995884774, 122.61527230769231)
		(18.106995884773664, 117.83492)
		(16.049382716049383, 112.95672307692308)
		(13.991769547325102, 115.96123230769231)
		(2.05761316872428, 117.56572153846153)
		(31.275720164609055, 117.8457323076923)
		(23.045267489711936, 112.8896)
		(12.345679012345679, 117.35646923076924)
		(68.72427983539094, 116.75647538461538)
		(44.8559670781893, 117.20996461538462)
		(50.20576131687243, 121.00435076923077)
		(18.930041152263374, 117.54514153846154)
		(0.411522633744856, 129.05818)
		(72.0164609053498, 123.35539692307692)
		(41.1522633744856, 121.17123538461539)
		(0.0, 116.45261384615385)
		(60.90534979423868, 120.45114615384615)
		(51.028806584362144, 116.96270923076924)
		(22.22222222222222, 119.40240615384616)
		(10.2880658436214, 118.9864)
		(6.172839506172839, 116.8913276923077)
		(62.139917695473244, 122.98940461538461)
		(53.086419753086425, 122.03403230769231)
		(32.92181069958848, 117.86491846153847)
		(17.28395061728395, 118.69502461538461)
		(11.11111111111111, 117.14130769230769)
		(25.925925925925924, 120.04929692307692)
		(27.160493827160494, 117.62664615384615)
	};

	\addplot+[Pattern A] coordinates{
		(51.85185185185185, 98.77684)
		(63.78600823045267, 98.84524)
		(43.20987654320987, 98.36232615384615)
		(46.913580246913575, 99.95576153846154)
		(69.95884773662551, 98.82914769230769)
		(72.8395061728395, 98.60330153846154)
		(39.09465020576132, 99.50878307692308)
		(62.96296296296296, 99.41400307692308)
		(39.91769547325103, 98.46678461538461)
		(27.983539094650205, 97.0733923076923)
		(25.102880658436217, 97.87683538461539)
		(58.8477366255144, 98.54415076923077)
		(34.15637860082305, 96.27592307692308)
		(20.98765432098765, 96.25279846153846)
		(32.098765432098766, 99.20771692307693)
	};

	\addplot+[Pattern B] coordinates{
		(55.96707818930041, 98.38092307692308)
		(65.02057613168725, 100.26148153846154)
		(30.04115226337449, 97.88481076923077)
		(8.23045267489712, 96.70032615384615)
		(79.83539094650206, 98.08506461538461)
		(86.0082304526749, 98.42036769230769)
		(41.9753086419753, 96.1938876923077)
		(46.09053497942387, 98.90593076923076)
		(75.7201646090535, 100.33826923076923)
		(53.90946502057613, 97.73607384615384)
		(93.00411522633745, 100.62641538461538)
		(37.03703703703704, 100.16418615384616)
		(48.971193415637856, 99.28898615384615)
	};

	\addplot+[Pattern AB] coordinates{
		(74.8971193415638, 102.79937538461539)
		(36.21399176954733, 97.7412476923077)
		(34.97942386831276, 96.24038461538461)
		(81.06995884773663, 97.43668307692307)
		(87.65432098765432, 100.17044)
		(82.71604938271605, 100.61789846153846)
		(98.76543209876543, 101.01187692307693)
		(55.144032921810705, 97.86580615384615)
		(96.70781893004116, 99.71338153846153)
		(89.7119341563786, 100.00995692307693)
		(70.78189300411523, 97.49270769230769)
		(77.77777777777779, 100.27120923076923)
		(84.77366255144034, 99.67058)
		(88.88888888888889, 100.48995692307692)
		(81.89300411522635, 100.9443323076923)
		(91.76954732510289, 100.00746615384615)
		(97.94238683127571, 101.46202615384615)
		(86.83127572016461, 99.74354307692307)
		(67.90123456790124, 98.11441692307692)
		(58.0246913580247, 97.80864769230769)
		(93.82716049382715, 99.72120153846154)
		(83.9506172839506, 97.88067846153847)
		(67.07818930041152, 98.66171230769231)
		(74.07407407407408, 99.99728461538461)
		(94.65020576131687, 99.56739384615385)
		(79.01234567901234, 99.42199538461539)
		(90.94650205761316, 100.99156307692307)
		(65.84362139917695, 100.93916307692308)
		(95.88477366255144, 100.80978615384615)
		(48.148148148148145, 100.4532)
		(56.79012345679012, 99.00476461538462)
		(76.95473251028807, 101.22075538461539)
		(100.0, 101.4503723076923)
	};

}
\&
\DOUBLEAXIS{Pattern = {141.963998}, Column = {40} = {141.963998}{40}, title={\null\hfill{\small(243 Type Annotations)}}}{Pattern = {1.1629152408205454}, Column = {40}, Right Axis = {1.1629152408205454}}{
	\addplot+[] coordinates{
		(0.5, 116.74531384615385)
		(1.5, 116.91709538461538)
		(2.5, 117.62972)
		(3.5, 116.86439538461538)
		(4.5, 116.6936523076923)
		(5.5, 116.99404769230769)
		(6.5, 116.49197692307692)
		(7.5, 115.5123476923077)
		(8.5, 117.04472923076924)
		(9.5, 116.64264307692308)
		(10.5, 116.65655384615384)
		(11.5, 116.10929538461538)
		(12.5, 116.65045846153846)
		(13.5, 117.03309538461538)
		(14.5, 116.42543538461538)
		(15.5, 116.99313846153846)
		(16.5, 117.10740769230769)
		(17.5, 116.59944461538461)
		(18.5, 116.4218476923077)
		(19.5, 116.75588615384615)
		(20.5, 117.63690307692308)
		(21.5, 116.56406)
		(22.5, 116.30167846153846)
		(23.5, 117.34294923076924)
		(24.5, 117.0516)
		(25.5, 117.1055)
		(26.5, 116.29434923076923)
		(27.5, 116.46458)
		(28.5, 115.78954)
		(29.5, 115.9864076923077)
		(30.5, 116.50119692307692)
		(31.5, 116.50545692307692)
		(32.5, 117.18080615384615)
		(33.5, 116.21433692307693)
		(34.5, 116.6035323076923)
		(35.5, 116.72165076923076)
		(36.5, 116.20573846153846)
		(37.5, 116.60038)
		(38.5, 116.90259846153846)
		(39.5, 116.19876307692307)
		(40.5, 116.95658)
		(41.5, 116.77549384615385)
		(42.5, 116.78774461538461)
		(43.5, 116.89448153846153)
		(44.5, 116.10561846153846)
		(45.5, 116.25032153846153)
		(46.5, 116.16884615384616)
		(47.5, 116.91134153846154)
		(48.5, 116.85408153846154)
		(49.5, 116.74683076923077)
		(50.5, 117.22684461538462)
		(51.5, 117.60870153846153)
		(52.5, 117.29290153846154)
		(53.5, 117.48839076923078)
		(54.5, 117.16234615384616)
		(55.5, 116.27269846153847)
		(56.5, 116.19951692307693)
		(57.5, 117.33017692307692)
		(58.5, 116.00914769230769)
		(59.5, 115.98390923076923)
		(60.5, 116.48964307692307)
		(61.5, 117.76276923076924)
		(62.5, 116.18521846153845)
		(63.5, 116.44139692307692)
		(64.5, 117.5089)
		(65.5, 116.30252)
		(66.5, 116.26350923076923)
		(67.5, 116.91315076923077)
		(68.5, 116.02003846153846)
		(69.5, 115.83441692307693)
		(70.5, 116.75598615384615)
		(71.5, 116.91122923076924)
		(72.5, 117.04522615384616)
		(73.5, 117.39356461538462)
		(74.5, 116.53272)
		(75.5, 116.88692307692308)
		(76.5, 116.72968153846153)
		(77.5, 115.70713076923077)
		(78.5, 115.93962615384615)
		(79.5, 117.70244923076923)
		(80.5, 116.2454523076923)
		(81.5, 116.50393846153847)
		(82.5, 116.21592)
		(83.5, 116.58304)
		(84.5, 117.42432461538462)
		(85.5, 116.8801323076923)
		(86.5, 113.03246769230769)
		(87.5, 116.54850307692308)
		(88.5, 117.06728615384615)
		(89.5, 117.07531230769231)
		(90.5, 115.20932153846154)
		(91.5, 115.68052923076922)
		(92.5, 116.01358615384615)
		(93.5, 116.72745538461538)
		(94.5, 118.43676153846154)
		(95.5, 117.11923846153846)
		(96.5, 116.3212476923077)
		(97.5, 116.48765692307693)
		(98.5, 118.66870769230769)
		(99.5, 117.0947076923077)
		(100.5, 116.3757323076923)
		(101.5, 116.26395846153846)
		(102.5, 117.21062461538462)
		(103.5, 116.79518461538461)
		(104.5, 116.07431076923076)
		(105.5, 116.96640307692307)
		(106.5, 116.27111692307692)
		(107.5, 116.36065846153846)
		(108.5, 117.52408461538461)
		(109.5, 115.53221692307693)
		(110.5, 112.32858461538461)
		(111.5, 117.15568307692308)
		(112.5, 116.01894461538461)
		(113.5, 115.93277692307693)
		(114.5, 115.96279384615384)
		(115.5, 115.56723076923078)
		(116.5, 116.42185846153846)
		(117.5, 122.44274307692308)
		(118.5, 116.91816153846153)
		(119.5, 115.62161076923077)
		(120.5, 115.96619692307692)
		(121.5, 116.53319692307693)
		(122.5, 116.28338923076923)
		(123.5, 115.21599538461538)
		(124.5, 115.70314923076923)
		(125.5, 116.39666153846154)
		(126.5, 116.1154)
		(127.5, 116.70710923076923)
		(128.5, 117.55334)
		(129.5, 116.15221230769231)
		(130.5, 116.14959846153846)
		(131.5, 116.61056461538462)
		(132.5, 116.71892769230769)
		(133.5, 116.69499692307693)
		(134.5, 116.338)
		(135.5, 116.68499076923077)
		(136.5, 115.94873076923076)
		(137.5, 116.67588153846154)
		(138.5, 117.76650461538462)
		(139.5, 117.81882307692308)
		(140.5, 116.62837846153846)
		(141.5, 116.68456153846154)
		(142.5, 117.42545692307692)
		(143.5, 116.39792153846153)
		(144.5, 117.67091846153846)
		(145.5, 117.03749538461538)
		(146.5, 116.45286)
		(147.5, 117.79331846153846)
		(148.5, 117.10283230769231)
		(149.5, 116.80542461538461)
		(150.5, 117.68837538461538)
		(151.5, 116.88116615384615)
		(152.5, 117.08713076923077)
		(153.5, 97.13462769230769)
		(154.5, 96.87568307692308)
		(155.5, 117.4719123076923)
		(156.5, 117.53963846153846)
		(157.5, 115.88246307692307)
		(158.5, 115.96306615384616)
		(159.5, 117.17221692307692)
		(160.5, 115.70484923076923)
		(161.5, 116.16181384615385)
		(162.5, 116.66715692307692)
		(163.5, 115.77530461538461)
		(164.5, 116.39209846153847)
		(165.5, 117.48753846153846)
		(166.5, 116.93038461538461)
		(167.5, 116.58265230769231)
		(168.5, 116.51298461538461)
		(169.5, 116.15669538461539)
		(170.5, 117.32928769230769)
		(171.5, 116.51604)
		(172.5, 117.37683230769231)
		(173.5, 116.81863384615384)
		(174.5, 116.32079384615385)
		(175.5, 116.28411384615384)
		(176.5, 116.36102153846154)
		(177.5, 116.95430923076923)
		(178.5, 116.27462)
		(179.5, 117.46742307692308)
		(180.5, 116.56845230769231)
		(181.5, 116.53336923076922)
		(182.5, 121.88996307692308)
		(183.5, 117.32177538461538)
		(184.5, 117.1589076923077)
		(185.5, 116.47059846153846)
		(186.5, 116.53913538461538)
		(187.5, 117.76920153846154)
		(188.5, 117.53771692307693)
		(189.5, 116.53882615384616)
		(190.5, 116.58074153846154)
		(191.5, 116.28551846153846)
		(192.5, 116.58513692307692)
		(193.5, 116.32674615384616)
		(194.5, 121.79391846153847)
		(195.5, 116.51246)
		(196.5, 116.98455846153846)
		(197.5, 128.9586123076923)
		(198.5, 116.54375692307693)
		(199.5, 116.43876461538461)
		(200.5, 116.59389846153846)
		(201.5, 116.59151384615384)
		(202.5, 116.8215723076923)
		(203.5, 116.54613384615385)
		(204.5, 117.03205076923076)
		(205.5, 119.00658923076924)
		(206.5, 117.29753384615384)
		(207.5, 112.7146523076923)
		(208.5, 116.55990769230769)
		(209.5, 116.77825384615385)
		(210.5, 116.54641384615384)
		(211.5, 117.08562307692307)
		(212.5, 116.9752)
		(213.5, 117.12369230769231)
		(214.5, 116.15663692307692)
		(215.5, 115.61834923076923)
		(216.5, 115.65663538461538)
		(217.5, 117.33665538461538)
		(218.5, 115.59058461538461)
		(219.5, 115.94971692307692)
		(220.5, 116.15117846153846)
		(221.5, 115.67936923076923)
		(222.5, 115.77632461538461)
		(223.5, 116.12719384615384)
		(224.5, 116.55176)
		(225.5, 116.10273384615385)
		(226.5, 116.56520923076923)
		(227.5, 116.32727692307692)
		(228.5, 116.34687692307692)
		(229.5, 117.84875692307692)
		(230.5, 116.77964615384616)
		(231.5, 117.79575076923076)
		(232.5, 116.78849846153847)
		(233.5, 116.61203384615385)
		(234.5, 116.88545384615385)
		(235.5, 116.01693846153846)
		(236.5, 116.84227538461539)
		(237.5, 116.38898153846154)
		(238.5, 116.37299538461538)
		(239.5, 116.42685846153846)
		(240.5, 116.6950923076923)
		(241.5, 117.22067846153845)
		(242.5, 116.18841692307693)
		(243.5, 0)
		(0.5, 0)
	};

	\addplot+[Pattern A] coordinates{
		(153.5, 0)
		(153.5, 97.13462769230769)
		(154.5, 0)
		(153.5, 0)
	};

	\addplot+[Pattern B] coordinates{
		(154.5, 0)
		(154.5, 96.87568307692308)
		(155.5, 0)
		(154.5, 0)
	};

}
\&
\DOUBLEAXIS{Pattern = {267.975460923077}, Scatter = {267.975460923077}, title={\null\hfill Go}}{Pattern = {1.1403910642505535}, Scatter, Right Axis = {1.1403910642505535}}{
	\addplot+[] coordinates{
		(5.217391304347826, 235.86017692307692)
		(23.91304347826087, 236.05612153846153)
		(12.173913043478262, 224.07898923076922)
		(42.173913043478265, 242.67882)
		(32.17391304347826, 243.6140553846154)
		(23.043478260869566, 224.7242169230769)
		(36.08695652173913, 243.04395384615384)
		(3.0434782608695654, 236.27922923076923)
		(43.913043478260875, 242.9471523076923)
		(38.26086956521739, 223.37118)
		(9.130434782608695, 238.63282153846154)
		(6.086956521739131, 237.62395846153845)
		(20.0, 222.25060307692308)
		(14.347826086956522, 235.75727076923076)
		(1.3043478260869565, 236.12947384615384)
		(33.91304347826087, 221.95980153846153)
		(29.130434782608695, 222.4939123076923)
		(0.0, 234.98558461538462)
		(11.304347826086957, 235.50840461538462)
		(10.0, 236.74045384615386)
		(2.1739130434782608, 235.1877123076923)
		(0.43478260869565216, 236.22696923076924)
		(30.0, 235.20751076923077)
		(4.3478260869565215, 236.2188323076923)
		(46.08695652173913, 241.89788153846155)
		(13.043478260869565, 237.3012646153846)
		(16.956521739130434, 236.42621384615384)
		(16.08695652173913, 236.76494461538462)
		(30.869565217391305, 236.26694923076923)
		(15.217391304347828, 235.57470153846154)
		(7.391304347826087, 235.00985384615385)
		(8.26086956521739, 236.12082615384617)
	};

	\addplot+[Pattern A] coordinates{
		(50.0, 223.77532461538462)
		(90.8695652173913, 225.25820153846155)
		(70.0, 223.85701076923078)
		(28.26086956521739, 220.4592953846154)
		(63.91304347826087, 224.82604615384616)
		(50.8695652173913, 222.3208753846154)
		(86.95652173913044, 223.56521230769232)
		(59.130434782608695, 223.21814307692307)
		(57.826086956521735, 223.8374476923077)
		(45.21739130434783, 219.87266615384615)
		(19.130434782608695, 222.27543384615385)
		(25.217391304347824, 220.0284276923077)
		(64.78260869565217, 223.6128276923077)
		(26.08695652173913, 221.65232923076923)
		(35.21739130434783, 226.05219076923078)
		(43.04347826086957, 220.54209384615385)
		(26.956521739130434, 221.11829538461538)
	};

	\addplot+[Pattern B] coordinates{
		(46.95652173913044, 224.79925846153847)
		(40.869565217391305, 224.95568923076922)
		(77.82608695652173, 221.80032153846153)
		(49.130434782608695, 218.7477323076923)
		(61.73913043478261, 221.25410923076922)
		(53.04347826086957, 225.3541153846154)
		(53.91304347826087, 223.88956307692308)
		(21.304347826086957, 224.52041076923078)
		(36.95652173913043, 220.1883523076923)
		(33.04347826086956, 220.54861076923078)
		(85.65217391304348, 222.55481230769232)
		(22.17391304347826, 218.91295692307693)
		(60.86956521739131, 223.84949230769232)
		(47.82608695652174, 224.68816615384614)
		(18.26086956521739, 222.47198153846153)
		(97.82608695652173, 225.22459384615385)
	};

	\addplot+[Pattern AB] coordinates{
		(98.69565217391305, 222.84408461538462)
		(100.0, 225.30320615384616)
		(93.91304347826087, 226.74027846153845)
		(52.17391304347826, 218.8292153846154)
		(96.95652173913044, 224.5267723076923)
		(76.95652173913044, 222.79824461538462)
		(78.69565217391305, 225.05610153846155)
		(74.78260869565217, 224.6892969230769)
		(80.8695652173913, 225.48148615384616)
		(73.91304347826086, 223.7416323076923)
		(39.130434782608695, 225.27353692307693)
		(87.82608695652175, 224.17702)
		(81.73913043478261, 225.42404307692308)
		(83.04347826086956, 224.75323076923078)
		(40.0, 223.17001384615384)
		(88.69565217391305, 224.03955384615384)
		(56.086956521739125, 221.89136923076924)
		(91.73913043478261, 224.75514)
		(90.0, 225.0864723076923)
		(66.08695652173913, 224.30836153846153)
		(84.78260869565217, 242.00618461538463)
		(76.08695652173914, 225.25789076923076)
		(54.78260869565217, 224.2157923076923)
		(71.73913043478261, 242.5008553846154)
		(60.0, 225.49804153846154)
		(63.04347826086957, 223.35454615384614)
		(67.82608695652173, 225.48838615384616)
		(56.95652173913044, 219.44589692307693)
		(66.95652173913044, 225.12337076923077)
		(94.78260869565217, 226.03764)
		(83.91304347826087, 225.64304)
		(70.86956521739131, 221.65042)
		(95.65217391304348, 224.3283953846154)
		(69.1304347826087, 223.90974)
		(80.0, 224.5386630769231)
		(92.6086956521739, 226.14022615384616)
		(73.04347826086956, 223.19428)
	};

}
\&
\DOUBLEAXIS{Pattern = {267.975460923077}, Column = {35} = {267.975460923077}{35}, title={\null\hfill{\small(230 Type Annotations)}}}{Pattern = {1.1378096682818448}, Column = {35}, Right Axis, Column Right Label = {1.1378096682818448}}{
	\addplot+[] coordinates{
		(0.5, 236.89064461538462)
		(1.5, 236.27087384615385)
		(2.5, 235.34293692307693)
		(3.5, 235.96629692307692)
		(4.5, 236.23525846153845)
		(5.5, 236.29656615384616)
		(6.5, 236.5964476923077)
		(7.5, 235.9260646153846)
		(8.5, 236.33456615384614)
		(9.5, 236.45064769230768)
		(10.5, 236.19046461538463)
		(11.5, 235.67646153846155)
		(12.5, 236.11740153846154)
		(13.5, 235.67463692307692)
		(14.5, 235.53201692307692)
		(15.5, 235.66836)
		(16.5, 236.27374153846154)
		(17.5, 235.62035076923078)
		(18.5, 235.22205538461537)
		(19.5, 235.84882461538461)
		(20.5, 235.99161230769232)
		(21.5, 235.10746153846154)
		(22.5, 235.55518)
		(23.5, 235.86273538461538)
		(24.5, 236.96262153846155)
		(25.5, 235.07833076923077)
		(26.5, 235.9812476923077)
		(27.5, 236.40416769230768)
		(28.5, 235.73042)
		(29.5, 234.99714923076922)
		(30.5, 219.96183846153846)
		(31.5, 236.33851692307692)
		(32.5, 236.89566923076924)
		(33.5, 235.85834307692306)
		(34.5, 236.35378307692307)
		(35.5, 235.0189)
		(36.5, 237.3691323076923)
		(37.5, 236.6769153846154)
		(38.5, 236.77777384615385)
		(39.5, 236.96902615384616)
		(40.5, 235.60679076923077)
		(41.5, 236.09580461538462)
		(42.5, 236.45154153846153)
		(43.5, 236.78684153846154)
		(44.5, 235.50643846153847)
		(45.5, 235.14050153846154)
		(46.5, 220.10203692307692)
		(47.5, 236.7250753846154)
		(48.5, 236.07488615384617)
		(49.5, 236.15231538461538)
		(50.5, 235.62625692307694)
		(51.5, 236.74966)
		(52.5, 235.77403846153845)
		(53.5, 235.68604153846155)
		(54.5, 223.11598615384617)
		(55.5, 241.19812)
		(56.5, 235.62993076923078)
		(57.5, 236.34346461538462)
		(58.5, 234.25446)
		(59.5, 234.44961538461538)
		(60.5, 235.8733830769231)
		(61.5, 235.68974615384616)
		(62.5, 235.46057384615384)
		(63.5, 238.14490923076923)
		(64.5, 236.4987953846154)
		(65.5, 234.73140153846154)
		(66.5, 236.51441538461538)
		(67.5, 236.58014153846153)
		(68.5, 236.39909384615385)
		(69.5, 237.25688153846153)
		(70.5, 235.71928)
		(71.5, 238.50153538461538)
		(72.5, 235.73789846153846)
		(73.5, 237.17489384615385)
		(74.5, 237.61308)
		(75.5, 237.3321276923077)
		(76.5, 235.7520123076923)
		(77.5, 236.4042876923077)
		(78.5, 234.96827076923077)
		(79.5, 235.77756615384615)
		(80.5, 235.91546)
		(81.5, 236.77496153846153)
		(82.5, 236.43162)
		(83.5, 235.63578307692308)
		(84.5, 235.02170307692307)
		(85.5, 236.30510307692307)
		(86.5, 236.12716461538463)
		(87.5, 235.00788153846153)
		(88.5, 236.05309076923078)
		(89.5, 236.05414307692308)
		(90.5, 237.51633384615386)
		(91.5, 234.77000153846154)
		(92.5, 236.45764461538462)
		(93.5, 237.20498615384616)
		(94.5, 236.02906307692308)
		(95.5, 236.4654753846154)
		(96.5, 234.80436461538463)
		(97.5, 234.5931369230769)
		(98.5, 238.89852153846155)
		(99.5, 236.44789846153847)
		(100.5, 236.6537046153846)
		(101.5, 236.27882)
		(102.5, 237.65765692307693)
		(103.5, 236.07027846153846)
		(104.5, 236.0193723076923)
		(105.5, 236.34283692307693)
		(106.5, 236.03722307692308)
		(107.5, 236.54696923076924)
		(108.5, 235.41963846153845)
		(109.5, 235.09140615384615)
		(110.5, 235.45523384615385)
		(111.5, 236.01352)
		(112.5, 236.2856076923077)
		(113.5, 236.4956830769231)
		(114.5, 236.3788569230769)
		(115.5, 235.61625384615385)
		(116.5, 236.02133076923076)
		(117.5, 235.17346923076923)
		(118.5, 236.3805)
		(119.5, 237.13624615384614)
		(120.5, 235.95931384615383)
		(121.5, 235.33013384615384)
		(122.5, 235.46749846153847)
		(123.5, 237.27019692307692)
		(124.5, 235.65330461538463)
		(125.5, 234.91837846153845)
		(126.5, 236.8351353846154)
		(127.5, 235.8294369230769)
		(128.5, 236.18838615384615)
		(129.5, 237.0055523076923)
		(130.5, 235.00507846153846)
		(131.5, 237.09169384615385)
		(132.5, 235.94293692307693)
		(133.5, 235.8720476923077)
		(134.5, 236.3354923076923)
		(135.5, 237.53154307692307)
		(136.5, 236.00626923076922)
		(137.5, 235.48754)
		(138.5, 237.0439276923077)
		(139.5, 235.2352369230769)
		(140.5, 236.17994)
		(141.5, 235.47776923076924)
		(142.5, 236.31482153846153)
		(143.5, 235.5879723076923)
		(144.5, 236.45006307692307)
		(145.5, 236.9568046153846)
		(146.5, 237.4349)
		(147.5, 234.79338615384614)
		(148.5, 236.34752153846154)
		(149.5, 235.48453230769232)
		(150.5, 235.88390307692308)
		(151.5, 236.4657523076923)
		(152.5, 236.2226923076923)
		(153.5, 236.83425846153847)
		(154.5, 235.39055692307693)
		(155.5, 235.12796)
		(156.5, 236.73123076923076)
		(157.5, 236.2983076923077)
		(158.5, 235.43473846153847)
		(159.5, 236.49432923076924)
		(160.5, 235.32750307692308)
		(161.5, 235.8275323076923)
		(162.5, 235.10152923076924)
		(163.5, 236.6135123076923)
		(164.5, 235.69703076923076)
		(165.5, 236.16966)
		(166.5, 238.01134769230768)
		(167.5, 235.11224307692308)
		(168.5, 237.03528615384616)
		(169.5, 234.83803538461538)
		(170.5, 236.53014461538461)
		(171.5, 236.79397076923078)
		(172.5, 234.93290153846155)
		(173.5, 235.75572461538462)
		(174.5, 237.31569384615383)
		(175.5, 236.06622923076924)
		(176.5, 235.29993692307693)
		(177.5, 235.66665384615385)
		(178.5, 235.4128)
		(179.5, 236.94898923076923)
		(180.5, 235.4718723076923)
		(181.5, 235.67999076923076)
		(182.5, 235.5298646153846)
		(183.5, 236.57187846153846)
		(184.5, 236.39338)
		(185.5, 235.2083876923077)
		(186.5, 234.41998615384617)
		(187.5, 239.99885230769232)
		(188.5, 235.2414723076923)
		(189.5, 238.74944769230768)
		(190.5, 237.74747846153846)
		(191.5, 236.3470523076923)
		(192.5, 237.4567923076923)
		(193.5, 235.60256307692308)
		(194.5, 235.97926307692308)
		(195.5, 235.93274923076922)
		(196.5, 235.6754169230769)
		(197.5, 235.96374307692307)
		(198.5, 236.26713076923076)
		(199.5, 236.42140307692307)
		(200.5, 237.31732923076922)
		(201.5, 235.7019723076923)
		(202.5, 235.73902)
		(203.5, 236.48043076923076)
		(204.5, 235.60639384615385)
		(205.5, 237.01498153846154)
		(206.5, 237.21524)
		(207.5, 236.20040461538463)
		(208.5, 236.60107384615384)
		(209.5, 235.74049076923077)
		(210.5, 236.40209692307693)
		(211.5, 236.52432461538461)
		(212.5, 235.82976923076924)
		(213.5, 236.00383384615384)
		(214.5, 236.30242)
		(215.5, 236.0916923076923)
		(216.5, 235.15206307692307)
		(217.5, 235.63386307692306)
		(218.5, 236.4616246153846)
		(219.5, 236.01522923076922)
		(220.5, 237.44778461538462)
		(221.5, 235.87928153846153)
		(222.5, 236.33400461538463)
		(223.5, 235.93169230769232)
		(224.5, 239.8333846153846)
		(225.5, 236.50861384615385)
		(226.5, 235.34423384615386)
		(227.5, 235.98711076923078)
		(228.5, 237.27336461538462)
		(229.5, 236.40862307692308)
		(230.5, 0)
		(0.5, 0)
	};

	\addplot+[Pattern A] coordinates{
		(30.5, 0)
		(30.5, 219.96183846153846)
		(31.5, 0)
		(30.5, 0)
	};

	\addplot+[Pattern B] coordinates{
		(46.5, 0)
		(46.5, 220.10203692307692)
		(47.5, 0)
		(46.5, 0)
	};

}
\\
\DOUBLEAXIS{Pattern = {487.8821284615385}, Scatter, Scatter Left Label = {487.8821284615385}, title={\null\hfill List}}{Pattern = {2.2729455968334378}, Scatter, Right Axis = {2.2729455968334378}}{
	\addplot+[] coordinates{
		(39.130434782608695, 203.76336153846154)
		(13.043478260869565, 204.96308923076924)
		(4.3478260869565215, 205.62648615384614)
		(13.043478260869565, 205.5540553846154)
		(4.3478260869565215, 204.70628923076924)
		(34.78260869565217, 205.61570461538463)
		(43.47826086956522, 213.50244307692307)
		(34.78260869565217, 209.87120307692308)
		(52.17391304347826, 218.45641076923076)
		(4.3478260869565215, 212.5166123076923)
		(34.78260869565217, 211.06125076923078)
		(8.695652173913043, 205.6064076923077)
		(4.3478260869565215, 205.85359692307694)
		(13.043478260869565, 204.50179076923078)
		(26.08695652173913, 209.40336153846152)
		(47.82608695652174, 213.70174615384616)
		(21.73913043478261, 209.31984615384616)
		(39.130434782608695, 213.87380153846155)
		(8.695652173913043, 208.15599076923078)
		(43.47826086956522, 209.81009076923078)
		(13.043478260869565, 205.26062461538461)
		(65.21739130434783, 218.41380307692307)
		(30.434782608695656, 208.08602769230768)
		(17.391304347826086, 205.4463323076923)
		(52.17391304347826, 204.90848615384616)
		(0.0, 214.64751692307692)
		(52.17391304347826, 209.32162923076922)
		(26.08695652173913, 212.80659846153847)
		(21.73913043478261, 205.4845723076923)
		(17.391304347826086, 205.39220153846153)
		(8.695652173913043, 205.6112523076923)
		(21.73913043478261, 206.71018)
		(43.47826086956522, 209.51488153846154)
		(8.695652173913043, 205.51459384615384)
		(17.391304347826086, 207.17219076923078)
		(17.391304347826086, 203.89332923076924)
		(26.08695652173913, 209.65139846153846)
		(56.52173913043478, 208.76297692307693)
		(47.82608695652174, 208.63745384615385)
	};

	\addplot+[Pattern A] coordinates{
		(30.434782608695656, 301.67285692307695)
		(30.434782608695656, 299.45758615384614)
		(26.08695652173913, 298.4554676923077)
		(82.6086956521739, 313.68094)
		(21.73913043478261, 302.4377123076923)
		(43.47826086956522, 301.66147384615385)
		(78.26086956521739, 303.14302615384617)
		(39.130434782608695, 298.39793692307694)
		(65.21739130434783, 302.0162723076923)
		(60.86956521739131, 302.72734153846153)
		(26.08695652173913, 302.15180923076923)
		(8.695652173913043, 300.79152153846155)
		(60.86956521739131, 316.606)
	};

	\addplot+[Pattern B] coordinates{
		(56.52173913043478, 294.11142615384614)
		(4.3478260869565215, 285.2347246153846)
		(65.21739130434783, 279.92569692307694)
		(56.52173913043478, 276.1561923076923)
		(95.65217391304348, 281.86659692307694)
		(73.91304347826086, 283.2683430769231)
		(43.47826086956522, 292.5950323076923)
		(52.17391304347826, 272.7143384615385)
		(17.391304347826086, 275.9406153846154)
		(82.6086956521739, 281.6327923076923)
		(60.86956521739131, 277.5932507692308)
		(56.52173913043478, 284.55373692307694)
		(52.17391304347826, 286.3656384615385)
		(30.434782608695656, 275.90955384615387)
		(82.6086956521739, 278.40741846153844)
		(73.91304347826086, 280.3063446153846)
		(69.56521739130434, 281.52176)
		(34.78260869565217, 284.3918261538461)
		(86.95652173913044, 289.7698753846154)
	};

	\addplot+[Pattern AB] coordinates{
		(65.21739130434783, 414.2744723076923)
		(95.65217391304348, 432.19901846153846)
		(82.6086956521739, 443.5292076923077)
		(73.91304347826086, 422.14082)
		(69.56521739130434, 422.4920584615385)
		(82.6086956521739, 426.7074030769231)
		(39.130434782608695, 421.6899507692308)
		(91.30434782608695, 434.93765076923074)
		(69.56521739130434, 422.15309692307693)
		(95.65217391304348, 432.1258323076923)
		(86.95652173913044, 432.1335446153846)
		(73.91304347826086, 427.2881246153846)
		(91.30434782608695, 432.2348984615385)
		(95.65217391304348, 426.98387692307693)
		(78.26086956521739, 432.3689123076923)
		(86.95652173913044, 432.3203169230769)
		(91.30434782608695, 418.42740153846154)
		(100.0, 433.9553723076923)
		(34.78260869565217, 422.1372430769231)
		(91.30434782608695, 426.43911692307694)
		(78.26086956521739, 424.6643523076923)
		(91.30434782608695, 433.4522969230769)
		(56.52173913043478, 432.34235384615386)
		(47.82608695652174, 424.3952830769231)
		(78.26086956521739, 436.7692215384615)
		(73.91304347826086, 431.49852923076924)
		(60.86956521739131, 418.4613846153846)
		(47.82608695652174, 421.9823030769231)
		(69.56521739130434, 422.20660615384617)
		(65.21739130434783, 419.67258923076923)
		(86.95652173913044, 435.5484276923077)
	};

}
\&
\DOUBLEAXIS{Pattern = {487.8821284615385}, Column = {5} = {487.8821284615385}{5}, title={\null\hfill{\small(23 Type Annotations)}}}{Pattern = {2.3741491715343184}, Column = {5}, Right Axis = {2.3741491715343184}}{
	\addplot+[] coordinates{
		(0.5, 205.34465076923078)
		(1.5, 205.9239246153846)
		(2.5, 203.81086461538462)
		(3.5, 205.51077692307692)
		(4.5, 205.42346153846154)
		(5.5, 203.69912461538462)
		(6.5, 205.11846153846153)
		(7.5, 203.63581846153846)
		(8.5, 205.76394)
		(9.5, 205.86509846153845)
		(10.5, 204.8784830769231)
		(11.5, 205.31700307692307)
		(12.5, 206.52830615384616)
		(13.5, 205.8049353846154)
		(14.5, 207.88240307692308)
		(15.5, 206.37279846153845)
		(16.5, 301.3789046153846)
		(17.5, 299.5947815384615)
		(18.5, 207.4952123076923)
		(19.5, 205.30317692307693)
		(20.5, 208.22801076923076)
		(21.5, 208.70558923076922)
		(22.5, 205.0413)
		(23.5, 0)
		(0.5, 0)
	};

	\addplot+[Pattern A] coordinates{
		(16.5, 0)
		(16.5, 301.3789046153846)
		(17.5, 0)
		(16.5, 0)
	};

	\addplot+[Pattern B] coordinates{
		(17.5, 0)
		(17.5, 299.5947815384615)
		(18.5, 0)
		(17.5, 0)
	};

}
\&
\DOUBLEAXIS{Pattern = {737.2800803076924}, Scatter = {737.2800803076924}, title={\null\hfill Richards}}{Pattern = {1.7896774262069879}, Scatter, Right Axis = {1.7896774262069879}}{
	\addplot+[] coordinates{
		(44.06779661016949, 443.7914476923077)
		(8.47457627118644, 418.9058923076923)
		(31.07344632768362, 413.02338153846154)
		(22.033898305084744, 433.82622769230767)
		(61.016949152542374, 478.88476)
		(15.254237288135593, 397.02209230769233)
		(29.943502824858758, 425.8875123076923)
		(2.2598870056497176, 409.80076153846153)
		(16.94915254237288, 402.7624246153846)
		(29.37853107344633, 411.3420446153846)
		(41.24293785310734, 444.98053076923077)
		(36.15819209039548, 455.2175646153846)
		(0.5649717514124294, 409.27085384615384)
		(4.519774011299435, 428.37446923076925)
		(25.423728813559322, 425.7896507692308)
		(24.293785310734464, 425.57146153846156)
		(67.79661016949152, 470.47513692307695)
		(3.389830508474576, 411.21206615384614)
		(6.214689265536723, 414.99601384615386)
		(0.0, 411.9625523076923)
		(46.89265536723164, 437.62101538461536)
		(7.344632768361582, 416.0288676923077)
		(14.124293785310735, 412.0237507692308)
		(66.10169491525424, 455.1298061538462)
		(50.847457627118644, 472.2157446153846)
		(27.11864406779661, 440.8407476923077)
		(9.03954802259887, 415.4103215384615)
		(5.084745762711865, 409.3059153846154)
		(10.16949152542373, 429.6085953846154)
		(1.1299435028248588, 407.94353538461536)
		(45.76271186440678, 447.66892461538464)
	};

	\addplot+[Pattern A] coordinates{
		(20.903954802259886, 546.2992246153847)
		(54.80225988700565, 558.5911630769231)
		(59.887005649717516, 543.8835415384615)
		(51.9774011299435, 532.7223984615384)
		(23.163841807909606, 514.4926307692308)
		(88.70056497175142, 559.4826923076923)
		(38.983050847457626, 538.9156415384615)
		(32.20338983050847, 670.2546184615385)
		(41.80790960451977, 532.1499538461538)
		(20.33898305084746, 508.6414323076923)
		(64.97175141242938, 566.348076923077)
		(62.71186440677966, 578.1479938461539)
		(53.10734463276836, 581.0554507692308)
		(89.83050847457628, 615.2387015384616)
		(75.14124293785311, 601.6462923076923)
		(50.282485875706215, 560.1288384615384)
		(49.152542372881356, 576.8831292307692)
		(25.98870056497175, 579.5762184615385)
		(42.93785310734463, 555.3567876923076)
		(28.24858757062147, 521.8492446153846)
		(90.96045197740112, 596.9399169230769)
		(37.28813559322034, 545.5940215384616)
		(79.09604519774011, 597.7494830769231)
		(12.994350282485875, 523.7277953846154)
		(18.07909604519774, 537.4592184615385)
	};

	\addplot+[Pattern B] coordinates{
		(19.2090395480226, 430.73608461538464)
		(12.429378531073446, 414.0604769230769)
		(33.33333333333333, 449.57924615384616)
		(66.66666666666666, 462.14758153846157)
		(68.92655367231639, 438.82138)
		(58.75706214689266, 475.82073384615387)
		(16.38418079096045, 417.8806030769231)
		(80.7909604519774, 477.85305999999997)
		(35.02824858757062, 420.3649461538462)
		(37.85310734463277, 434.8993261538462)
		(70.05649717514125, 477.4141107692308)
		(77.96610169491525, 500.93047538461536)
		(95.48022598870057, 491.86513384615387)
		(58.19209039548022, 465.6170569230769)
		(40.11299435028249, 469.9168246153846)
		(63.84180790960452, 478.30451846153846)
		(45.19774011299435, 463.3425846153846)
		(11.299435028248588, 410.50579076923077)
		(83.61581920903954, 479.42617538461536)
		(84.7457627118644, 461.5028123076923)
	};

	\addplot+[Pattern AB] coordinates{
		(91.52542372881356, 504.05951230769233)
		(62.14689265536724, 488.02977384615383)
		(76.8361581920904, 497.25998153846155)
		(100.0, 514.5723876923076)
		(92.65536723163842, 519.7150861538462)
		(71.75141242937853, 492.34334615384614)
		(85.87570621468926, 508.7282784615385)
		(33.89830508474576, 459.4782)
		(97.74011299435028, 510.97612153846154)
		(87.00564971751412, 504.8343953846154)
		(48.0225988700565, 463.5241446153846)
		(75.70621468926554, 485.0088123076923)
		(83.05084745762711, 468.17091384615384)
		(70.62146892655367, 515.2256676923076)
		(74.01129943502825, 473.71236153846155)
		(79.66101694915254, 489.4486076923077)
		(87.57062146892656, 515.3765123076923)
		(55.932203389830505, 465.5353276923077)
		(72.88135593220339, 532.7232446153846)
		(54.23728813559322, 474.4570907692308)
		(81.92090395480226, 485.3327815384615)
		(98.87005649717514, 492.9808323076923)
		(57.06214689265536, 452.8882323076923)
		(96.61016949152543, 501.57664923076925)
		(94.91525423728814, 499.26756)
		(93.78531073446328, 517.9437307692308)
	};

}
\&
\DOUBLEAXIS{Pattern = {737.2800803076924}, Column = {25} = {737.2800803076924}{25}, title={\null\hfill{\small(177 Type Annotations)}}}{Pattern = {1.784892851969946}, Column = {25}, Right Axis, Column Right Label = {1.784892851969946}}{
	\addplot+[] coordinates{
		(0.5, 410.3935646153846)
		(1.5, 410.1452353846154)
		(2.5, 417.33332461538464)
		(3.5, 409.59511692307694)
		(4.5, 409.25806153846156)
		(5.5, 412.89140769230767)
		(6.5, 412.9776876923077)
		(7.5, 410.5904876923077)
		(8.5, 410.7358476923077)
		(9.5, 410.5000753846154)
		(10.5, 400.1843276923077)
		(11.5, 410.6544446153846)
		(12.5, 410.4719215384615)
		(13.5, 410.0076246153846)
		(14.5, 409.8279707692308)
		(15.5, 404.36568923076925)
		(16.5, 402.7539723076923)
		(17.5, 404.14727846153846)
		(18.5, 409.9141492307692)
		(19.5, 410.06829230769233)
		(20.5, 409.85063384615387)
		(21.5, 410.0155846153846)
		(22.5, 410.2853846153846)
		(23.5, 426.76424769230766)
		(24.5, 412.6493676923077)
		(25.5, 410.00299538461536)
		(26.5, 410.3270907692308)
		(27.5, 410.8093292307692)
		(28.5, 411.09415384615386)
		(29.5, 413.2809938461538)
		(30.5, 410.4224169230769)
		(31.5, 410.13386)
		(32.5, 409.9743446153846)
		(33.5, 410.3934892307692)
		(34.5, 402.48292153846154)
		(35.5, 409.82025076923077)
		(36.5, 409.56770615384613)
		(37.5, 410.60736153846153)
		(38.5, 428.22206307692306)
		(39.5, 409.98990769230767)
		(40.5, 414.68785692307694)
		(41.5, 410.53562923076925)
		(42.5, 412.4959846153846)
		(43.5, 416.6845584615385)
		(44.5, 410.69522615384614)
		(45.5, 409.9580184615385)
		(46.5, 409.3726446153846)
		(47.5, 416.4169892307692)
		(48.5, 414.65479846153846)
		(49.5, 412.26395076923075)
		(50.5, 399.9667415384615)
		(51.5, 410.32817846153847)
		(52.5, 413.0051384615385)
		(53.5, 412.11783692307694)
		(54.5, 410.17575230769233)
		(55.5, 409.7891092307692)
		(56.5, 410.27063846153845)
		(57.5, 410.3237369230769)
		(58.5, 400.8904338461538)
		(59.5, 412.7759)
		(60.5, 410.8456553846154)
		(61.5, 410.6720369230769)
		(62.5, 409.5221430769231)
		(63.5, 413.60922)
		(64.5, 410.42977384615386)
		(65.5, 413.4504338461538)
		(66.5, 411.0375707692308)
		(67.5, 410.73995076923075)
		(68.5, 409.59380923076924)
		(69.5, 413.25622153846155)
		(70.5, 404.4750369230769)
		(71.5, 410.62515538461537)
		(72.5, 410.8607076923077)
		(73.5, 410.10006615384617)
		(74.5, 412.8934076923077)
		(75.5, 410.31636615384616)
		(76.5, 412.88462615384617)
		(77.5, 411.2134369230769)
		(78.5, 410.7647984615385)
		(79.5, 411.79508)
		(80.5, 405.3774015384615)
		(81.5, 410.5775169230769)
		(82.5, 410.58350615384614)
		(83.5, 412.34174461538464)
		(84.5, 410.62273846153846)
		(85.5, 409.9734676923077)
		(86.5, 401.7282630769231)
		(87.5, 412.2349938461538)
		(88.5, 409.9274030769231)
		(89.5, 416.4278476923077)
		(90.5, 412.9753507692308)
		(91.5, 414.4101553846154)
		(92.5, 414.53583692307694)
		(93.5, 410.3179707692308)
		(94.5, 412.30498153846156)
		(95.5, 400.0740523076923)
		(96.5, 405.9701138461538)
		(97.5, 410.4082846153846)
		(98.5, 409.4258646153846)
		(99.5, 412.6379138461538)
		(100.5, 399.5957846153846)
		(101.5, 410.58379692307693)
		(102.5, 409.9771676923077)
		(103.5, 410.3954692307692)
		(104.5, 410.82378)
		(105.5, 409.8308430769231)
		(106.5, 415.02957692307695)
		(107.5, 410.2507415384615)
		(108.5, 410.69648)
		(109.5, 403.2450276923077)
		(110.5, 410.76700307692306)
		(111.5, 388.16206)
		(112.5, 400.5129476923077)
		(113.5, 410.3535907692308)
		(114.5, 411.37550923076924)
		(115.5, 411.59574153846154)
		(116.5, 395.58865692307694)
		(117.5, 410.1741415384615)
		(118.5, 410.00234461538463)
		(119.5, 412.56339846153844)
		(120.5, 411.9394107692308)
		(121.5, 410.71868615384614)
		(122.5, 409.6107123076923)
		(123.5, 411.0191046153846)
		(124.5, 410.1184492307692)
		(125.5, 410.3011723076923)
		(126.5, 410.1524753846154)
		(127.5, 410.4251323076923)
		(128.5, 411.80179846153845)
		(129.5, 412.17672)
		(130.5, 409.43382153846153)
		(131.5, 411.5917123076923)
		(132.5, 409.8424415384615)
		(133.5, 413.93752)
		(134.5, 409.5070692307692)
		(135.5, 409.25404769230767)
		(136.5, 410.4214846153846)
		(137.5, 409.85442923076926)
		(138.5, 410.76613076923076)
		(139.5, 413.1138584615385)
		(140.5, 413.88375846153843)
		(141.5, 400.0314430769231)
		(142.5, 410.23710153846156)
		(143.5, 410.0435892307692)
		(144.5, 411.53149538461537)
		(145.5, 410.0815276923077)
		(146.5, 409.2790476923077)
		(147.5, 409.53003384615386)
		(148.5, 410.0379261538462)
		(149.5, 409.82891384615385)
		(150.5, 410.06028153846154)
		(151.5, 410.21454923076925)
		(152.5, 410.24440615384617)
		(153.5, 410.2934507692308)
		(154.5, 410.0093876923077)
		(155.5, 411.1184646153846)
		(156.5, 409.9064646153846)
		(157.5, 410.0383969230769)
		(158.5, 409.9878276923077)
		(159.5, 409.4781538461539)
		(160.5, 411.1131046153846)
		(161.5, 411.3579107692308)
		(162.5, 409.7501030769231)
		(163.5, 409.8183)
		(164.5, 410.69946923076924)
		(165.5, 415.00427076923074)
		(166.5, 401.37291230769233)
		(167.5, 411.0138969230769)
		(168.5, 410.4005707692308)
		(169.5, 406.75050153846155)
		(170.5, 414.0770430769231)
		(171.5, 540.7485846153846)
		(172.5, 411.9242246153846)
		(173.5, 410.89756)
		(174.5, 410.90150153846156)
		(175.5, 411.5746646153846)
		(176.5, 410.70273846153844)
		(177.5, 0)
		(0.5, 0)
	};

	\addplot+[Pattern A] coordinates{
		(171.5, 0)
		(171.5, 540.7485846153846)
		(172.5, 0)
		(171.5, 0)
	};

	\addplot+[Pattern B] coordinates{
		(172.5, 0)
		(172.5, 411.9242246153846)
		(173.5, 0)
		(172.5, 0)
	};

}
\\
\DOUBLEAXIS{Pattern = {212.05917615384615}, Scatter, Scatter Left Label, Scatter Bottom Label = {212.05917615384615}, title={\null\hfill Json}}{Pattern = {1.2189819042402108}, Scatter, Right Axis = {1.2189819042402108}}{
	\addplot+[] coordinates{
		(35.07462686567165, 172.28333692307692)
		(5.223880597014925, 172.68620153846155)
		(10.44776119402985, 171.83396307692308)
		(46.26865671641791, 172.77547384615386)
		(50.74626865671642, 168.83480307692307)
		(50.0, 173.22181076923076)
		(58.95522388059702, 177.52718153846155)
		(23.134328358208954, 169.47055384615385)
		(20.8955223880597, 173.48855230769232)
		(5.970149253731343, 172.10772307692307)
		(17.16417910447761, 170.55124615384617)
		(67.16417910447761, 171.68151230769232)
		(70.1492537313433, 169.4971476923077)
		(2.2388059701492535, 173.38218307692307)
		(17.91044776119403, 173.72397230769232)
		(8.955223880597014, 170.38774615384617)
		(47.76119402985074, 173.77804615384616)
		(43.28358208955223, 174.24379846153846)
		(28.35820895522388, 170.3916046153846)
		(23.88059701492537, 173.2905153846154)
		(53.73134328358209, 170.83348923076923)
		(7.462686567164178, 170.00080923076922)
		(29.850746268656714, 168.18070153846153)
		(0.7462686567164178, 173.58076461538462)
		(14.925373134328357, 174.84640769230768)
		(4.477611940298507, 172.9343353846154)
		(49.25373134328358, 171.3912476923077)
		(1.4925373134328357, 170.19082)
		(40.298507462686565, 174.0615723076923)
		(91.04477611940298, 171.71981384615384)
		(82.83582089552239, 170.4570369230769)
		(11.194029850746269, 171.60086615384614)
		(38.059701492537314, 168.46174923076924)
		(97.01492537313433, 171.6187276923077)
		(56.71641791044776, 168.76458)
		(11.940298507462686, 172.90889846153846)
		(19.402985074626866, 168.58201846153847)
		(68.65671641791045, 170.31847692307693)
		(26.119402985074625, 171.41591384615384)
		(2.9850746268656714, 172.6440276923077)
		(29.1044776119403, 173.28739076923077)
		(52.98507462686567, 169.87476307692307)
		(32.83582089552239, 173.28686461538462)
		(58.2089552238806, 169.83102461538462)
		(0.0, 173.96417076923078)
		(16.417910447761194, 170.57913384615384)
		(26.865671641791046, 173.92400923076923)
		(41.7910447761194, 169.5297676923077)
		(31.343283582089555, 172.63255846153845)
		(14.17910447761194, 174.31824307692307)
		(8.208955223880597, 172.6991046153846)
		(79.1044776119403, 170.96677692307694)
		(55.970149253731336, 174.57297692307694)
	};

	\addplot+[Pattern A] coordinates{
		(55.223880597014926, 175.71049384615384)
		(92.53731343283582, 171.53264153846155)
		(89.55223880597015, 171.94824307692306)
		(97.76119402985076, 171.90346153846153)
		(85.07462686567165, 170.21673846153845)
		(62.68656716417911, 170.99116153846154)
		(71.64179104477611, 171.15657692307693)
		(20.149253731343283, 172.25893076923077)
		(37.3134328358209, 170.0299553846154)
		(44.02985074626866, 171.7505676923077)
		(73.13432835820896, 171.08624615384616)
		(70.8955223880597, 172.49303076923076)
		(88.80597014925374, 171.37408615384615)
		(59.70149253731343, 170.16109230769231)
		(76.86567164179104, 171.12082615384617)
		(61.940298507462686, 170.19189846153847)
		(64.92537313432835, 171.6184246153846)
		(86.56716417910447, 172.13789538461538)
		(73.88059701492537, 172.22969846153845)
		(85.82089552238806, 172.88388923076923)
		(77.61194029850746, 175.41718307692307)
		(80.59701492537313, 173.47786923076924)
		(83.5820895522388, 174.09923076923076)
		(61.19402985074627, 167.54895692307693)
		(82.08955223880598, 173.48941846153846)
		(94.77611940298507, 169.68675384615383)
		(65.67164179104478, 169.16839846153846)
		(47.01492537313433, 172.11491076923076)
		(95.52238805970148, 172.71231538461538)
		(94.02985074626866, 170.01346)
		(44.776119402985074, 172.36557846153846)
		(25.37313432835821, 174.68568153846155)
		(35.82089552238806, 168.75591076923078)
		(64.17910447761194, 172.26682615384615)
		(79.8507462686567, 171.07154461538462)
		(13.432835820895523, 173.4628646153846)
		(38.80597014925373, 173.3625676923077)
		(91.7910447761194, 169.78995076923076)
		(76.11940298507463, 169.97938461538462)
		(98.50746268656717, 172.76446)
		(67.91044776119402, 173.86280923076924)
		(100.0, 170.72024615384615)
		(41.04477611940299, 170.14858615384614)
		(52.23880597014925, 171.1361446153846)
		(74.6268656716418, 173.55657076923077)
		(88.05970149253731, 170.97635076923078)
		(22.388059701492537, 170.43658307692309)
		(34.32835820895522, 172.73087384615386)
		(32.08955223880597, 173.75718307692307)
	};

	\addplot+[Pattern B] coordinates{
	};

	\addplot+[Pattern AB] coordinates{
	};

}
\&
\DOUBLEAXIS{Pattern = {212.05917615384615}, Column = {20}, Column Bottom Label = {212.05917615384615}{20}, title={\null\hfill{\small(134 Type Annotations)}}}{Pattern = {1.2374376505013052}, Column = {20}, Right Axis = {1.2374376505013052}}{
	\addplot+[] coordinates{
		(0.5, 171.39390615384616)
		(1.5, 173.4138153846154)
		(2.5, 170.61360769230768)
		(3.5, 171.0169923076923)
		(4.5, 174.13434307692307)
		(5.5, 171.35723846153846)
		(6.5, 173.34031076923077)
		(7.5, 171.4403676923077)
		(8.5, 172.07709538461538)
		(9.5, 174.94990615384614)
		(10.5, 172.22272615384617)
		(11.5, 171.07729384615385)
		(12.5, 172.3132030769231)
		(13.5, 174.04801846153845)
		(14.5, 171.38190615384616)
		(15.5, 173.82652923076924)
		(16.5, 174.08601692307693)
		(17.5, 174.66852307692307)
		(18.5, 171.3227523076923)
		(19.5, 172.12921384615385)
		(20.5, 170.3377923076923)
		(21.5, 168.95014)
		(22.5, 170.0242153846154)
		(23.5, 171.1847676923077)
		(24.5, 171.34396615384617)
		(25.5, 171.69378307692307)
		(26.5, 171.61189692307693)
		(27.5, 171.7062646153846)
		(28.5, 173.55301846153847)
		(29.5, 172.05697538461538)
		(30.5, 175.36038923076924)
		(31.5, 175.37595384615383)
		(32.5, 172.26860307692309)
		(33.5, 172.11142923076923)
		(34.5, 172.1168676923077)
		(35.5, 171.6715323076923)
		(36.5, 171.3028)
		(37.5, 172.73757846153845)
		(38.5, 170.34234615384617)
		(39.5, 174.08749076923078)
		(40.5, 173.3443630769231)
		(41.5, 173.44526615384615)
		(42.5, 173.30650307692306)
		(43.5, 171.24672615384614)
		(44.5, 170.39602923076924)
		(45.5, 171.8203276923077)
		(46.5, 172.19196)
		(47.5, 171.96192307692309)
		(48.5, 174.06913538461538)
		(49.5, 176.84604153846155)
		(50.5, 171.22771230769231)
		(51.5, 171.82789692307693)
		(52.5, 174.15355384615384)
		(53.5, 181.55467692307693)
		(54.5, 171.38828615384617)
		(55.5, 172.5353323076923)
		(56.5, 171.62043384615384)
		(57.5, 172.62991538461537)
		(58.5, 171.84112769230768)
		(59.5, 170.4899076923077)
		(60.5, 170.7773723076923)
		(61.5, 172.50369076923076)
		(62.5, 173.52939692307692)
		(63.5, 171.37507384615384)
		(64.5, 171.81147384615386)
		(65.5, 172.03966461538462)
		(66.5, 173.22016769230768)
		(67.5, 173.82126769230769)
		(68.5, 173.7855646153846)
		(69.5, 173.9259846153846)
		(70.5, 175.58752153846154)
		(71.5, 171.83938923076923)
		(72.5, 170.53561846153846)
		(73.5, 173.2215723076923)
		(74.5, 171.34254)
		(75.5, 176.92297076923077)
		(76.5, 168.56272307692308)
		(77.5, 172.0992523076923)
		(78.5, 171.91488)
		(79.5, 172.29225076923078)
		(80.5, 172.9049246153846)
		(81.5, 173.25976307692306)
		(82.5, 170.8096876923077)
		(83.5, 174.52111692307693)
		(84.5, 170.53688615384615)
		(85.5, 173.6887153846154)
		(86.5, 170.19787846153847)
		(87.5, 171.4802446153846)
		(88.5, 170.98162)
		(89.5, 172.65823538461538)
		(90.5, 171.8077153846154)
		(91.5, 170.11281692307693)
		(92.5, 170.60526615384615)
		(93.5, 172.82028923076922)
		(94.5, 171.1760230769231)
		(95.5, 174.81551384615383)
		(96.5, 172.1102476923077)
		(97.5, 171.74101846153846)
		(98.5, 174.6339923076923)
		(99.5, 171.62539384615386)
		(100.5, 170.72225384615385)
		(101.5, 175.2031630769231)
		(102.5, 172.73494153846153)
		(103.5, 174.34018461538463)
		(104.5, 173.26057076923078)
		(105.5, 172.42387692307693)
		(106.5, 173.62660461538462)
		(107.5, 174.64854923076922)
		(108.5, 169.77868923076923)
		(109.5, 171.85114)
		(110.5, 172.59952)
		(111.5, 192.78106923076922)
		(112.5, 172.4763123076923)
		(113.5, 173.9787630769231)
		(114.5, 170.97575076923076)
		(115.5, 173.61701384615384)
		(116.5, 174.05645076923076)
		(117.5, 169.15539076923076)
		(118.5, 169.42369846153846)
		(119.5, 172.75758461538462)
		(120.5, 174.58386307692308)
		(121.5, 173.53821846153846)
		(122.5, 171.37838461538462)
		(123.5, 174.17401692307692)
		(124.5, 174.63966)
		(125.5, 172.8189476923077)
		(126.5, 173.75946615384615)
		(127.5, 174.3486723076923)
		(128.5, 171.34242923076923)
		(129.5, 173.59248)
		(130.5, 180.33236923076922)
		(131.5, 173.2842723076923)
		(132.5, 176.32364307692308)
		(133.5, 171.2113646153846)
		(134.5, 0)
		(0.5, 0)
	};

	\addplot+[Pattern A] coordinates{
		(111.5, 0)
		(111.5, 192.78106923076922)
		(112.5, 0)
		(111.5, 0)
	};

}
\&
\DOUBLEAXIS{Pattern = {65.71047492307693}, Scatter, Scatter Bottom Label = {65.71047492307693}, title={\null\hfill PyStone}}{Pattern = {1.1760959870666312}, Scatter, Right Axis = {1.1760959870666312}}{
	\addplot+[] coordinates{
		(15.294117647058824, 56.660033846153844)
		(43.529411764705884, 56.40105692307692)
		(29.411764705882355, 58.848078461538464)
		(34.11764705882353, 58.13297076923077)
		(56.470588235294116, 58.69872615384615)
		(24.705882352941178, 57.57697846153846)
		(28.235294117647058, 59.37404)
		(30.58823529411765, 57.064676923076924)
		(40.0, 58.21980153846154)
		(18.823529411764707, 56.09611384615385)
		(25.882352941176475, 56.89674307692308)
		(7.0588235294117645, 57.661316923076924)
		(37.64705882352941, 57.744843076923075)
		(32.94117647058823, 57.49186923076923)
		(41.17647058823529, 56.14072923076923)
		(18.823529411764707, 55.878126153846154)
		(17.647058823529413, 57.535556923076925)
		(12.941176470588237, 58.41391076923077)
		(31.76470588235294, 56.39568461538462)
		(55.294117647058826, 57.73047230769231)
		(23.52941176470588, 58.36691230769231)
		(45.88235294117647, 58.68052615384615)
		(4.705882352941177, 56.16973230769231)
		(25.882352941176475, 57.873004615384616)
		(1.1764705882352942, 55.84576461538462)
		(0.0, 55.871693846153846)
		(3.5294117647058822, 56.30575384615385)
		(2.3529411764705883, 55.91780307692308)
		(60.0, 58.14136307692308)
		(11.76470588235294, 56.12747230769231)
		(42.35294117647059, 56.60390769230769)
		(5.88235294117647, 57.76413230769231)
		(51.76470588235295, 57.21867692307692)
		(58.82352941176471, 58.29754153846154)
		(80.0, 56.24801076923077)
		(35.294117647058826, 57.688324615384616)
		(37.64705882352941, 58.10487538461538)
		(65.88235294117646, 59.39486)
		(47.05882352941176, 59.19590615384615)
		(7.0588235294117645, 58.35161230769231)
		(50.588235294117645, 57.31403076923077)
		(38.82352941176471, 58.694367692307694)
		(20.0, 56.35365846153846)
		(1.1764705882352942, 55.33909538461538)
		(81.17647058823529, 59.62351230769231)
		(64.70588235294117, 56.14512)
		(31.76470588235294, 57.342815384615385)
		(21.176470588235293, 57.50318769230769)
		(8.235294117647058, 57.202804615384615)
		(16.470588235294116, 59.458395384615386)
	};

	\addplot+[Pattern A] coordinates{
		(84.70588235294117, 58.87713230769231)
		(68.23529411764706, 56.657747692307694)
		(63.52941176470588, 55.79986615384615)
		(96.47058823529412, 56.51936769230769)
		(36.470588235294116, 58.00651538461538)
		(49.411764705882355, 59.18167846153846)
		(50.588235294117645, 58.45326615384615)
		(75.29411764705883, 56.50503846153846)
		(9.411764705882353, 56.80349846153846)
		(88.23529411764706, 57.69730615384616)
		(78.82352941176471, 57.33108615384615)
		(74.11764705882354, 59.55146)
		(12.941176470588237, 57.07210307692308)
		(75.29411764705883, 57.503483076923075)
		(62.35294117647059, 57.423864615384616)
		(43.529411764705884, 57.19272153846154)
		(76.47058823529412, 56.03307538461539)
		(56.470588235294116, 56.744883076923074)
		(67.05882352941175, 56.51126769230769)
		(85.88235294117646, 58.50896461538461)
		(27.058823529411764, 57.073107692307694)
		(54.11764705882353, 59.248584615384615)
		(87.05882352941177, 59.736795384615384)
		(44.70588235294118, 56.42778153846154)
		(10.588235294117647, 57.410586153846154)
		(72.94117647058823, 56.57508769230769)
		(98.82352941176471, 59.69126923076923)
		(91.76470588235294, 59.54349846153846)
		(95.29411764705881, 59.66524153846154)
		(87.05882352941177, 56.599369230769234)
		(61.1764705882353, 57.718583076923075)
		(68.23529411764706, 57.794570769230766)
		(52.94117647058824, 57.28815384615385)
		(89.41176470588236, 56.764738461538464)
		(97.6470588235294, 59.53561538461538)
		(92.94117647058823, 57.245404615384615)
		(77.64705882352942, 59.20310615384616)
		(92.94117647058823, 56.10924461538462)
		(94.11764705882352, 59.46309230769231)
		(70.58823529411765, 57.02556153846154)
		(14.117647058823529, 57.05501692307692)
		(48.23529411764706, 55.70439538461538)
		(100.0, 59.63024769230769)
		(83.52941176470588, 57.94184461538462)
		(81.17647058823529, 57.17773692307692)
		(71.76470588235294, 59.10808)
		(57.647058823529406, 57.51346461538461)
		(22.35294117647059, 56.72327846153846)
		(69.41176470588235, 59.0571)
		(90.58823529411765, 57.11436)
		(82.35294117647058, 57.1653)
		(62.35294117647059, 56.44223846153846)
	};

	\addplot+[Pattern B] coordinates{
	};

	\addplot+[Pattern AB] coordinates{
	};

}
\&
\DOUBLEAXIS{Pattern = {65.71047492307693}, Column = {10}, Column Bottom Label = {65.71047492307693}{10}, title={\null\hfill{\small(85 Type Annotations)}}}{Pattern = {1.1867701735022866}, Column = {10}, Right Axis, Column Right Label = {1.1867701735022866}}{
	\addplot+[] coordinates{
		(0.5, 55.68036461538462)
		(1.5, 56.429718461538464)
		(2.5, 56.51241538461539)
		(3.5, 57.34346461538462)
		(4.5, 56.035164615384616)
		(5.5, 58.07991230769231)
		(6.5, 57.03559076923077)
		(7.5, 56.47412615384616)
		(8.5, 56.40279384615385)
		(9.5, 57.139023076923074)
		(10.5, 55.758364615384615)
		(11.5, 56.874533846153845)
		(12.5, 56.699806153846154)
		(13.5, 56.00562)
		(14.5, 55.57033846153846)
		(15.5, 58.22897692307692)
		(16.5, 56.69441538461538)
		(17.5, 56.00709230769231)
		(18.5, 55.26326615384615)
		(19.5, 55.37102307692308)
		(20.5, 55.48418153846154)
		(21.5, 55.565204615384616)
		(22.5, 56.957093846153846)
		(23.5, 55.812623076923074)
		(24.5, 56.09097538461538)
		(25.5, 56.5951)
		(26.5, 55.89578923076923)
		(27.5, 55.82186307692308)
		(28.5, 56.79867384615385)
		(29.5, 56.697963076923074)
		(30.5, 55.27834307692308)
		(31.5, 55.08627846153846)
		(32.5, 57.21192)
		(33.5, 55.704670769230766)
		(34.5, 56.40157846153846)
		(35.5, 56.04115076923077)
		(36.5, 55.473475384615384)
		(37.5, 55.558672307692305)
		(38.5, 56.51954307692308)
		(39.5, 55.679933846153844)
		(40.5, 56.29488923076923)
		(41.5, 56.03475538461539)
		(42.5, 55.76776769230769)
		(43.5, 56.65791230769231)
		(44.5, 55.46746769230769)
		(45.5, 55.69953230769231)
		(46.5, 57.119458461538464)
		(47.5, 56.79805692307692)
		(48.5, 57.955226153846155)
		(49.5, 56.05410923076923)
		(50.5, 55.95924307692308)
		(51.5, 56.12654461538462)
		(52.5, 56.14745692307692)
		(53.5, 55.72630153846154)
		(54.5, 55.53099538461539)
		(55.5, 56.356346153846154)
		(56.5, 56.66045384615385)
		(57.5, 56.918123076923074)
		(58.5, 56.75323230769231)
		(59.5, 56.384312307692305)
		(60.5, 56.29353538461538)
		(61.5, 56.298546153846154)
		(62.5, 54.67920461538461)
		(63.5, 57.359507692307695)
		(64.5, 56.58062)
		(65.5, 55.74772461538461)
		(66.5, 55.81007384615385)
		(67.5, 56.007004615384616)
		(68.5, 57.28462)
		(69.5, 55.87768923076923)
		(70.5, 55.65443230769231)
		(71.5, 55.61888)
		(72.5, 55.11486769230769)
		(73.5, 55.852003076923076)
		(74.5, 55.755556923076924)
		(75.5, 55.59046923076923)
		(76.5, 56.17762923076923)
		(77.5, 56.25727538461538)
		(78.5, 55.605373846153846)
		(79.5, 55.76552923076923)
		(80.5, 56.14806461538461)
		(81.5, 55.918103076923074)
		(82.5, 55.60980615384615)
		(83.5, 56.394216923076925)
		(84.5, 55.571083076923074)
		(85.5, 0)
		(0.5, 0)
	};

	\addplot+[Pattern A] coordinates{
		(15.5, 0)
		(15.5, 58.22897692307692)
		(16.5, 0)
		(15.5, 0)
	};

}
}
 	\caption{Pairs of colour coded scatter and column graphs. The scatter graphs represent the performance of a sample of the typing lattices. The column graphs show the performance of every configuration with only one type annotation. The scatter plots and column graphs are colour coded based on whether a particular type annotation or two are present in the source code.}
	\label{f:pattern}		
\end{figure*}

We hypothesise that the previously identified bi/tri-modal performance behaviour
as seen in the graphs for Permute and others (cf. figure~\ref{f:full}) are caused by only a few type annotations: i.e.\ there are a very few annotations that determine each benchmark's performance.

To verify this hypothesis 
for each type annotation we measured one additional
configuration, with only that single type annotation present.
We did this to compare the overhead of
each type annotation in isolation against the no-typecheck baseline.

Figure \ref{f:pattern} shows the results of these experiments.
It shows a pair of graphs for a selection of ten benchmarks: it's associated typing lattice scatter plot and a column graph showing the results of these single type annotation experiments.

The column graphs, to the right of the corresponding scatter plot, show the execution time of configurations with only a single type annotation, the x-axis indicates the index of this annotation, thus the first column represents the first annotation, and the last column represents the last one (in the order they appear in the source code). The y-axes for the column graphs are the same as the associated scatter graphs.

For each benchmark we highlighted one or two type annotations that seem to show a pattern for the typing lattice performance scatter plots, or in the case of SpectralNorm and Json had higher than usual columns. The identified types are those represented by the red and blue columns. The scatter plots are colour-coded accordingly: a red or blue circle represents configurations with the given type annotation present, but not both, purple circles represent configurations with both type annotations present, and grey circles represent those with neither. Though we exhaustively inspected such colour coded scatter plots for all 1,775 type annotations across all 21 benchmarks, the only patterns we noticed are those shown in \ref{f:pattern}.

As can be seen, most of the patterns we found in the typing lattice graphs correspond to outliers in the single type annotation column graphs. For the Snake, Towers and CD benchmarks, there is clearly a pattern where an individual type annotation (highlighted in red) appears responsible for the upper/slowest half of the typing lattice graphs. For Permute there is actually a significant performance \emph{increase}, indicated by the lower half of the lattice being highlighted in red. DeltaBlue and Go also show a slight increase in performance, but here this is caused by two type annotations (in red and blue); this effect is not cumulative, however: the purple dots are about the same height as the red and blue ones, so \emph{either} annotation is sufficient for the full benefit.

List is interesting as it demonstrates that the red and blue type annotations both cause a significant decrease in performance, which is even greater when both are present. Richards is particularly odd as it shows a performance decrease that occurs when the red type annotation is present, but not the blue one. We had previously identified this benchmark and the aforementioned type annotations \cite{roberts-and-co-ecoop-2019} as being the worst case for Moth.

For Json and PyStone, although we found outliers in the column graphs (such as types \#112 and \#16, highlighted in red), we were unable to find a matching pattern in the configuration performance.
Finally, for DeltaBlue there is a noticeable spike in the column graphs (at \#198), however we again found no apparent relation between this outlier and the overall performance.

In particular, observe that the grey dots for Snake, Towers, DeltaBlue, Go, and List are mostly flat horizontal lines, this indicates that by simply deleting the red and blue type annotations, the performance impact of transient type checking becomes negligible. However for CD and Richards, the transient type checking overhead of the grey dots is roughly linear, albeit still less than with the red and blue type annotations present. Thus we can observe that usually, only a couple of type annotations are responsible for the overhead caused by Moth's transient typed checking, whereas the rest are ``free''.

The remaining benchmarks, which we have not shown in figure \ref{f:pattern}, have even flatter column graphs than PyStone, and we could not identify any patterns in the typing lattices. Of those we did not show, the greatest difference in single-type configurations relative to the untyped configuration was $5.07\%$ (for the Float benchmark).

\section{Related Work}
\label{s-rel}

The high-performance computing community has been investigating how
tools and visualisations can help developers to utilise their systems
more efficiently \citep{Papenhausen:2016:IVT,daSilva:2019:PSV}.
Their focus is typically on parallelisation opportunities,
guided by run-time feedback, cost models, or heuristics. 
Their large body of work \citep{Isaacs:2014:PerfViz} uses various approaches,
though, we are not aware of work that has used an approach similar to ours.

At the moment, our approach to identifying type annotations that cause
performance anomalies is not integrated into a development environment.
Though, for instance Optimization Coaching
\citep{St-Amour:2012:OCO} is a promising direction.
Optimization Coaching uses feedback from the runtime to guide developers
to insert or change type declarations 
to enable a compiler to generate a more optimal program.
In this spirit, we would eventually want to achieve the same, although
in our case,
we need to run full experiments to get the necessary information.

\section{Discussion and Conclusion}
\label{s-concl}

In this paper we have investigated how benchmarks can be
repurposed to determine precisely which transient type checks are
likely to cause performance effects,
and are currently not optimised away by the just-in-time compiler.
However, detailed analysis will be needed in order to identify exactly what causes such performance effects: we previously undertook such an analysis for the List benchmark \cite{roberts-and-co-ecoop-2019}.

We observed that many of our benchmark results
conducted under the Takikawa protocol 
showed a characteristic bimodal performance profile: some
of the benchmark configurations ran significantly slower than
the remaining configurations. We also observed trimodal profiles, as well as performance increases when type annotations where added.

By inspecting graphs of the performance of where only one type annotation is present, we can easily identify type annotations that likely have a significant effect on performance. By then inspecting the typing lattice graphs colour coded based on whether such type annotation was present or not, we were often able to notice patterns.
In particular, these patterns suggested the just one or two type annotations are likely responsible for the bimodal performance and most of the overhead caused by Moth's transient type checking.
Though every type annotation that showed a significant effect across the typing lattice also showed a significant performance effect when no other typing annotations where present, the converse did not hold.

This is preliminary work, in particular we have not yet identified exactly why these type annotations show an effect on performance.
There are also a number of threats to validity. 
Regarding construct validity, 
our underlying
implementation may contain undetected bugs that affect the semantics
or performance of the gradual typing checks. Regarding internal
validity,
our benchmarking harness runs on the same implementation
and therefore is subject to the same issues.
Regarding external validity, 
Moth is built
on the Truffle and Graal toolchain, so we expect
to resemble other Graal
VMs doing similar AST-based optimizations of transient type checks.
Because we rely on common techniques, 
we expect our results to be transferable to other JIT implementations as well.

Finally, it is not clear how our results would transfer
to other gradually typed-languages or other semantics for gradual typing.
Our benchmarks do not depend on any features of Grace
that are not common in other object-oriented
languages, but as Grace lacks a large corpus of programs the
benchmarks are necessarily small and artificial.
The advantage of Grace for this research is
that their relative simplicity means we have been able to build an
implementation that features competitive performance with significantly less
effort than would be required for larger and more complex languages.

In the future, we hope to investigate statistical techniques to
determine the significance of each type annotation's contribution to a
programs overall performance. We would also like to investigate 
whether this approach can assist with optimisations for programmers' day-to-day  development, or help VM engineers identifying performance bugs in the underlying virtual machines.

\begin{acks}
The authors would like to thank the anonymous reviewers for their
valuable comments and helpful suggestions. This work is supported
in part by the \grantsponsor{Marsden Fund}{Royal Society of New Zealand (Te Ap\={a}rangi) Marsden Fund (Te P\={u}tea Rangahau a Marsden)}{https://royalsociety.org.nz/what-we-do/funds-and-opportunities/marsden/} under grant \grantnum{Marsden Fund}{VUW1815}.
\end{acks}

\bibliography{references}


\begin{thebibliography}{43}


\ifx \showCODEN    \undefined \def \showCODEN     #1{\unskip}     \fi
\ifx \showDOI      \undefined \def \showDOI       #1{#1}\fi
\ifx \showISBNx    \undefined \def \showISBNx     #1{\unskip}     \fi
\ifx \showISBNxiii \undefined \def \showISBNxiii  #1{\unskip}     \fi
\ifx \showISSN     \undefined \def \showISSN      #1{\unskip}     \fi
\ifx \showLCCN     \undefined \def \showLCCN      #1{\unskip}     \fi
\ifx \shownote     \undefined \def \shownote      #1{#1}          \fi
\ifx \showarticletitle \undefined \def \showarticletitle #1{#1}   \fi
\ifx \showURL      \undefined \def \showURL       {\relax}        \fi
\providecommand\bibfield[2]{#2}
\providecommand\bibinfo[2]{#2}
\providecommand\natexlab[1]{#1}
\providecommand\showeprint[2][]{arXiv:#2}

\bibitem[\protect\citeauthoryear{Abadi, Cardelli, Pierce, and Plotkin}{Abadi
  et~al\mbox{.}}{1991}]%
        {AbadiTOPLAS1991}
\bibfield{author}{\bibinfo{person}{Mart{\'{\i}}n Abadi}, \bibinfo{person}{Luca
  Cardelli}, \bibinfo{person}{Benjamin~C. Pierce}, {and}
  \bibinfo{person}{Gordon~D. Plotkin}.} \bibinfo{year}{1991}\natexlab{}.
\newblock \showarticletitle{Dynamic Typing in a Statically Typed Language}.
\newblock \bibinfo{journal}{\emph{{ACM} Trans. Program. Lang. Syst.}}
  \bibinfo{volume}{13}, \bibinfo{number}{2} (\bibinfo{year}{1991}),
  \bibinfo{pages}{237--268}.
\newblock


\bibitem[\protect\citeauthoryear{Barrett, Bolz-Tereick, Killick, Mount, and
  Tratt}{Barrett et~al\mbox{.}}{2017}]%
        {Barrett:2017:VMW}
\bibfield{author}{\bibinfo{person}{Edd Barrett},
  \bibinfo{person}{Carl~Friedrich Bolz-Tereick}, \bibinfo{person}{Rebecca
  Killick}, \bibinfo{person}{Sarah Mount}, {and} \bibinfo{person}{Laurence
  Tratt}.} \bibinfo{year}{2017}\natexlab{}.
\newblock \showarticletitle{{Virtual Machine Warmup Blows Hot and Cold}}.
\newblock \bibinfo{journal}{\emph{Proc. ACM Program. Lang.}}
  \bibinfo{volume}{1}, \bibinfo{number}{OOPSLA}, Article
  \bibinfo{articleno}{52} (\bibinfo{date}{Oct.} \bibinfo{year}{2017}),
  \bibinfo{numpages}{27}~pages.
\newblock
\showISSN{2475-1421}


\bibitem[\protect\citeauthoryear{Bauman, Bolz-Tereick, Siek, and
  Tobin-Hochstadt}{Bauman et~al\mbox{.}}{2017}]%
        {Bauman2017}
\bibfield{author}{\bibinfo{person}{Spenser Bauman},
  \bibinfo{person}{Carl~Friedrich Bolz-Tereick}, \bibinfo{person}{Jeremy Siek},
  {and} \bibinfo{person}{Sam Tobin-Hochstadt}.}
  \bibinfo{year}{2017}\natexlab{}.
\newblock \showarticletitle{Sound Gradual Typing: Only Mostly Dead}.
\newblock \bibinfo{journal}{\emph{Proc. ACM Program. Lang.}}
  \bibinfo{volume}{1}, \bibinfo{number}{OOPSLA}, Article
  \bibinfo{articleno}{54} (\bibinfo{date}{Oct.} \bibinfo{year}{2017}),
  \bibinfo{numpages}{24}~pages.
\newblock
\showISSN{2475-1421}


\bibitem[\protect\citeauthoryear{Bentley and Floyd}{Bentley and Floyd}{1987}]%
        {Bentley:1987:PPS:30401.315746}
\bibfield{author}{\bibinfo{person}{Jon Bentley} {and} \bibinfo{person}{Bob
  Floyd}.} \bibinfo{year}{1987}\natexlab{}.
\newblock \showarticletitle{Programming Pearls: A Sample of Brilliance}.
\newblock \bibinfo{journal}{\emph{Commun. ACM}} \bibinfo{volume}{30},
  \bibinfo{number}{9} (\bibinfo{date}{Sept.} \bibinfo{year}{1987}),
  \bibinfo{pages}{754--757}.
\newblock
\showISSN{0001-0782}
\urldef\tempurl%
\url{https://doi.org/10.1145/30401.315746}
\showDOI{\tempurl}


\bibitem[\protect\citeauthoryear{Black, Bruce, Homer, and Noble}{Black
  et~al\mbox{.}}{2012}]%
        {graceOnward12}
\bibfield{author}{\bibinfo{person}{Andrew~P. Black}, \bibinfo{person}{Kim~B.
  Bruce}, \bibinfo{person}{Michael Homer}, {and} \bibinfo{person}{James
  Noble}.} \bibinfo{year}{2012}\natexlab{}.
\newblock \showarticletitle{{Grace}: the absence of (inessential) difficulty}.
  In \bibinfo{booktitle}{\emph{Onward! '12: Proceedings 12th SIGPLAN Symp. on
  New Ideas in Programming and Reflections on Software}}.
  \bibinfo{publisher}{ACM}, \bibinfo{address}{New York, NY},
  \bibinfo{pages}{85--98}.
\newblock


\bibitem[\protect\citeauthoryear{Black, Hutchinson, Jul, and Levy}{Black
  et~al\mbox{.}}{2007}]%
        {Black2007-emeraldHOPL}
\bibfield{author}{\bibinfo{person}{Andrew~P. Black}, \bibinfo{person}{Norman~C.
  Hutchinson}, \bibinfo{person}{Eric Jul}, {and} \bibinfo{person}{Henry~M.
  Levy}.} \bibinfo{year}{2007}\natexlab{}.
\newblock \showarticletitle{The development of the {E}merald programming
  language}. In \bibinfo{booktitle}{\emph{Proceedings of the Third {ACM}
  {SIGPLAN} History of Programming Languages Conference (HOPL-III), San Diego,
  California, USA, 9-10 June 2007}}. \bibinfo{pages}{1--51}.
\newblock


\bibitem[\protect\citeauthoryear{Bloom, Field, Nystrom, \"{O}stlund, Richards,
  Strni\v{s}a, Vitek, and Wrigstad}{Bloom et~al\mbox{.}}{2009}]%
        {Bloom2009}
\bibfield{author}{\bibinfo{person}{Bard Bloom}, \bibinfo{person}{John Field},
  \bibinfo{person}{Nathaniel Nystrom}, \bibinfo{person}{Johan \"{O}stlund},
  \bibinfo{person}{Gregor Richards}, \bibinfo{person}{Rok Strni\v{s}a},
  \bibinfo{person}{Jan Vitek}, {and} \bibinfo{person}{Tobias Wrigstad}.}
  \bibinfo{year}{2009}\natexlab{}.
\newblock \showarticletitle{Thorn: Robust, Concurrent, Extensible Scripting on
  the {JVM}}. In \bibinfo{booktitle}{\emph{Proceedings of the ACM
  SIGPLAN-SIGACT Symposium on Principles of Programming Languages {(POPL)}}}.
  \bibinfo{pages}{117--136}.
\newblock


\bibitem[\protect\citeauthoryear{Bracha}{Bracha}{2004}]%
        {GiladPluggable2004}
\bibfield{author}{\bibinfo{person}{Gilad Bracha}.}
  \bibinfo{year}{2004}\natexlab{}.
\newblock \bibinfo{title}{{Pluggable Type Systems}}.
\newblock \bibinfo{howpublished}{{OOPSLA} Workshop on Revival of Dynamic
  Languages}. , \bibinfo{numpages}{6}~pages.
\newblock


\bibitem[\protect\citeauthoryear{Bruce, Black, Homer, Noble, Ruskin, and
  Yannow}{Bruce et~al\mbox{.}}{2013}]%
        {graceSigcse13}
\bibfield{author}{\bibinfo{person}{Kim Bruce}, \bibinfo{person}{Andrew Black},
  \bibinfo{person}{Michael Homer}, \bibinfo{person}{James Noble},
  \bibinfo{person}{Amy Ruskin}, {and} \bibinfo{person}{Richard Yannow}.}
  \bibinfo{year}{2013}\natexlab{}.
\newblock \showarticletitle{Seeking {Grace}: a new object-oriented language for
  novices}. In \bibinfo{booktitle}{\emph{Proceedings 44th SIGCSE Technical
  Symposium on Computer Science Education}}. \bibinfo{publisher}{ACM},
  \bibinfo{pages}{129--134}.
\newblock


\bibitem[\protect\citeauthoryear{Chung, Li, Nardelli, and Vitek}{Chung
  et~al\mbox{.}}{2018}]%
        {kafka18}
\bibfield{author}{\bibinfo{person}{Benjamin Chung}, \bibinfo{person}{Paley Li},
  \bibinfo{person}{Francesco~Zappa Nardelli}, {and} \bibinfo{person}{Jan
  Vitek}.} \bibinfo{year}{2018}\natexlab{}.
\newblock \showarticletitle{{KafKa:} Gradual Typing for Objects}. In
  \bibinfo{booktitle}{\emph{32nd European Conference on Object-Oriented
  Programming, {ECOOP} 2018, July 16-21, 2018, Amsterdam, The Netherlands}}.
  \bibinfo{pages}{12:1--12:24}.
\newblock
\urldef\tempurl%
\url{https://doi.org/10.4230/LIPIcs.ECOOP.2018.12}
\showDOI{\tempurl}


\bibitem[\protect\citeauthoryear{da~Silva, Cunha, Silva, de~A.~Furtunato, and
  Xavier-de Souza}{da~Silva et~al\mbox{.}}{2019}]%
        {daSilva:2019:PSV}
\bibfield{author}{\bibinfo{person}{Anderson B.~N. da Silva},
  \bibinfo{person}{Daniel A.~M. Cunha}, \bibinfo{person}{Vitor R.~G. Silva},
  \bibinfo{person}{Alex~F. de A.~Furtunato}, {and} \bibinfo{person}{Samuel
  Xavier-de Souza}.} \bibinfo{year}{2019}\natexlab{}.
\newblock \showarticletitle{{PaScal Viewer: A Tool for the Visualization of
  Parallel Scalability Trends}}. In \bibinfo{booktitle}{\emph{Programming and
  Performance Visualization Tools}}, \bibfield{editor}{\bibinfo{person}{Abhinav
  Bhatele}, \bibinfo{person}{David Boehme}, \bibinfo{person}{Joshua~A. Levine},
  \bibinfo{person}{Allen~D. Malony}, {and} \bibinfo{person}{Martin Schulz}}
  (Eds.). \bibinfo{publisher}{Springer International Publishing},
  \bibinfo{address}{Cham}, \bibinfo{pages}{250--264}.
\newblock
\showISBNx{978-3-030-17872-7}


\bibitem[\protect\citeauthoryear{Daloze, Marr, Bonetta, and
  M{\"o}ssenb{\"o}ck}{Daloze et~al\mbox{.}}{2016}]%
        {Daloze2016}
\bibfield{author}{\bibinfo{person}{Benoit Daloze}, \bibinfo{person}{Stefan
  Marr}, \bibinfo{person}{Daniele Bonetta}, {and} \bibinfo{person}{Hanspeter
  M{\"o}ssenb{\"o}ck}.} \bibinfo{year}{2016}\natexlab{}.
\newblock \showarticletitle{{Efficient and Thread-Safe Objects for
  Dynamically-Typed Languages}}. In \bibinfo{booktitle}{\emph{Proceedings of
  the 2016 ACM International Conference on Object Oriented Programming Systems
  Languages \& Applications}} \emph{(\bibinfo{series}{OOPSLA'16})}.
  \bibinfo{publisher}{ACM}, \bibinfo{pages}{642--659}.
\newblock


\bibitem[\protect\citeauthoryear{Goldberg and Robson}{Goldberg and
  Robson}{1983}]%
        {bluebook}
\bibfield{author}{\bibinfo{person}{Adele Goldberg} {and} \bibinfo{person}{David
  Robson}.} \bibinfo{year}{1983}\natexlab{}.
\newblock \bibinfo{booktitle}{\emph{{S}malltalk-80: The Language and its
  Implementation}}.
\newblock \bibinfo{publisher}{Addison-Wesley}.
\newblock


\bibitem[\protect\citeauthoryear{Greenman and Felleisen}{Greenman and
  Felleisen}{2018}]%
        {bensurvey18icfp}
\bibfield{author}{\bibinfo{person}{Ben Greenman} {and}
  \bibinfo{person}{Matthias Felleisen}.} \bibinfo{year}{2018}\natexlab{}.
\newblock \showarticletitle{A spectrum of type soundness and performance}.
\newblock \bibinfo{journal}{\emph{{PACMPL}}} \bibinfo{volume}{2},
  \bibinfo{number}{{ICFP}} (\bibinfo{year}{2018}),
  \bibinfo{pages}{71:1--71:32}.
\newblock
\urldef\tempurl%
\url{https://doi.org/10.1145/3236766}
\showDOI{\tempurl}


\bibitem[\protect\citeauthoryear{Greenman and Migeed}{Greenman and
  Migeed}{2018}]%
        {Greenman2018}
\bibfield{author}{\bibinfo{person}{Ben Greenman} {and} \bibinfo{person}{Zeina
  Migeed}.} \bibinfo{year}{2018}\natexlab{}.
\newblock \showarticletitle{On the Cost of Type-Tag Soundness}. In
  \bibinfo{booktitle}{\emph{Proceedings of the ACM SIGPLAN Workshop on Partial
  Evaluation and Program Manipulation}} \emph{(\bibinfo{series}{PEPM'18})}.
  \bibinfo{publisher}{ACM}, \bibinfo{pages}{30--39}.
\newblock
\showISBNx{978-1-4503-5587-2}


\bibitem[\protect\citeauthoryear{Greenman, Takikawa, New, Feltey, Findler,
  Vitek, and Felleisen}{Greenman et~al\mbox{.}}{2019}]%
        {Greenman2019jfp}
\bibfield{author}{\bibinfo{person}{Ben Greenman}, \bibinfo{person}{Asumu
  Takikawa}, \bibinfo{person}{Max~S. New}, \bibinfo{person}{Daniel Feltey},
  \bibinfo{person}{Robert~Bruce Findler}, \bibinfo{person}{Jan Vitek}, {and}
  \bibinfo{person}{Matthias Felleisen}.} \bibinfo{year}{2019}\natexlab{}.
\newblock \showarticletitle{How to evaluate the performance of gradual type
  systems}.
\newblock \bibinfo{journal}{\emph{Journal of Functional Programming}}
  \bibinfo{volume}{29} (\bibinfo{year}{2019}), \bibinfo{pages}{45}.
\newblock
\urldef\tempurl%
\url{https://doi.org/10.1017/S0956796818000217}
\showDOI{\tempurl}


\bibitem[\protect\citeauthoryear{Hölzle, Chambers, and Ungar}{Hölzle
  et~al\mbox{.}}{1991}]%
        {Hoelzle:91:PIC}
\bibfield{author}{\bibinfo{person}{Urs Hölzle}, \bibinfo{person}{Craig
  Chambers}, {and} \bibinfo{person}{David Ungar}.}
  \bibinfo{year}{1991}\natexlab{}.
\newblock \showarticletitle{{Optimizing Dynamically-Typed Object-Oriented
  Languages With Polymorphic Inline Caches}}. In
  \bibinfo{booktitle}{\emph{ECOOP '91: European Conference on Object-Oriented
  Programming}} \emph{(\bibinfo{series}{LNCS})}, Vol.~\bibinfo{volume}{512}.
  \bibinfo{publisher}{Springer}, \bibinfo{pages}{21--38}.
\newblock
\showISBNx{3-540-54262-0}
\urldef\tempurl%
\url{https://doi.org/10.1007/BFb0057013}
\showDOI{\tempurl}


\bibitem[\protect\citeauthoryear{Isaacs, Giménez, Jusufi, Gamblin, Bhatele,
  Schulz, Hamann, and Bremer}{Isaacs et~al\mbox{.}}{2014}]%
        {Isaacs:2014:PerfViz}
\bibfield{author}{\bibinfo{person}{Katherine~E. Isaacs},
  \bibinfo{person}{Alfredo Giménez}, \bibinfo{person}{Ilir Jusufi},
  \bibinfo{person}{Todd Gamblin}, \bibinfo{person}{Abhinav Bhatele},
  \bibinfo{person}{Martin Schulz}, \bibinfo{person}{Bernd Hamann}, {and}
  \bibinfo{person}{Peer-Timo Bremer}.} \bibinfo{year}{2014}\natexlab{}.
\newblock \showarticletitle{{State of the Art of Performance Visualization}}.
  In \bibinfo{booktitle}{\emph{EuroVis - STARs}},
  \bibfield{editor}{\bibinfo{person}{R.~Borgo},
  \bibinfo{person}{R.~Maciejewski}, {and} \bibinfo{person}{I.~Viola}} (Eds.).
  \bibinfo{publisher}{The Eurographics Association}.
\newblock
\showISBNx{978-3-03868-028-4}
\urldef\tempurl%
\url{https://doi.org/10.2312/eurovisstar.20141177}
\showDOI{\tempurl}


\bibitem[\protect\citeauthoryear{Jones}{Jones}{2017}]%
        {TimJonesThesis}
\bibfield{author}{\bibinfo{person}{Timothy Jones}.}
  \bibinfo{year}{2017}\natexlab{}.
\newblock \emph{\bibinfo{title}{Classless Object Semantics}}.
\newblock \bibinfo{thesistype}{Ph.D. Dissertation}. \bibinfo{school}{Victoria
  University of Wellington}.
\newblock


\bibitem[\protect\citeauthoryear{Jones, Homer, Noble, and Bruce}{Jones
  et~al\mbox{.}}{2016}]%
        {JonesECOOP2016}
\bibfield{author}{\bibinfo{person}{Timothy Jones}, \bibinfo{person}{Michael
  Homer}, \bibinfo{person}{James Noble}, {and} \bibinfo{person}{Kim Bruce}.}
  \bibinfo{year}{2016}\natexlab{}.
\newblock \showarticletitle{Object Inheritance Without Classes}. In
  \bibinfo{booktitle}{\emph{30th European Conference on Object-Oriented
  Programming (ECOOP 2016)}}, Vol.~\bibinfo{volume}{56}.
  \bibinfo{pages}{13:1--13:26}.
\newblock


\bibitem[\protect\citeauthoryear{Lehrmann~Madsen, M{\o}ller-Pedersen, and
  Nygaard}{Lehrmann~Madsen et~al\mbox{.}}{1993}]%
        {betabook}
\bibfield{author}{\bibinfo{person}{Ole Lehrmann~Madsen},
  \bibinfo{person}{Birger M{\o}ller-Pedersen}, {and} \bibinfo{person}{Kristen
  Nygaard}.} \bibinfo{year}{1993}\natexlab{}.
\newblock \bibinfo{booktitle}{\emph{Object-Oriented Programming in the {BETA}
  Programming Language}}.
\newblock \bibinfo{publisher}{Addison-Wesley}.
\newblock


\bibitem[\protect\citeauthoryear{Marr}{Marr}{2018a}]%
        {ReBench:2018}
\bibfield{author}{\bibinfo{person}{Stefan Marr}.}
  \bibinfo{year}{2018}\natexlab{a}.
\newblock \bibinfo{title}{ReBench: Execute and Document Benchmarks
  Reproducibly}.
\newblock
\newblock
\urldef\tempurl%
\url{https://doi.org/10.5281/zenodo.1311762}
\showDOI{\tempurl}
\newblock
\shownote{Version 1.0.}


\bibitem[\protect\citeauthoryear{Marr}{Marr}{2018b}]%
        {SOMns}
\bibfield{author}{\bibinfo{person}{Stefan Marr}.}
  \bibinfo{year}{2018}\natexlab{b}.
\newblock \bibinfo{title}{{SOMns:} A Newspeak for Concurrency Research}.
\newblock
\newblock
\urldef\tempurl%
\url{https://doi.org/10.5281/zenodo.3270908}
\showDOI{\tempurl}


\bibitem[\protect\citeauthoryear{Marr, Daloze, and M\"{o}ssenb\"{o}ck}{Marr
  et~al\mbox{.}}{2016}]%
        {Marr2016}
\bibfield{author}{\bibinfo{person}{Stefan Marr}, \bibinfo{person}{Benoit
  Daloze}, {and} \bibinfo{person}{Hanspeter M\"{o}ssenb\"{o}ck}.}
  \bibinfo{year}{2016}\natexlab{}.
\newblock \showarticletitle{{Cross-Language Compiler Benchmarking---Are We Fast
  Yet?}}. In \bibinfo{booktitle}{\emph{Proceedings of the 12th Symposium on
  Dynamic Languages}} \emph{(\bibinfo{series}{DLS'16})}.
  \bibinfo{publisher}{ACM}, \bibinfo{pages}{120--131}.
\newblock
\showISBNx{978-1-4503-4445-6}


\bibitem[\protect\citeauthoryear{Muehlboeck and Tate}{Muehlboeck and
  Tate}{2017}]%
        {Muehlboeck2017}
\bibfield{author}{\bibinfo{person}{Fabian Muehlboeck} {and}
  \bibinfo{person}{Ross Tate}.} \bibinfo{year}{2017}\natexlab{}.
\newblock \showarticletitle{Sound Gradual Typing is Nominally Alive and Well}.
\newblock \bibinfo{journal}{\emph{Proc. ACM Program. Lang.}}
  \bibinfo{volume}{1}, \bibinfo{number}{OOPSLA}, Article
  \bibinfo{articleno}{56} (\bibinfo{date}{Oct.} \bibinfo{year}{2017}),
  \bibinfo{numpages}{30}~pages.
\newblock
\showISSN{2475-1421}


\bibitem[\protect\citeauthoryear{{Papenhausen}, {Mueller}, {Langston},
  {Meister}, and {Lethin}}{{Papenhausen} et~al\mbox{.}}{2016}]%
        {Papenhausen:2016:IVT}
\bibfield{author}{\bibinfo{person}{E. {Papenhausen}}, \bibinfo{person}{K.
  {Mueller}}, \bibinfo{person}{M.~H. {Langston}}, \bibinfo{person}{B.
  {Meister}}, {and} \bibinfo{person}{R. {Lethin}}.}
  \bibinfo{year}{2016}\natexlab{}.
\newblock \showarticletitle{An Interactive Visual Tool for Code Optimization
  and Parallelization Based on the Polyhedral Model}. In
  \bibinfo{booktitle}{\emph{45th International Conference on Parallel
  Processing Workshops}} \emph{(\bibinfo{series}{ICPPW'16})}.
  \bibinfo{pages}{309--318}.
\newblock
\showISSN{2332-5690}
\urldef\tempurl%
\url{https://doi.org/10.1109/ICPPW.2016.52}
\showDOI{\tempurl}


\bibitem[\protect\citeauthoryear{Richards, Arteca, and Turcotte}{Richards
  et~al\mbox{.}}{2017}]%
        {Richards2017}
\bibfield{author}{\bibinfo{person}{Gregor Richards}, \bibinfo{person}{Ellen
  Arteca}, {and} \bibinfo{person}{Alexi Turcotte}.}
  \bibinfo{year}{2017}\natexlab{}.
\newblock \showarticletitle{The {VM} Already Knew That: Leveraging Compile-time
  Knowledge to Optimize Gradual Typing}.
\newblock \bibinfo{journal}{\emph{Proc. ACM Program. Lang.}}
  \bibinfo{volume}{1}, \bibinfo{number}{OOPSLA}, Article
  \bibinfo{articleno}{55} (\bibinfo{date}{Oct.} \bibinfo{year}{2017}),
  \bibinfo{numpages}{27}~pages.
\newblock
\showISSN{2475-1421}


\bibitem[\protect\citeauthoryear{Richards, Nardelli, and Vitek}{Richards
  et~al\mbox{.}}{2015}]%
        {concrete15}
\bibfield{author}{\bibinfo{person}{Gregor Richards},
  \bibinfo{person}{Francesco~Zappa Nardelli}, {and} \bibinfo{person}{Jan
  Vitek}.} \bibinfo{year}{2015}\natexlab{}.
\newblock \showarticletitle{Concrete Types for {TypeScript}}. In
  \bibinfo{booktitle}{\emph{29th European Conference on Object-Oriented
  Programming, {ECOOP} 2015, July 5-10, 2015, Prague, Czech Republic}}.
  \bibinfo{pages}{76--100}.
\newblock
\urldef\tempurl%
\url{https://doi.org/10.4230/LIPIcs.ECOOP.2015.76}
\showDOI{\tempurl}


\bibitem[\protect\citeauthoryear{Roberts, Marr, Homer, and Noble}{Roberts
  et~al\mbox{.}}{2017}]%
        {Roberts2017}
\bibfield{author}{\bibinfo{person}{Richard Roberts}, \bibinfo{person}{Stefan
  Marr}, \bibinfo{person}{Michael Homer}, {and} \bibinfo{person}{James Noble}.}
  \bibinfo{year}{2017}\natexlab{}.
\newblock \showarticletitle{Toward Virtual Machine Adaption Rather than
  Reimplementation}. In \bibinfo{booktitle}{\emph{MoreVMs'17: 1st International
  Workshop on Workshop on Modern Language Runtimes, Ecosystems, and VMs at
  <Programming> 2017}}.
\newblock
\newblock
\shownote{Presentation.}


\bibitem[\protect\citeauthoryear{Roberts, Marr, Homer, and Noble}{Roberts
  et~al\mbox{.}}{2019}]%
        {roberts-and-co-ecoop-2019}
\bibfield{author}{\bibinfo{person}{Richard Roberts}, \bibinfo{person}{Stefan
  Marr}, \bibinfo{person}{Michael Homer}, {and} \bibinfo{person}{James Noble}.}
  \bibinfo{year}{2019}\natexlab{}.
\newblock \showarticletitle{Transient Typechecks are (Almost) Free}. In
  \bibinfo{booktitle}{\emph{{ECOOP}}}.
\newblock


\bibitem[\protect\citeauthoryear{Siek and Taha}{Siek and Taha}{2006}]%
        {Siek2006}
\bibfield{author}{\bibinfo{person}{Jeremy~G. Siek} {and} \bibinfo{person}{Walid
  Taha}.} \bibinfo{year}{2006}\natexlab{}.
\newblock \showarticletitle{Gradual typing for functional languages}. In
  \bibinfo{booktitle}{\emph{Seventh Workshop on Scheme and Functional
  Programming}}, Vol.~\bibinfo{volume}{Technical Report TR-2006-06}.
  \bibinfo{publisher}{University of Chicago}, \bibinfo{pages}{81--92}.
\newblock


\bibitem[\protect\citeauthoryear{Siek and Taha}{Siek and Taha}{2007}]%
        {Siek2007}
\bibfield{author}{\bibinfo{person}{Jeremy~G. Siek} {and} \bibinfo{person}{Walid
  Taha}.} \bibinfo{year}{2007}\natexlab{}.
\newblock \showarticletitle{Gradual Typing for Objects}. In
  \bibinfo{booktitle}{\emph{{ECOOP} 2007 - Object-Oriented Programming, 21st
  European Conference, Berlin, Germany, July 30 - August 3, 2007,
  Proceedings}}. \bibinfo{pages}{2--27}.
\newblock


\bibitem[\protect\citeauthoryear{Siek, Vitousek, Cimini, and Boyland}{Siek
  et~al\mbox{.}}{2015a}]%
        {XXXSiek2015}
\bibfield{author}{\bibinfo{person}{Jeremy~G. Siek}, \bibinfo{person}{Michael~M.
  Vitousek}, \bibinfo{person}{Matteo Cimini}, {and} \bibinfo{person}{John~Tang
  Boyland}.} \bibinfo{year}{2015}\natexlab{a}.
\newblock \showarticletitle{{Refined Criteria for Gradual Typing}}. In
  \bibinfo{booktitle}{\emph{1st Summit on Advances in Programming Languages
  (SNAPL 2015)}} \emph{(\bibinfo{series}{Leibniz International Proceedings in
  Informatics (LIPIcs)})}, \bibfield{editor}{\bibinfo{person}{Thomas Ball},
  \bibinfo{person}{Rastislav Bodik}, \bibinfo{person}{Shriram Krishnamurthi},
  \bibinfo{person}{Benjamin~S. Lerner}, {and} \bibinfo{person}{Greg Morrisett}}
  (Eds.), Vol.~\bibinfo{volume}{32}. \bibinfo{publisher}{Schloss
  Dagstuhl--Leibniz-Zentrum fuer Informatik}, \bibinfo{pages}{274--293}.
\newblock
\showISBNx{978-3-939897-80-4}
\showISSN{1868-8969}


\bibitem[\protect\citeauthoryear{Siek, Vitousek, Cimini, Tobin{-}Hochstadt, and
  Garcia}{Siek et~al\mbox{.}}{2015b}]%
        {monotonic2015}
\bibfield{author}{\bibinfo{person}{Jeremy~G. Siek}, \bibinfo{person}{Michael~M.
  Vitousek}, \bibinfo{person}{Matteo Cimini}, \bibinfo{person}{Sam
  Tobin{-}Hochstadt}, {and} \bibinfo{person}{Ronald Garcia}.}
  \bibinfo{year}{2015}\natexlab{b}.
\newblock \showarticletitle{Monotonic References for Efficient Gradual Typing}.
  In \bibinfo{booktitle}{\emph{European Symposium on Programming {(ESOP)}}}.
  \bibinfo{pages}{432--456}.
\newblock


\bibitem[\protect\citeauthoryear{St-Amour, Tobin-Hochstadt, and
  Felleisen}{St-Amour et~al\mbox{.}}{2012}]%
        {St-Amour:2012:OCO}
\bibfield{author}{\bibinfo{person}{Vincent St-Amour}, \bibinfo{person}{Sam
  Tobin-Hochstadt}, {and} \bibinfo{person}{Matthias Felleisen}.}
  \bibinfo{year}{2012}\natexlab{}.
\newblock \showarticletitle{{Optimization Coaching: Optimizers Learn to
  Communicate with Programmers}}. In \bibinfo{booktitle}{\emph{Proceedings of
  the ACM International Conference on Object Oriented Programming Systems
  Languages and Applications}} \emph{(\bibinfo{series}{OOPSLA'12})}.
  \bibinfo{publisher}{ACM}, \bibinfo{pages}{163--178}.
\newblock
\showISBNx{978-1-4503-1561-6}
\urldef\tempurl%
\url{https://doi.org/10.1145/2384616.2384629}
\showDOI{\tempurl}


\bibitem[\protect\citeauthoryear{Takikawa, Feltey, Greenman, New, Vitek, and
  Felleisen}{Takikawa et~al\mbox{.}}{2016}]%
        {Takikawa2016}
\bibfield{author}{\bibinfo{person}{Asumu Takikawa}, \bibinfo{person}{Daniel
  Feltey}, \bibinfo{person}{Ben Greenman}, \bibinfo{person}{Max~S. New},
  \bibinfo{person}{Jan Vitek}, {and} \bibinfo{person}{Matthias Felleisen}.}
  \bibinfo{year}{2016}\natexlab{}.
\newblock \showarticletitle{{Is Sound Gradual Typing Dead?}}. In
  \bibinfo{booktitle}{\emph{Proceedings of the 43rd Annual ACM SIGPLAN-SIGACT
  Symposium on Principles of Programming Languages}}
  \emph{(\bibinfo{series}{POPL'16})}. \bibinfo{publisher}{ACM},
  \bibinfo{pages}{456--468}.
\newblock
\showISBNx{978-1-4503-3549-2}


\bibitem[\protect\citeauthoryear{Vitousek, Kent, Siek, and Baker}{Vitousek
  et~al\mbox{.}}{2014}]%
        {reticPython2014}
\bibfield{author}{\bibinfo{person}{Michael~M. Vitousek},
  \bibinfo{person}{Andrew~M. Kent}, \bibinfo{person}{Jeremy~G. Siek}, {and}
  \bibinfo{person}{Jim Baker}.} \bibinfo{year}{2014}\natexlab{}.
\newblock \showarticletitle{Design and evaluation of gradual typing for
  {P}ython}. In \bibinfo{booktitle}{\emph{DLS'14, Proceedings of the 10th {ACM}
  Symposium on Dynamic Languages, part of {SPLASH} 2014, Portland, OR, USA,
  October 20-24, 2014}}. \bibinfo{pages}{45--56}.
\newblock


\bibitem[\protect\citeauthoryear{Vitousek, Siek, and Chaudhuri}{Vitousek
  et~al\mbox{.}}{2019}]%
        {vitousek-transient-arXive-2019}
\bibfield{author}{\bibinfo{person}{Michael~M. Vitousek},
  \bibinfo{person}{Jeremy~G. Siek}, {and} \bibinfo{person}{Avik Chaudhuri}.}
  \bibinfo{year}{2019}\natexlab{}.
\newblock \showarticletitle{Optimizing and Evaluating Transient Gradual
  Typing}.
\newblock \bibinfo{journal}{\emph{CoRR}}  \bibinfo{volume}{abs/1902.07808}
  (\bibinfo{year}{2019}).
\newblock
\showeprint[arxiv]{1902.07808}
\urldef\tempurl%
\url{http://arxiv.org/abs/1902.07808}
\showURL{%
\tempurl}


\bibitem[\protect\citeauthoryear{Vitousek, Swords, and Siek}{Vitousek
  et~al\mbox{.}}{2017}]%
        {Vitousek2017}
\bibfield{author}{\bibinfo{person}{Michael~M. Vitousek},
  \bibinfo{person}{Cameron Swords}, {and} \bibinfo{person}{Jeremy~G. Siek}.}
  \bibinfo{year}{2017}\natexlab{}.
\newblock \showarticletitle{{Big Types in Little Runtime: Open-world Soundness
  and Collaborative Blame for Gradual Type Systems}}. In
  \bibinfo{booktitle}{\emph{Proceedings of the 44th ACM SIGPLAN Symposium on
  Principles of Programming Languages}} \emph{(\bibinfo{series}{POPL'17})}.
  \bibinfo{publisher}{ACM}, \bibinfo{pages}{762--774}.
\newblock
\showISBNx{978-1-4503-4660-3}


\bibitem[\protect\citeauthoryear{W\"{o}\ss, Wirth, Bonetta, Seaton, Humer, and
  M\"{o}ssenb\"{o}ck}{W\"{o}\ss et~al\mbox{.}}{2014}]%
        {woss2014object}
\bibfield{author}{\bibinfo{person}{Andreas W\"{o}\ss},
  \bibinfo{person}{Christian Wirth}, \bibinfo{person}{Daniele Bonetta},
  \bibinfo{person}{Chris Seaton}, \bibinfo{person}{Christian Humer}, {and}
  \bibinfo{person}{Hanspeter M\"{o}ssenb\"{o}ck}.}
  \bibinfo{year}{2014}\natexlab{}.
\newblock \showarticletitle{{An Object Storage Model for the Truffle Language
  Implementation Framework}}. In \bibinfo{booktitle}{\emph{Proceedings of the
  2014 International Conference on Principles and Practices of Programming on
  the Java Platform: Virtual Machines, Languages, and Tools}}
  \emph{(\bibinfo{series}{PPPJ'14})}. \bibinfo{publisher}{ACM},
  \bibinfo{pages}{133--144}.
\newblock
\showISBNx{978-1-4503-2926-2}


\bibitem[\protect\citeauthoryear{W\"{u}rthinger, Wimmer, Humer, W\"{o}\ss,
  Stadler, Seaton, Duboscq, Simon, and Grimmer}{W\"{u}rthinger
  et~al\mbox{.}}{2017}]%
        {Wurthinger:2017:PPE}
\bibfield{author}{\bibinfo{person}{Thomas W\"{u}rthinger},
  \bibinfo{person}{Christian Wimmer}, \bibinfo{person}{Christian Humer},
  \bibinfo{person}{Andreas W\"{o}\ss}, \bibinfo{person}{Lukas Stadler},
  \bibinfo{person}{Chris Seaton}, \bibinfo{person}{Gilles Duboscq},
  \bibinfo{person}{Doug Simon}, {and} \bibinfo{person}{Matthias Grimmer}.}
  \bibinfo{year}{2017}\natexlab{}.
\newblock \showarticletitle{{Practical Partial Evaluation for High-performance
  Dynamic Language Runtimes}}. In \bibinfo{booktitle}{\emph{Proceedings of the
  38th ACM SIGPLAN Conference on Programming Language Design and
  Implementation}} \emph{(\bibinfo{series}{PLDI'17})}.
  \bibinfo{publisher}{ACM}, \bibinfo{pages}{662--676}.
\newblock
\showISBNx{978-1-4503-4988-8}


\bibitem[\protect\citeauthoryear{W\"{u}rthinger, Wimmer, W\"{o}\ss, Stadler,
  Duboscq, Humer, Richards, Simon, and Wolczko}{W\"{u}rthinger
  et~al\mbox{.}}{2013}]%
        {Wurthinger2013}
\bibfield{author}{\bibinfo{person}{Thomas W\"{u}rthinger},
  \bibinfo{person}{Christian Wimmer}, \bibinfo{person}{Andreas W\"{o}\ss},
  \bibinfo{person}{Lukas Stadler}, \bibinfo{person}{Gilles Duboscq},
  \bibinfo{person}{Christian Humer}, \bibinfo{person}{Gregor Richards},
  \bibinfo{person}{Doug Simon}, {and} \bibinfo{person}{Mario Wolczko}.}
  \bibinfo{year}{2013}\natexlab{}.
\newblock \showarticletitle{{One VM to Rule Them All}}. In
  \bibinfo{booktitle}{\emph{Proceedings of the 2013 ACM International Symposium
  on New Ideas, New Paradigms, and Reflections on Programming \& Software}}
  \emph{(\bibinfo{series}{Onward! 2013})}. \bibinfo{publisher}{ACM},
  \bibinfo{pages}{187--204}.
\newblock
\showISBNx{978-1-4503-2472-4}


\bibitem[\protect\citeauthoryear{W\"{u}rthinger, W\"{o}\ss{}, Stadler, Duboscq,
  Simon, and Wimmer}{W\"{u}rthinger et~al\mbox{.}}{2012}]%
        {Wurthinger:2012:SelfOptAST}
\bibfield{author}{\bibinfo{person}{Thomas W\"{u}rthinger},
  \bibinfo{person}{Andreas W\"{o}\ss{}}, \bibinfo{person}{Lukas Stadler},
  \bibinfo{person}{Gilles Duboscq}, \bibinfo{person}{Doug Simon}, {and}
  \bibinfo{person}{Christian Wimmer}.} \bibinfo{year}{2012}\natexlab{}.
\newblock \showarticletitle{{Self-Optimizing AST Interpreters}}. In
  \bibinfo{booktitle}{\emph{Proceedings of the 8th Dynamic Languages
  Symposium}} \emph{(\bibinfo{series}{DLS'12})}. \bibinfo{pages}{73--82}.
\newblock
\showISBNx{978-1-4503-1564-7}
\urldef\tempurl%
\url{https://doi.org/10.1145/2384577.2384587}
\showDOI{\tempurl}


\end{thebibliography}
\end{document}